\documentclass[sigconf, nonacm=true]{acmart}

\usepackage{booktabs} 

\usepackage{amsmath,amssymb,amsfonts}
\usepackage{algorithmic}
\usepackage{graphicx}
\usepackage[T1]{fontenc}
\usepackage[toc,page]{appendix}
\setcopyright{rightsretained}

\usepackage{makecell, tabularx, pifont}
\usepackage{multirow}

\begin{document}
\title{All One Needs to Know about Fog Computing and Related Edge Computing Paradigms}
\subtitle{A Complete Survey}
\subtitlenote{A complete list of conferences, journals, and magazines that publish state-of-the-art research papers on fog computing and its related edge computing paradigms is available at {\footnotesize https://anrlutdallas.github.io/resource/projects/fog-computing-conferences.html}. We have included papers from the above list in this survey.}
\subtitlenote{The data (categories and features/objectives of the papers) of this survey are available at {\footnotesize https://github.com/ashkan-software/fog-survey-data}}

\author{Ashkan~Yousefpour}
\authornote{ashkan.y@berkeley.edu}
\affiliation{%
  \institution{UC Berkeley}
}

\author{Caleb~Fung}
\affiliation{%
  \institution{UT Dallas}
}

\author{Tam~Nguyen}
\affiliation{%
  \institution{Georgia Tech}
}

\author{Krishna~Kadiyala}
\affiliation{%
  \institution{UT Dallas}
}

\author{Fatemeh~Jalali}
\affiliation{%
  \institution{IBM Research}
}

\author{Amirreza~Niakanlahiji}
\affiliation{%
  \institution{UNC Charlotte}
}

\author{Jian~Kong}
\affiliation{%
  \institution{UT Dallas}
}

\author{Jason~P.~Jue}
\affiliation{%
  \institution{UT Dallas}
}

\renewcommand{\shortauthors}{A. Yousefpour et al.}

\begin{abstract}
With the Internet of Things (IoT) becoming part of our daily life and our environment, we expect rapid growth in the number of connected devices. IoT is expected to connect billions of devices and humans to bring promising advantages for us. With this growth, fog computing, along with its related edge computing paradigms, such as multi-access edge computing (MEC) and cloudlet, are seen as promising solutions for handling the large volume of security-critical and time-sensitive data that is being produced by the IoT. In this paper, we first provide a tutorial on fog computing and its related computing paradigms, including their similarities and differences. Next, we provide a taxonomy of research topics in fog computing, and through a comprehensive survey, we summarize and categorize the efforts on fog computing and its related computing paradigms. Finally, we provide challenges and future directions for research in fog computing.
\end{abstract}

%
%

\keywords{
Fog Computing, Edge Computing, Cloud Computing, Internet of Things (IoT), Cloudlet, Mobile Edge Computing, Multi-access Edge Computing, Mist Computing
}

\maketitle


\section{Introduction} \label{introduction}
In today's information technology age, data is the main commodity, and possessing more data typically generates more value in data-driven businesses. According to the International Data Corporation (IDC), the amount of digital data generated surpassed 1 zettabyte in 2010 \cite{intro-IDC}. Furthermore, 2.5 exabytes of new data is generated each day since 2012 \cite{intro-Harvard}. Cisco estimates that there will be around 50 billion connected devices by 2020 \cite{cisco2020}. These connected devices constitute the Internet of Things (IoT) and possibly generate a massive amount of data. With this astronomical amount of data, the current mobile network architectures will have trouble managing the momentum and magnitude of data. In current implementations of cloud-based applications, most data that needs storage, analysis, and decision making is sent to the data centers in the cloud \cite{ravandi2017self}. 

As the data velocity and volume increases, moving the big data from IoT devices to the cloud might not be efficient, or might be even infeasible in some cases due to bandwidth constraints. On the other hand, as time-sensitive and location-aware applications emerge (such as patient monitoring, real-time manufacturing, self-driving cars, flocks of drones, or cognitive assistance), the distant cloud will not be able to satisfy the ultra-low latency requirements of these applications, provide location-aware services, or scale to the magnitude of the data that these applications produce \cite{Cloud-is-Not-Enough}. Moreover, in some applications, sending the data to the cloud may not be a feasible solution due to privacy concerns. 

In order to address the issues of high-bandwidth, geographically-dispersed, ultra-low latency, and privacy-sensitive applications, there is a quintessential need for a computing paradigm that takes place closer to connected devices. Fog computing has been proposed by both industry and academia \cite{29, fog-cisco} to address the above issues and to quench the need for computing paradigm closer to connected devices. Fog computing bridges the gap between the cloud and IoT devices by enabling computing, storage, networking, and data management on the network nodes within the close vicinity of IoT devices. Therefore, computation, storage, networking, decision making, and data management occur along the path between IoT devices and the cloud, as data moves to the cloud from the IoT devices. Other similar computing paradigms to fog computing such as edge computing, mist computing, cloud of things, and cloudlets, have been proposed by the research community to address the mentioned issues. In this survey, we compare fog computing with other related computing paradigms, and argue that fog computing is a more general form of computing, mainly due to its comprehensive definition scope and flexibility. 

\begin{figure}[!t]
\includegraphics[width=\linewidth]{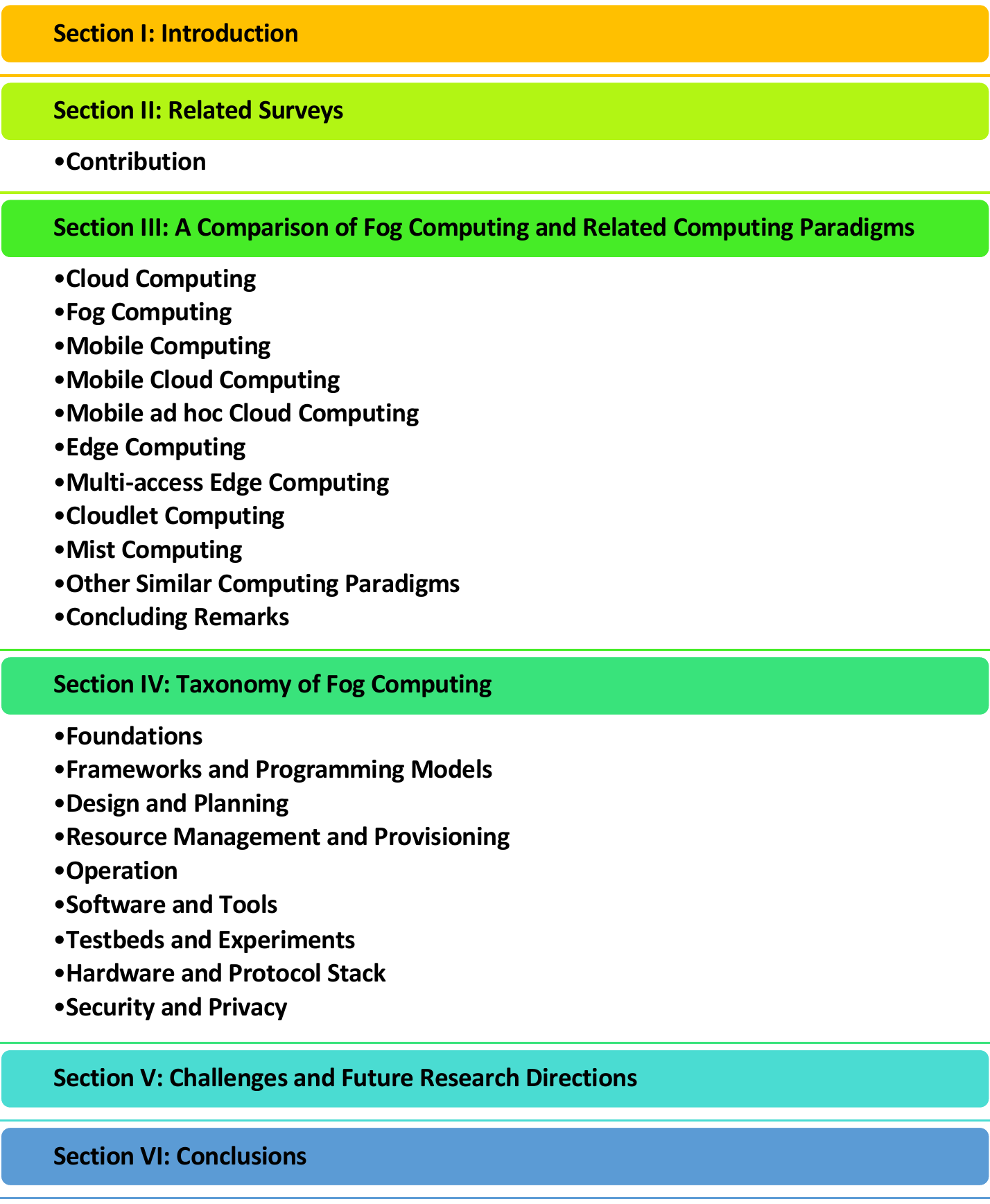}
\caption{The structure of the survey.}
\label{fig:TOC}
\end{figure}

In this article, we present a comprehensive survey on fog computing, and discuss how fog computing can meet the growing demand of applications with strict latency, privacy, and bandwidth requirements. A comparison of the related survey papers in the area of fog computing is included in Section \ref{related}. To gain a thorough understanding of fog computing, in Section \ref{comparison}, we will first look at cloud computing, then discuss how fog computing extends cloud computing to address the above issues of cloud, and finally, compare fog computing to other similar computing paradigms. Next, in Section \ref{taxonomy}, we describe our taxonomy of research topics in fog computing. Later, in a comprehensive survey, we summarize and categorize the efforts on fog computing and its related computing paradigms. In Section \ref{future-section}, we present the challenges and limitations in the fog computing area and provide future directions and potential starting points for those challenges. Finally, Section \ref{conclusion} concludes the paper. Fig. \ref{fig:TOC} shows the structure of the survey and a reading map for the reader.

\section{Related Surveys} \label{related}

There are are some existing related studies in the area of fog computing that have attempted to provide a survey of the papers in the field of fog computing, edge computing, or MEC. The authors in \cite{61} present a comprehensive review of current literature in fog computing with a focus on architectures and algorithms in fog systems. They further sketch the prospects of fog computing in terms of emerging technologies with a focus on Tactile Internet. Cihat et al. \cite{78} state the importance of cooperation between edge and cloud computing, and motivate how edge computing can benefit from Software Defined Networking (SDN). They note the technical challenges in edge computing and propose using SDN as a solution for implementing edge computing infrastructure. The authors survey publications primarily on edge computing and SDN to support their argument and give future directions for SDN developments. The authors in \cite{mukherjee2018survey} compile a comprehensive survey of recent efforts in fog-enabled network architectures, and provide various network applications of fog computing.

The recent survey in \cite{fog-smart-city-survey} focuses on connectivity and device configuration aspects of the fog computing and identifies major features that fog computing platforms need to build infrastructure for smart city applications. They further review existing approaches that have been proposed to tackle the challenges in fog computing for building such smart city infrastructure. Comparably, the authors of \cite{li2018edge} focus on architecture design and system management of edge computing to provide a detailed and focused survey in the edge computing field. They also characterize fog and edge computing by comparing a list of related computing concepts, including peer-to-peer computing, mobile grid computing, and mobile crowd computing. The authors in \cite{86} take a closer look at fog-assisted IoT applications, discuss security and privacy challenges in fog computing, and review and analyze promising techniques to resolve security and privacy issues in fog-assisted IoT applications. 

There are a number of surveys in the area of MEC that also discuss similar concepts to fog computing and summarize papers applicable to fog computing research. The survey in \cite{MEC-orchestration} introduces a survey on MEC and focuses on the fundamental key enabling technologies in MEC. The paper also analyzes the MEC reference architecture, overviews the current standardization activities, and introduces main deployment scenarios. Similarly, the survey in \cite{mao2017survey} provides a survey of the recent state of MEC research with a focus on joint radio and computation resource management. 

\subsection{Contribution}

Different from the mentioned surveys, the contribution of this paper is three-fold: (1) We provide a detailed tutorial on fog computing, how it is defined, and how it is related to or different from other similar computing paradigms, such as cloud computing, cloudlets, edge computing, and MEC (2) We propose an exhaustive taxonomy of research topics in the area of fog computing, and present a comprehensive survey on fog computing.\footnote{We compiled a comprehensive list of conferences, journals, and magazines that publish state-of-the-art research papers on fog computing and its related edge computing paradigms The list is available at \\{\footnotesize https://anrlutdallas.github.io/resource/projects/fog-computing-conferences.html}} (3) We have compiled a list of challenges and future directions for research in fog computing.

\section{A Comparison of Fog Computing and Related Computing Paradigms} \label{comparison}
This section focuses on the comparison of fog computing and related computing paradigms to demonstrate the value of fog computing in a variety of use cases. Moreover, this section provides a better understanding of how these computing paradigms can benefit the current and future landscape of connected devices. We compare fog computing with cloud computing as well as other related computing paradigms and summarize this information in Tables \ref{tab:comparison-table-1} and \ref{tab:comparison-table-2}. 

\subsection{Cloud Computing}
Cloud computing has been instrumental in expanding the reach and capabilities of computing, storage, and networking infrastructure to the applications. The National Institute of Standards and Technology (NIST) defines cloud computing as a model that promotes ubiquitous, on-demand network access to shared computing resources \cite{cloud-definition}. Cloud data centers are large pools of highly accessible virtualized resources that can be dynamically reconfigured for a scalable workload; this reconfigurability is beneficial for clouds services that are offered with a pay-as-you-go cost model \cite{cloud-pool}. The pay-as-you-go cost model allows users to conveniently access remote computing resources and data management services, while only being charged for the amount of resources they use. Cloud providers, such as Google, IBM, Microsoft, and Amazon provide and provision large data centers to host these cloud-based resources.

\subsubsection{Cloud Services}
Cloud offers infrastructure, platform, and software as services (IaaS, PaaS, SaaS). Application developers can use a variety of these services depending on the needs of the applications they develop. Infrastructure as a service (IaaS) allows cloud consumers to directly access IT infrastructures for processing, storage, and networking resources \cite{cloud-services}. Suppose Sam wants to set up a high-tech agricultural system that utilizes IoT devices to monitor the condition of crops. Sam contacts a cloud provider and acquires an IaaS for development of his system. Sam now can configure the IaaS (often offered as a standalone VM) in terms of hardware and corresponding software for his need. Control over infrastructure (IaaS) allows Sam to customize hardware configuration, such as the number of CPU cores and RAM capacity, in addition to systems-level software. Sam can obtain an IaaS from Amazon Web Services (AWS), Microsoft Azure, or Google Compute Engine (GCE).

On the other hand, platform as a service (PaaS) allows cloud consumers to develop software and fully supports software lifecycle -- often with the help of a middleware -- for software management and configuration. If Sam does not need to configure the infrastructure of the cloud, managing and configuration of hardware and software may detract from the productivity of Sam's business. Now, Sam could consider using PaaS offered by Apache Stratos, Azure App Services, or Google App Engine for his business. PaaS manages the underlying low-level processes and allows Sam to focus on managing software for his IoT-specific interactions. Moreover, PaaS providers often include tools for convenient management of databases and scaling applications. 

Now suppose Sam is willing to spend more money and likes to get full software packages, and he does not want to take care of software issues, such as database scalability, socket management, etc. Software as a service (SaaS) provides Sam an environment to centrally host his applications and removes the need for him to install software manually. Sam's client software now can be hosted on Google Apps or as a Web application. 

\begin{figure}[!t]
\includegraphics[width=\linewidth]{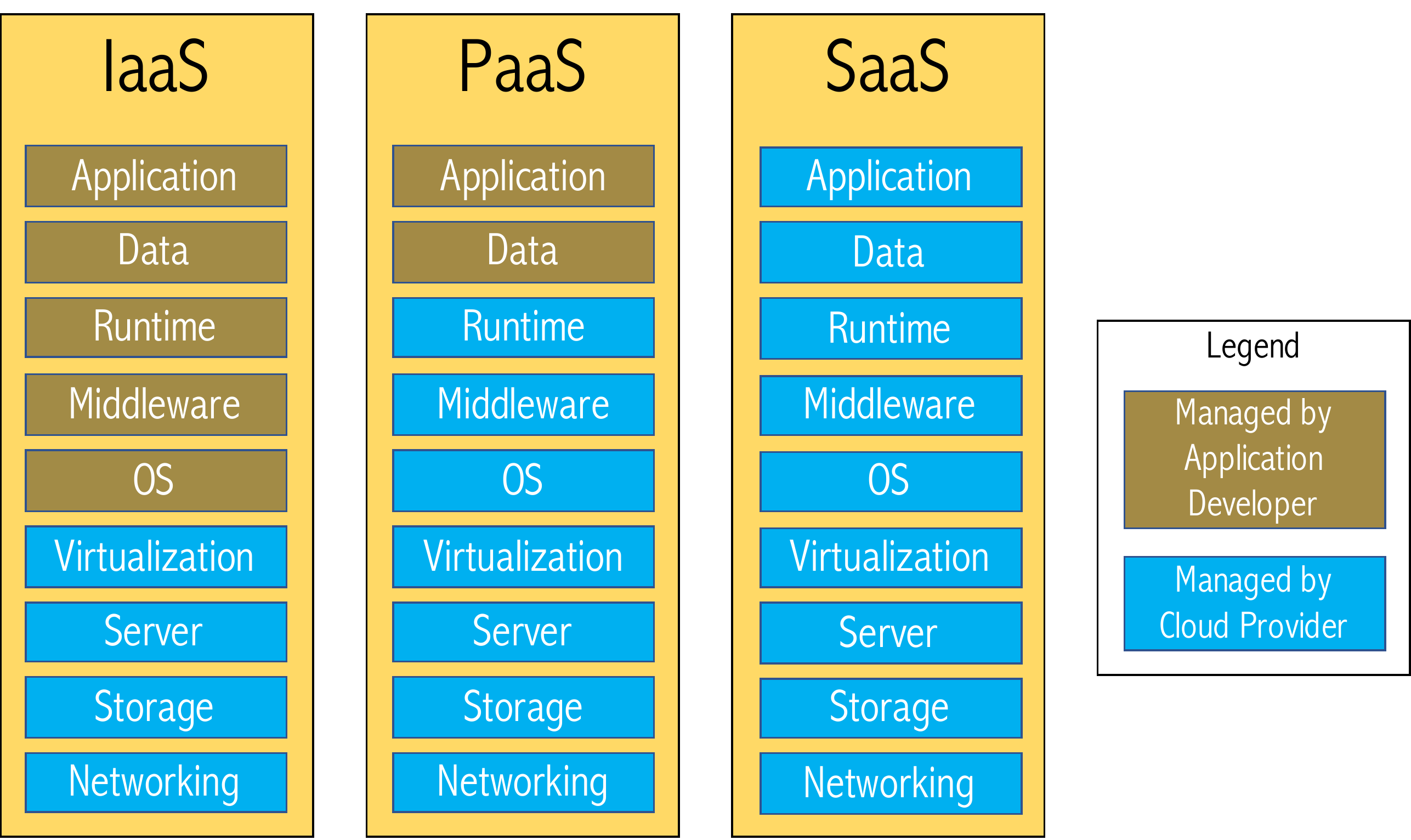}
\caption{Common cloud service models and their classifications relative what portion of the application stack is managed by cloud providers.}
\label{fig:cloud-services}
\end{figure}

As demonstrated by these examples, cloud services can be utilized for distinct use cases for a variety of end users. Figure \ref{fig:cloud-services} illustrates the relationship among IaaS, PaaS, and SaaS with the underlying cloud infrastructure, and illustrates what portion of the application stack is managed by cloud providers. 

\subsubsection{Cloud Resource Provisioning}
\begin{figure}[!t]
\centering
\includegraphics[width=\linewidth]{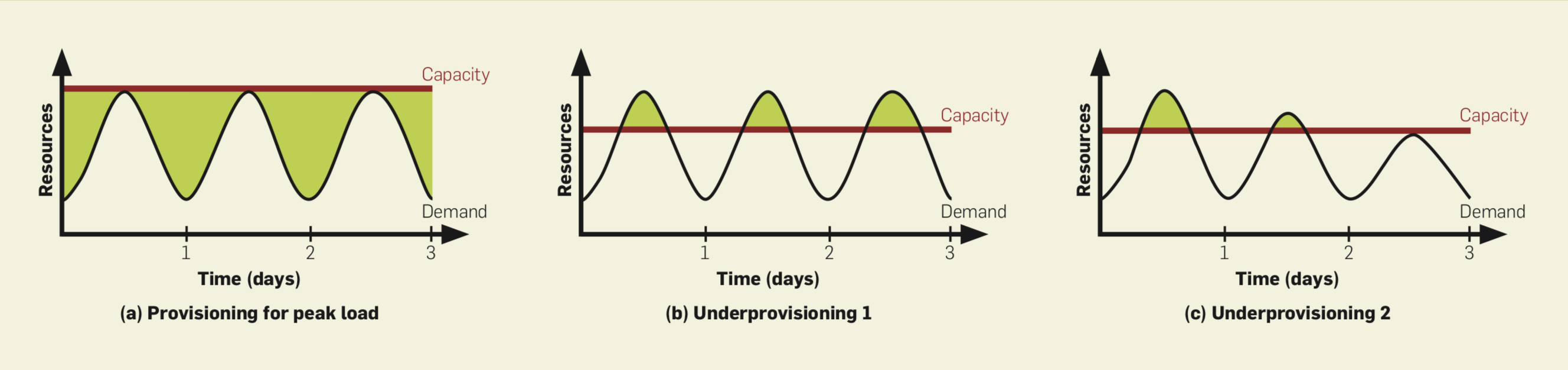}
\caption{Cloud provisioning is done based on the application demand \cite{cloud-provisioning}.}
\label{fig:cloud-provisioning}
\end{figure}

Since the demand for cloud resources is not fixed and can change over time, setting a fixed amount of resources results in either over-provisioning or under-provisioning, as depicted in Fig. \ref{fig:cloud-provisioning}. A foundation of cloud computing is based on provisioning only the required resources for the demand. This includes the use of virtualization for on-demand application deployment, and the use of resource provisioning to manage hardware and software in cloud data centers. Provisioning resources is an important topic in cloud computing that is widely explored. Since it is difficult to estimate service usage from tenants, most cloud providers have a pay-as-you-go payment scheme. As a result, providers can be more flexible on how to provision resources, and clients only pay for the amount of resources they actually use.

\subsubsection{Types of Cloud}
There are four types of cloud deployments: private cloud, community cloud, public cloud, and hybrid cloud \cite{cloud-definition}. Private clouds are designed for use by a singular entity and ensure high privacy and configurability. Private clouds are a good choice for organizations that require an infrastructure for their applications. This type of deployment is similar to traditional company-owned server farms and often do not benefit from a pay-as-you-go cost model. Community clouds are used by a community of users, and the infrastructure is shared between several organizations. A community cloud results in decentralized ownership of the cloud by multiple organizations within the community without relying on a large cloud vendor for the IT infrastructure. Public clouds are the typical model of cloud computing, where the cloud services are offered by cloud service providers, such as Amazon, IBM, Google, Microsoft, etc. Public clouds are generally more popular, easy-to-maintain, and cost-effective compared to private clouds. In contrast to private clouds, public clouds may benefit from the pay-as-you-go pricing model. However, public clouds do not always offer users complete customization of hardware, middleware, network, and security settings. Hybrid clouds are simply a combination of the cloud types mentioned above. Hybrid clouds allow users to have finer control over virtualized infrastructure, and combining the capabilities from different types of cloud deployments is accomplished through standardized or proprietary technology \cite{cloud-hybrid}.

The cloud computing paradigm was initially established to allow users to access a pool of computing resources for ubiquitous computing. Even though cloud computing has helped bring forth accessible computing, the time required to access cloud-based applications may be too high and may not be practical for some mission-critical applications, or applications with ultra-low latency requirements. Also, the rapid growth in the amount of data generated at the network edge by an increasing number of connected devices requires cloud resources to be closer to where the data is generated. Greater demand for high-bandwidth, geographically-dispersed, low-latency, and privacy-sensitive data processing has emerged -- a quintessential need for computing paradigms that take place closer to connected devices and that support low-latency, high-bandwidth, decentralized applications. To address these needs, fog computing has been proposed by both industry and academia \cite{29, fog-cisco}. In order to provide a detailed comparison among fog computing related paradigms, we introduce various computing paradigms, starting with fog computing.
\subsection{Fog Computing}
Fog computing bridges the gap between the cloud and end devices (e.g., IoT nodes) by enabling computing, storage, networking, and data management on network nodes within the close vicinity of IoT devices. Consequentially, computation, storage, networking, decision making, and data management not only occur in the cloud, but also occur along the IoT-to-Cloud path as data traverses to the cloud (preferably close to the IoT devices). For instance, compressing the GPS data can happen at the edge before transmission to the cloud in Intelligent Transportation Systems (ITS) \cite{316}. Fog computing is defined by the OpenFog Consortium \cite{29} as ``a horizontal system-level architecture that distributes computing, storage, control and networking functions closer to the users along a cloud-to-thing continuum.'' The ``horizontal'' platform in fog computing allows computing functions to be distributed between different platforms and industries, whereas a vertical platform promotes siloed applications \cite{fog-horizontal}. A vertical platform may provide strong support for a single type of application (silo), but it does not account for platform-to-platform interaction in other vertically focused platforms. In addition to facilitating a horizontal architecture, fog computing provides a flexible platform to meet the data-driven needs of operators and users. Fog computing is intended to provide strong support for the Internet of Things.

\subsubsection{Fog vs. Cloud}
A common example that is often used to distinguished fog and cloud computing is whether latency-sensitive applications can be supported while maintaining satisfactory quality of service (QoS). Fog nodes can be placed close to IoT source nodes, allowing latency to be noticeably reduced compared to traditional cloud computing. While this example gives an intuitive motivation for fog, latency-sensitive applications are only one of the many applications that warrant the need for fog computing. Nodes in fog computing are generally deployed in less centralized locations compared to centralized cloud data centers. Fog nodes are wide-spread and geographically available in large numbers. In fog computing, security must be provided at the edge or in the dedicated locations of fog nodes, as opposed to the centrally-developed security mechanisms in dedicated buildings for cloud data centers. The decentralized nature of fog computing allows devices to either serve as fog computing nodes themselves (e.g. a car acts as a fog node for onboard sensors) or use fog resources as the clients of the fog. 

The majority of differences between cloud and fog computing are attributed to the scale of hardware components associated with these computing paradigms. Cloud computing provides high availability of computing resources at relatively high power consumption, whereas fog computing provides moderate availability of computing resources at lower power consumption \cite{67}. Cloud computing typically utilizes large data centers, whereas fog computing utilizes small servers, routers, switches, gateways, set-top boxes, or access points. Since hardware for fog computing occupies much less space than that of cloud computing, hardware can be located closer to users. Fog computing can be accessed through connected devices from the edge of the network to the network core, whereas cloud computing must be accessed through the network core. Moreover, continuous Internet connectivity is not essential for the fog-based services to work. That is, the services can work independently with low or no Internet connectivity and send necessary updates to the cloud whenever the connection is available. Cloud computing, on the other hand, requires devices to be connected when the cloud service is in progress. 

\begin{figure*}[ht]
\centering
\includegraphics[width=0.8\textwidth]{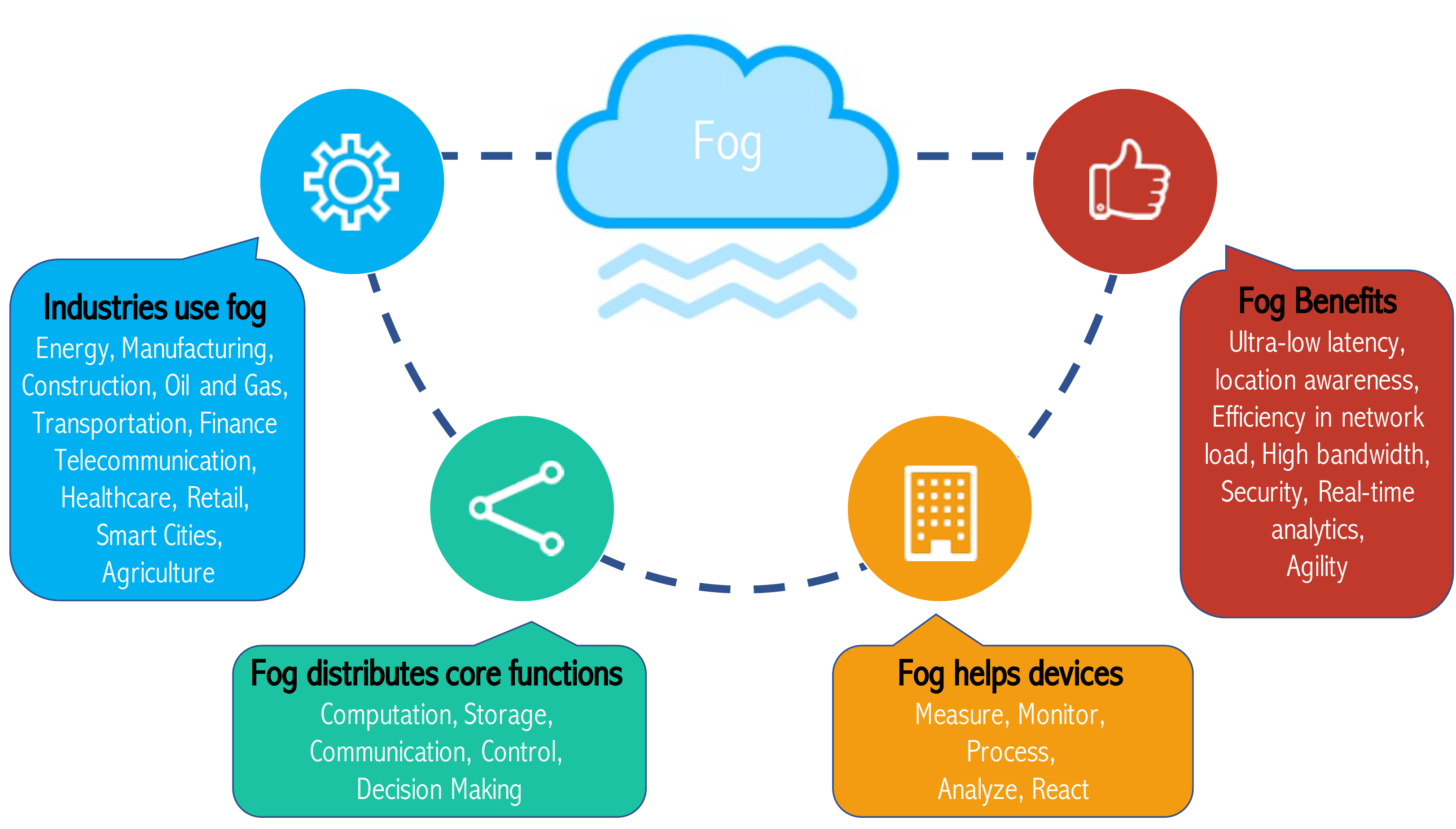}
\caption{Fog brings several benefits for the application developers, applications, and different industries by distributing the core functions.}
\label{fig:fog-benefits}
\end{figure*}

Fog helps devices measure, monitor, process, analyze, and react, and distributes computation, communication, storage, control, and decision making closer to IoT devices \cite{29} (refer to fig. \ref{fig:fog-benefits}). Many industries could use fog to their benefit: energy, manufacturing, transportation, healthcare, smart cities, to mention a few. 

\subsubsection{Fog-Cloud Federation}
There are clear differences and trade-offs between cloud and fog computing, and one might ask which one to choose. However, fog and cloud complement each other; one cannot replace the need of the other. By coupling cloud and fog computing, the services that connected devices use can be optimized even further. Federation between fog and cloud allows enhanced capabilities for data aggregation, processing, and storage. For instance, in a stream processing application, the fog could filter, preprocess, and aggregate traffic streams from source devices, while queries with heavy analytical processing, or archival results could be sent to the cloud. An orchestrator could handle the cooperation between cloud and fog. Specifically, a fog orchestrator could provide an interoperable resource pool, deploy and schedule resources to application workflows, and control QoS \cite{62}. Through the use of SDN, fog service providers will have greater control over how the network is configured with a large number of fog nodes that transfer data between the cloud and IoT devices.

\subsubsection{Fog RAN}
Fog computing can be integrated into mobile technologies in the form of radio access networks (RAN), to form what is referred to as fog RAN (F-RAN). Computing resources on F-RANs may be used for caching at the edge of the network, which enables faster retrieval of content and a lower burden on the front-haul. F-RAN can be implemented through 5G related mobile technologies \cite{fran-main}. On the other hand, cloud RAN (C-RAN) provides centralized control over F-RAN nodes. C-RAN takes advantage of virtualization, and decouples the base stations within a cell of the mobile network from its baseband functions by virtualizing those functions \cite{C-RAN}. In C-RAN a large number of low-cost Remote Radio Heads (RRHs) are randomly deployed and connected to the Base Band Unit (BBU) pool through the front-haul links. Both F-RAN and C-RAN are suited for mobile networks with base stations and are candidates for 5G deployments. Also, the use of F-RAN and C-RAN brings a more energy efficient form of network operation. We encourage the motivated reader to refer to reference \cite{69} for more information about F-RAN.

\begin{figure}[!t]
\includegraphics[width=0.5\textwidth]{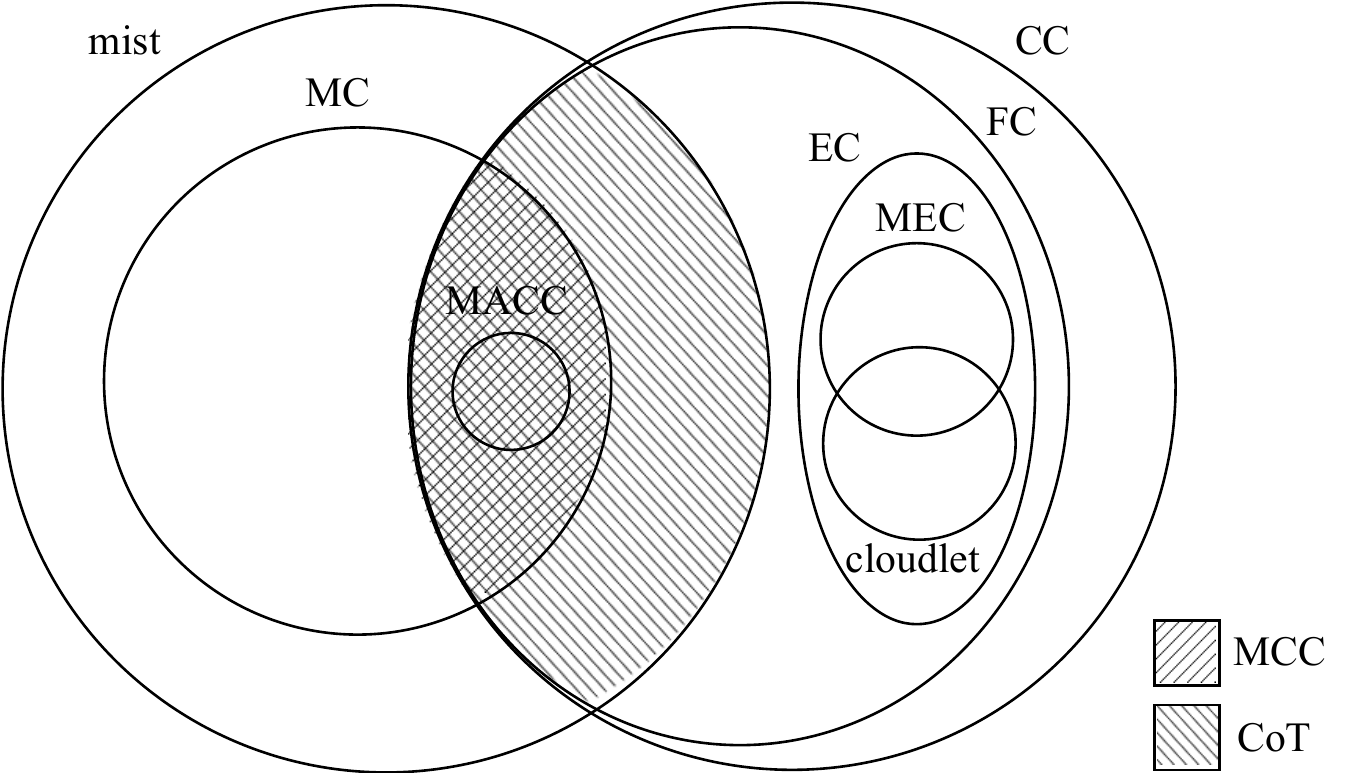}
\caption{A classification of scope of fog computing and its related computing paradigms. (Intersection of cloud computing and mist computing is Cloud of Things, and intersection of mobile computing and cloud computing is mobile cloud computing.)}
\label{fig:venn-diagram}
\end{figure}

Figure \ref{fig:venn-diagram} shows a classification of computing paradigms related to fog computing and their overlap in terms of their scope. The figure illustrates our comparison of fog computing and its related computing paradigms. Table \ref{tab:acronyms} lists the acronyms used for this figure and in the paper. We discuss the related computing paradigms in the order of their trend and show how some paradigms resulted in the emergence of others. 

\subsection{Mobile Computing}
The advancement in fog and cloud computing is influenced by the groundwork set forth by the development of mobile computing. Mobile computing, or nomadic computing, is when computing is performed via mobile, portable devices, such as laptops, tablets, or mobile phones. Mobile computing can be utilized to create pervasive context-aware applications, such as location-based reminders.

At the heart of mobile computing is the vision for adaptation in an environment of low processing power and intermittent, sparse network connectivity. The peak of mobile computing technologies precedes cloud computing. A large number of fundamental challenges (such as user mobility, network heterogeneity, and low bandwidth) in mobile computing have been addressed in the literature before 2000. These issues have been addressed by advancements such as robust caching, transmission hardware and protocols, and compression algorithms \cite{MC-forman}. Due to the evolving requirements of connected consumer devices, mobile computing alone is not suitable for many recent computing challenges. 

With fog and cloud computing, computation is no longer tied to a local network; fog and cloud computing expand the scale and scope of mobile computing. The only type of hardware that mobile computing requires are mobile devices, which can be connected through Bluetooth, WiFi, ZigBee, and other cellular protocols. In contrast, fog and cloud computing require more resource-rich hardware with virtualization capabilities. Security in mobile computing must be provided on the mobile device itself. Compared to fog and cloud computing, mobile computing is more resource-constrained, but in recent years, advancements in mobile hardware and wireless protocols have significantly reduced this gap. 

The power of mobile computing is from its distributed computing architecture. Distributed applications benefit from this architecture because mobile machines do not need a centralized location to operate. Mobile computing, however, comes with many drawbacks such as poor-resource constraints, the balance between autonomy and interdependence (prevalent in all distributed architectures), communication latency, and the need for mobile clients to efficiently adapt to changing environments \cite{MC-challenges}. These drawbacks often make mobile computing unsuitable for current applications that require low-latency or robustness, or that need large amounts of data to be generated, processed, and stored on devices.

\begin{table}
\begin{minipage}{0.96\linewidth}
    \renewcommand{\arraystretch}{1}
    
    \centering
    \caption{A list of computing paradigm acronyms used in this section.}
    \label{tab:acronyms}
    \begin{tabularx}{\textwidth}{lXlX}
        IoT & Internet of Things     & CC   & Cloud Computing \\ 
        MC  & Mobile Computing       & FC   & Fog Computing \\
        EC  & Edge Computing         & MEC  & \makecell[l]{Multi-access \\ Edge Computing} \\
        MCC & Mobile Cloud Computing & MACC & Mobile ad hoc Cloud Computing \\
        CoT  & Cloud of Things  & mist & Mist Computing
    \end{tabularx}
\end{minipage}
\end{table}

\subsection{Mobile Cloud Computing}
As cloud computing matured, it became a valuable complement to mobile computing. This combination resulted in mobile cloud computing (MCC), which is defined as an infrastructure where both the data storage and data processing occur outside of the mobile device, bringing mobile computing applications to not just smartphone users but a much broader range of mobile subscribers \cite{MCC-survey}. NIST extends this definition to include mobile devices: cloud computing is the synergy between IoT devices, mobile devices, and cloud computing that enables data-intensive and CPU-intensive applications for IoT environments \cite{MCC-NIST}. Some of these applications in MCC include crowdsourcing, healthcare, sensor data processing (such as optical character recognition and image processing), and task offloading \cite{MCC-crowdsourcing, MCC-healthcare}. Mobile applications can be partitioned at runtime so that computationally intensive components of the application can be handled through adaptive offloading \cite{MCC-application}. 

In mobile cloud computing, resource contained mobile devices can leverage resource-rich cloud services. MCC shifts the majority of computation from mobile devices to the cloud. MCC helps to run computation-intensive applications and to increase the battery life of mobile devices. MCC shares a blend of capabilities and characteristics in mobile computing and cloud computing. By adopting a combination of mobile computing and cloud computing objectives, high availability of computing resources is present in MCC as opposed to resource-constrained mobile computing. This allows for the emergence of high computation applications, such as mobile augmented reality. Also, the availability of cloud-based services in MCC is considerably higher than that of mobile computing. Similar to cloud computing and fog computing, MCC relies on cloud services for operating high-computation services. Computation in MCC can also be operated by mobile devices. Similar to cloud computing, security in MCC must be provisioned in both mobile devices and in the cloud. The authors in \cite{mcc-android-app} design and implement an Android app that helps drivers find parking space availability using MCC.

MCC also suffers from the same limitations of mobile computing and cloud computing. First, while a centralized architecture in MCC is great for sharing a pool of computation resources, this may not be well suited for applications where pervasiveness of devices is desired. Second, since both cloud computing and MCC require cloud-based services, and as access to those services is through the network core by WAN connection, applications running on these platforms require connection to the Internet all the time. MCC shifts the majority of computation from mobile devices to the cloud, and this introduces connectivity challenges that were not present in mobile computing. Finally, offloading computation to the cloud causes the latency to be relatively high for delay-sensitive applications. The authors in \cite{mcc-food} design a food recognition system based on MCC that distributes the data analytics between the mobile devices and the servers in the cloud. Mobile phones can perform light-weight computation on food images for food recognition, which allows the system to overcome some inherent limitations of traditional MCC paradigm, such as high latency and low battery life of mobile devices.

\subsection{Mobile ad hoc Cloud Computing}
Despite the pervasive nature of MCC, this computing paradigm is not always suitable for scenarios in which there is a lack of infrastructure or a centralized cloud. An ad hoc mobile network consists of nodes that form a temporary, dynamic network through routing and transport protocols; it is the most decentralized form of a network \cite{MACC-hubaux}. Mobile devices in an ad hoc mobile network form a highly dynamic network topology; the network formed by the mobile devices is highly dynamic and must accommodate for devices that continuously join or leave the network. Ad hoc mobile devices can form clouds that can be used for networking, storage, and computing. MACC could include use cases such as disaster relief, group live video streaming, and unmanned vehicular systems. 

\subsubsection{MACC vs. Cloud Computing}
Mobile ad hoc cloud computing (MACC) is fundamentally different from cloud computing, mainly due to the ad hoc nature of the resources. MACC involves mobile devices that function as data providers, storage, and processing devices. Mobile devices in a mobile ad hoc cloud network are also responsible for routing traffic among themselves, because of the lack of network infrastructure. By pooling local mobile resources to form an ad hoc cloud, MACC offers reasonably high computation. These attributes differ from the target users, architecture, and connectivity in cloud computing. In a study done by researchers of Carnegie Mellon University \cite{drolia2013case}, there is a tradeoff between offloading computation to distant clouds (labeled as ``infrastructure cloud'') versus running them on nearby mobile devices (labeled as ``mobile edge-clouds,'' but in this paper we call them ``mobile ad hoc clouds''). The authors compare the performance of executing some applications on a traditional infrastructure cloud versus running them in mobile ad hoc clouds. 

\subsubsection{MACC vs. MCC}
MACC is also different from MCC in the hardware, service access method, and the distance from users, since computation is done on mobile devices in MACC, whereas it is far from mobile devices in MCC. MACC only requires mobile devices to operate, whereas MCC requires large-scale data centers used for cloud computing in addition to mobile devices. This results in high computation power, but also higher latency in MCC. Security in MACC must be provided only in mobile devices, but ensuring trust may be challenging in MACC without a secure collaboration framework. Finally, in MACC, services are only accessed through mobile devices that are connected via Bluetooth, WiFi, and other cellular protocols.

\subsubsection{MACC vs. Fog}
Although fog computing can be performed across a variety of resource-rich and resource deficient devices, mobile ad hoc cloud computing is better suited for highly decentralized, dynamic network topologies in which Internet connection is not guaranteed. Connected devices in MACC are more decentralized compared to fog computing, and this allows the devices to form a more dynamic network in places of sparely connected devices or a constantly changing network. An example of this is an ad hoc network for peer-to-peer file sharing \cite{MACC-application}. 

\subsubsection{MACC vs. MANET}
A similar concept to MACC is a mobile ad hoc network (MANET). MANETs consist of mobile host devices that are connected to each other with single hop without base stations \cite{MANET-definition}. MANET devices form dynamic networks but do not necessarily form a cloud. In other words, the computing or storage resource pools are not necessary for MANETs. However, many solutions to MANETS, such as redundancy and broadcasting, can be applied to MACC. In a resource-constrained environment, peers may want to pool resources together to achieve a computationally demanding task that may not be feasible on a single mobile device. A use case for this is an unmanned vehicular system that consists of multiple unmanned vehicles and traffic devices.

\subsection{Edge Computing}
Similar to how MCC extends the capabilities of mobile devices, edge computing also enhances the management, storage, and processing power of data generated by connected devices. Unlike MCC, edge computing is located at the edge of the network close to IoT devices; note that the edge is not located on the IoT devices, but as close as one hop to them. It is worth noting that the edge can be more than one hop away from IoT devices in the local IoT network. OpenEdge Computing defines edge computing as computation done at the edge of the network through small data centers that are close to users \cite{EC-definition}. The original vision for edge computing is to provide compute and storage resources close to the user in open standards and ubiquitous manner \cite{EC-definition}. Edge computing is a crucial computing paradigm in the current landscape of IoT devices; it integrates the IoT devices with the cloud by filtering, preprocessing, and aggregating IoT data intelligently via cloud services deployed close to IoT devices \cite{EC-IBM}.

\begin{figure*}[ht]
\centering
\includegraphics[width=0.7\textwidth]{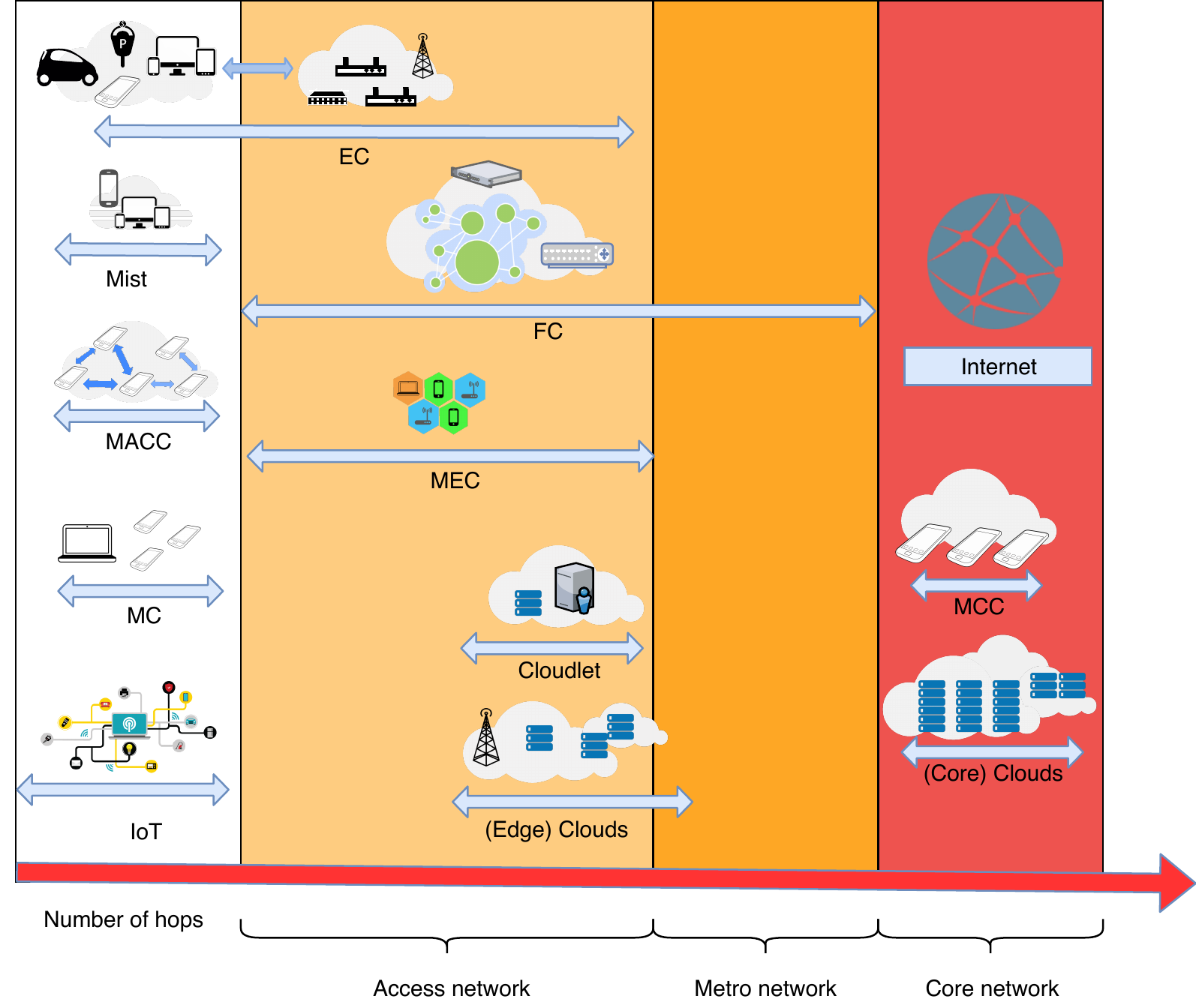}
\caption{Comparison of fog computing and its related computing paradigms in terms of their location and distance from the core clouds. }
\label{fig:fog-location}
\end{figure*}

Some issues that edge computing is well equipped to handle are privacy, latency, and connectivity. Due to its proximity to the users, latency in edge computing is typically lower than in MCC and cloud computing, if enough local computation power is provided; latency in edge computing can be slower than cloud or MCC if the local computation unit is not powerful enough. Service availability is also higher in edge computing because connected devices do not have to wait for a highly centralized platform to provide a service, nor are connected devices limited by the limited resources in traditional mobile computing. Compared to MACC, edge computing has small data centers, whereas MACC fundamentally does not need data centers. As a result, edge computing has higher service availability. Edge computing also can expand with broader computing capabilities than MACC by forming hybrid architectures with peer-to-peer and cloud computing models \cite{6}. 

\subsubsection{Edge Computing vs. Fog Computing}
Although fog computing and edge computing both move the computation and storage to the edge of the network and closer to end-nodes, these paradigms are not identical. In fact, the OpenFog Consortium states that edge computing is often erroneously called fog computing; OpenFog Consortium makes the distinction that fog computing is hierarchical and it provides computing, networking, storage, control, and acceleration anywhere from cloud to things; while, edge computing tends to be limited to computing at the edge \cite{29}. (Refer to Fig. \ref{fig:fog-location}.) Moreover, in a tutorial article \cite{clarifying-fog} about fog and edge, the authors explain that ``fog is inclusive of cloud, core, metro, edge, clients, and things,'' and ``fog seeks to realize a seamless continuum of computing services from the cloud to the things rather than treating the network edges as isolated computing platforms,'' and ``fog envisions a horizontal platform that will support the common fog computing functions for multiple industries and application domains, including but not limited to traditional telco services.'' \cite{clarifying-fog}

\subsubsection{Where is edge?}
It is worth mentioning that edge computing, cloudlets, fog computing, and mist computing (to be discussed in Section \ref{mist-computing}) are used interchangeably in some papers, as they all have ``edge'' as a common term. The term edge used by the telecommunications industry usually refers to 4G/5G base stations, RANs, and ISP (Internet Service Provider) access/edge networks. Yet, the term edge that is recently used in the IoT landscape \cite{EC-GE, EC-IBM} refers to the local network where sensors and IoT devices are located. In other words, the edge is the immediate first hop from the IoT devices (not the IoT nodes themselves), such as the WiFi access points or gateways. If the computation is done on IoT devices themselves, this computing paradigm is referred to as mist computing (see Section \ref{mist-computing}). General Electric notes that fog computing focuses on interactions between edge devices (e.g., RANs, base stations, or edge routers), whereas edge computing focuses on the technology attached to the connected things (e.g., WiFi access points) \cite{EC-GE}.

\subsection{Multi-access Edge Computing}
Mobile cloud computing is an extension of mobile computing through cloud computing. Analogously, multi-access edge (MEC) computing is an extension of mobile computing through edge computing. ETSI defines MEC as a platform that provides IT and cloud-computing capabilities within the Radio Access Network (RAN) in 4G and 5G, in close proximity to mobile subscribers \cite{MEC-definition}. Multi-access edge computing was previously referred to as ``mobile edge computing,'' but the paradigm has been expanded to include a broader range of applications beyond mobile device-specific tasks. Examples of multi-access edge computing applications include video analytics, connected vehicles, health monitoring, and augmented reality.

MEC extends edge computing by providing compute and storage resources near low energy, low resource mobile devices. MEC allows RAN operators to add edge computing functionality to existing base stations. Similar to edge computing, small-scale data centers with virtualization capacity can also be used in MEC. Due to underlying hardware used in MEC and edge computing, available computing resources is moderate, in comparison to cloud computing. Furthermore, low-latency applications can be supported in MEC. MEC applications can benefit from real-time radio and network information hence can offer a personalized and contextualized experience to the mobile subscriber.

Both edge computing and MEC computing services operate from the edge of the Internet and can function with little to no Internet connectivity. MEC, however, establishes connectivity through a WAN, WiFi, and cellular connections, whereas edge computing generally can establish any form of connectivity (e.g., LAN, WiFi, cellular). MEC also primarily differs from MCC in its operations: Research in MCC focuses on the relationship between cloud service users (on mobile devices) and cloud service providers, whereas research in MEC focuses on (RAN-based) network infrastructure providers. MEC is expected to benefit significantly from the up-and-coming 5G platform \cite{MEC-5G}. Likewise, 5G is seen as an enabler of MEC as it allows for lower latency and higher bandwidth among mobile devices, and it supports a wide range of mobile devices with finer granularity.

MEC allows edge computing to be accessible to a wide range of mobile devices with reduced latency and more efficient mobile core networks \cite{MEC-orchestration}. MEC also allows for mission-critical delay-sensitive applications over the mobile network \cite{MEC-5G}. It has also incorporated the use of SDN and network function virtualization (NFV) capabilities, in addition to 5G technologies. SDN allows for virtual networking devices to be easily managed through software APIs \cite{kadiyala2017inter}, and NFV allows for reduced deployment times for networking services through virtualized infrastructure. Moreover, through SDN and NFV, network engineers and possibly enterprise application developers can develop their own orchestrator, whose goal is to coordinate the resource provisioning across multiple layers \cite{mirkhanzadeh2018sdn}.

\subsection{Cloudlet Computing}
Proposed by Carnegie Mellon University, cloudlet computing is another direction in mobile computing that shares many traits with MCC and MEC. In fact, it addresses some of the disadvantages of MCC. A cloudlet is a trusted resource-rich computer or a cluster of computers with strong connection to the Internet that is utilized by nearby mobile devices \cite{cloudlet-definition}. Cloudlets are small data centers (miniature clouds, as the name suggests) that are typically one hop away from mobile devices. The idea is to offload computation from mobile devices to VM-based cloudlets located on the network edge \cite{247}. Although academia mostly drives current studies in cloudlet computing, it has high potential in domains such as wearable cognitive assistance and web applications companies.

Cloudlet is the middle tier of a 3-tier continuum: mobile device-cloudlet-cloud. Given the nature of cloudlets as a small cloud close to mobile devices, operators for cloudlet computing could be cloud service providers who want their services to be accessible closer to mobile devices. Network infrastructure owners (e.g., AT\&T, Nokia, etc.) can enable cloudlets with virtualization capacity to be situated closer to mobile devices, in smaller hardware footprints compared to the massive data centers used in cloud computing. The small footprint of cloudlets result in more moderate computing resources, but lower latency and energy consumption compared to cloud computing. Cloudlet computing is intended to serve devices in the local area. 

Just as MACC greatly differs from cloud computing, it also highly differs from cloudlet computing. Cloudlet needs infrastructure with virtualization in the form of virtual machine (VM) capability, whereas MACC does not require such infrastructure. MACC and cloudlet computing both support mobility, but MACC is resource constrained and lacks virtualization support for real-time IoT applications. Cloudlets support local services for mobile clients by dividing tasks among cloudlet nodes in the proximity of mobile devices \cite{cloudlet-MCC}. Although cloudlet computing fits well with the mobile-cloudlet-cloud framework \cite{cloudlet-infrastructure}, fog computing offers a more generic alternative that natively supports large amounts of traffic, and allows resources to be anywhere along the thing-to-cloud continuum. The concept of {\em mobile cloudlets} is similar to cloudlets, in which the cloudlets are a group of nearby mobile devices that are connected wirelessly, e.g., using WiFi or Bluetooth \cite{298}. In mobile cloudlets, mobile devices can be providers as well as clients of computing service.

\subsection{Mist Computing} \label{mist-computing}
Recently, mist computing has been introduced to capture a more extreme edge -- the endpoints -- of connected devices \cite{mist-cisco}. This computing paradigm describes dispersed computing at the extreme edge (the IoT devices themselves) and has been proposed with future self-aware and autonomic systems in mind \cite{mist-aware}. Mist computing could be seen as the first computing location in the IoT-fog-cloud continuum; it can be informally labeled as ``IoT computing'' or ``things computing.'' An IoT device may be wearable, a mobile device, a smart watch, or a smart fridge. Mist computing extends compute, storage, and networking across the fog through the things. In a sense, mist computing is a superset of MACC; since in mist, the networking may not be necessarily ad hoc, and the devices may not be mobile devices (refer to Fig. \ref{fig:venn-diagram}).

The authors in \cite{mist1} introduce the idea of using mobile devices in the vicinity as a cloud computing environment for storage, caching, and computing purposes. They study the use of mist computing to reduce the load in traditional WiFi infrastructures for video dissemination applications. In this study, the spectators of a sport event organize themselves into WiFi-Direct groups and exchange video replays whenever possible, bypassing the central server and access points. This study is also another example of mist computing, in which IoT devices act not only as ``thin clients,'' but also as ``thin servers.'' Some other uses of mist computing are to preserve the privacy of users' data via local processing \cite{mist2}, and to efficiently deploy virtualized instances on single-board computers \cite{mist3}.

\subsection{Other Similar Computing Paradigms}
\subsubsection{Micro Data Center}
Cloudlet is sometimes referred to as micro data center (MDC) in some studies \cite{bahl2015mdc}. The term micro data center (MDC) was proposed by Microsoft Research in 2015 \cite{bahl2015mdc} and is defined as ``an extension of traditional data centers used in cloud computing.'' An MDC can be an edge node or a cloudlet that is deployed between IoT devices and the cloud. 


\subsubsection{Cloud of Things}
Another similar concept to mist computing is the Cloud of Things (CoT) \cite{95}, where IoT devices form a virtualized cloud infrastructure. In mist computing computation is done on IoT devices, possibly via message exchange, and not necessarily in a cloud of pooled resources. However, in Cloud of Things, computation is done over the cloud that is formed by pooled resources of IoT devices. Abdelwahab et al. \cite{95} introduce the notion of Cloud of Things for sensing-as a service, which uses edge nodes as cloud agents sitting close to IoT nodes. The authors propose the idea of dynamically scaling up existing cloud resources (compute, storage, and network) by using the sensing capability of IoT devices. Edge nodes are used as cloud agents near the edge to discover, virtualize, and form a cloud network of IoT devices (CoT). This network is a geographically distributed infrastructure, in which cloud agents constantly discover resources of IoT devices and pool them as cloud resources. CoT enables remote sensing and in-network distributed processing of data. For instance, a cloud user can view pollution levels in cities from real-time temperature and $CO_2$ concentration sensors in vehicles with defined accuracy. The CoT framework is scalable to IoT networks, supports heterogeneity of IoT devices and edge computing nodes, and provides a foundation of sensing-as a service using fog computing. 

Similar to CoT, the authors in \cite{pcloud} propose the concept of PClouds (personal clouds), which are distributed networked resources that are from both local/personal and remote/public devices and machines. PCloud can service end users even when remote cloud resources are not present or difficult to access due to insufficient network connectivity. Another novel idea similar to Cloud of Things and MACC is the work of the authors in \cite{cloudrone}, where they propose Cloudrone, an idea of deploying ad hoc micro cloud infrastructures in the sky using low-cost drones, single-board computers, and lightweight OS virtualization technologies. The drones in this scheme form a cloud computing cluster in the sky, which provisions the cloud services nearer to the user, even in the absence of a terrestrial infrastructure to access the remote cloud. 

Similar to the concept of Cloud of Things, Femtoclouds have been proposed to tap into the computational capability and pervasiveness of underutilized mobile devices. Femtoclouds take advantage of clusters of devices that tend to be co-located in places such as schools, public transit, or malls. A hybrid edge-cloud workload management scheme is proposed in \cite{252} for management of resources and tasks in femtoclouds, to provide low latency. 

\begin{table*}
\small
\newcolumntype{Y}{>{\centering\arraybackslash}X} 
\renewcommand{\arraystretch}{1.5} 
\setlength\tabcolsep{3pt} 
\renewcommand\cellalign{tl} 
\caption{Attributes of fog-computing related paradigms}
\label{tab:comparison-table-1}
\begin{tabularx}{\textwidth}{|l@{}|Y|Y|Y|Y|Y|Y|Y|Y|Y|@{}}
    \hline
    
    \textbf{Attribute} & \textbf{CC} & \textbf{MC} & \textbf{FC} & \textbf{EC} & \textbf{MCC} & \textbf{MACC} & \textbf{MEC} & \textbf{cC} & \textbf{mist} \\ \hline
    
    \textbf{Users} & General & Mobile & General & General & Mobile & Mobile & Mobile & Mobile & General \\ \hline
    
    \makecell[l]{\textbf{General Use}\\ \textbf{Cases}} & Scalable data storage, virtualized apps, distributed computing for large data sets (Google MapReduce) & Mobile sales transactions, location dependent queries (travel recommendations), multimedia applications on mobile devices & IoT, Connected vehicles, smart grid/smart city, health care, smart delivery (high-scale package drone delivery), real-time subsurface imaging, video surveillance & Local video surveillance, video caching, traffic control & Social networking, sensor data processing, health care (tele-monitoring and tele-surgery) & Networking and computing for disaster relief, group live video streaming, unmanned vehicular system & Content Delivery, Video analytics, connected vehicles, health monitoring, augmented reality & Optical character recognition (OCR), wearable cognitive assistance (Google Glass) & Parallel computation on IoT devices, autonomous vehicles, privacy-preserving local processing\\ \hline
    
    \textbf{Operators} & Cloud service providers & Self-organized & Users and cloud service providers & Network infrastructure providers or local businesses & Users and cloud service providers & Self-organized & Network infrastructure providers (RAN-based) & Cloud service providers and network infrastructure providers & Self-organized or local businesses\\ \hline
    
    \textbf{Service Type} & Global & Local & Less global & Local & Local & Local & Less global & Local & Local \\ \hline
    
    \textbf{Hardware} & Large-scale data centers with devices with virtualization capacity & Mobile devices & Devices with virtualization capacity (servers, routers, switches, access points) & Edge devices with computing capability & Mobile devices or large-scale data centers with devices with virtual capability & Mobile devices & Small-scale data centers with devices with virtualization capacity, RAN in 4G and 5G & Devices with virtualization capability (micro and nano data centers) & IoT devices (e.g. sensors, cell phones, home appliance devices)\\ \hline
    
    \makecell[l]{\textbf{Available} \\ \textbf{Computing} \\ \textbf{Resources}} & High & Limited & Moderate & Moderate & High & Relatively less limited & Moderate & Moderate & Limited\\ \hline
    
    \textbf{Main Driver} & Academia/ industry & Academia & Academia/ industry & Academia/ industry & Academia & Academia & Academia/ industry & Academia & Academia\\ \hline
    
    \makecell[l]{\textbf{Distance} \\ \textbf{from Users}} & Far & Very close & Relatively close & Close & Far & Very close & Close & Close & Very close\\ \hline
    
	\makecell[l]{\textbf{Main} \\ \textbf{Standardization} \\ \textbf{Entity}}\hspace{3pt} & CSA, DMFT, NIST, OCC, GICTF & MobileInfo & OpenFog Consortium, IEEE & --- & NIST & --- & ETSI, 3GPP, ITU-T & OpenEdge & --- \\ \hline
    
    \makecell[l]{\textbf{Application} \\ \textbf{Type}} & Ample computation & Distributed and mobile processing & High computation with lower latency & Low latency computation & High computation & High computation with lower latency & Low latency computation & High computation with lower latency & Distributed processing on IoT devices\\ \hline
    
   \textbf{Architecture} & Centralized/ hierarchical & Distributed & Decentralized/ hierarchical & Localized/ distributed & Central cloud with distributed mobile devices & Distributed & Localized/ hierarchical & Localized & Localized/ distributed\\ \hline

    \end{tabularx}
\end{table*}

\begin{table*}
\small
\newcolumntype{Y}{>{\centering\arraybackslash}X} 
\renewcommand{\arraystretch}{1.5} 
\setlength\tabcolsep{3pt} 
\renewcommand\cellalign{tl} 
\begin{tabularx}{\textwidth}{|l@{}|Y|Y|Y|Y|Y|Y|Y|Y|Y|@{}}
    \hline
    
    \textbf{Attribute} & \textbf{CC} & \textbf{MC} & \textbf{FC} & \textbf{EC} & \textbf{MCC} & \textbf{MACC} & \textbf{MEC} & \textbf{cC} & \textbf{mist} \\ \hline

    \textbf{Availability} & High & Low & High & Average & High & Low & Average & High & Low\\ \hline
    
    \textbf{Latency} & Relatively high & Moderate & Low & Low & Relatively high & Moderate & Low & Low & Moderate\\ \hline
    
    \textbf{Security} & Must be provided along cloud-to-things continuum & Must be provided on mobile devices & Must be provided on participant nodes & Must be provided on edge devices & Must be provided along cloud-to-things continuum and on mobile devices & Must be provided on mobile devices & Must be provided on edge network equipment (RAN, AP) & Must be provided on participant nodes & Must be provided on IoT devices\\ \hline
    
    \makecell[l]{\textbf{Server} \\ \textbf{Location}} & Installed in large dedicated buildings & --- & Can be installed at the edge or in dedicated locations & Near edge devices & Installed in large dedicated buildings & --- & Can be installed at the edge & Near mobile devices & ---\\ \hline
    
    \makecell[l]{\textbf{Power} \\ \textbf{Consumption}} & Relatively high & --- & Low & Low & Low on mobile devices & Low & High & Moderate & Low\\ \hline

    \makecell[l]{\textbf{Internet} \\ \textbf{Connectivity}} & Must be connected to the Internet for the duration of services & Can operate with low or intermittent Internet connectivity & Can operate autonomously with no or intermittent Internet connectivity & Can operate autonomously with no or intermittent Internet connectivity & Requires Internet connection for offloading tasks or obtaining computation results from the cloud & Can operate autonomously with no or intermittent the Internet & May operate autonomously or connect to the Internet through RAN & Can operate with no or intermittent Internet connectivity; often requires connection to the Internet & Can operate with low or intermittent Internet connectivity\\ \hline
    
    \makecell[l]{\textbf{Hardware} \\ \textbf{Connectivity}} \hspace{3pt} & WAN & Bluetooth, WiFi, cellular, ZigBee & WAN, LAN, WLAN, WiFi, cellular & WAN, LAN, WLAN, WiFi, cellular, ZigBee & WAN & Bluetooth, WiFi, cellular, ZigBee & WAN, cellular & WAN, LAN, WLAN, WiFi, cellular & LAN, Bluetooth, WiFi, cellular, ZigBee \\ \hline
    
    \makecell[l]{\textbf{Service} \\ \textbf{Access}} & Through core & Through mobile devices & Through connected devices from the edge to the core & At the edge of the Internet & Through core & Through mobile devices & At the edge of the Internet & Through resource-rich computers at the edge of the Internet & Through IoT devices\\ \hline
    
    \end{tabularx}
\end{table*}

\begin{table*}
\small
\newcolumntype{Y}{>{\centering\arraybackslash}X} 
\renewcommand{\arraystretch}{1.5} 
\setlength\tabcolsep{3pt} 
\renewcommand\cellalign{tl} 
\caption{Features of fog-computing related paradigms}
\label{tab:comparison-table-2}
\begin{tabularx}{\textwidth}{|l@{}|Y|Y|Y|Y|Y|Y|Y|Y|Y|@{}}
    \hline
    
    \textbf{Feature} & \textbf{CC} & \textbf{MC} & \textbf{FC} & \textbf{EC} & \textbf{MCC} & \textbf{MACC} & \textbf{MEC} & \textbf{cC} & \textbf{mist} \\ \hline
    
    Heterogeneity support & \ding{51} & \ding{55} & \ding{51} & \ding{51} & \ding{51} & \ding{55} & \ding{55} & \ding{55} & \ding{51}\\ \hline
    
    Infrastructure need & \ding{51} & \ding{55} & \ding{51} & \ding{51} & \ding{51} & \ding{55} & \ding{51} & \ding{51} & \ding{51}\\ \hline
    
    Geographically distributed & \ding{55} & \ding{55} & \ding{51} & \ding{51} & \ding{55} & \ding{55} & \ding{51} & \ding{51} & \ding{51}\\ \hline
    
    Location awareness & \ding{55} & \ding{51} & \ding{51} & \ding{51} & \ding{55} & \ding{51} & \ding{51} & \ding{51} & \ding{51}\\ \hline
    
    Ultra-low latency & \ding{55} & \ding{55} & \ding{51} & \ding{51} & \ding{55} & \ding{55} & \ding{51} & \ding{51} & \ding{51}\\ \hline
    
    Mobility support & \ding{55} & \ding{51} & \ding{51} & \ding{51} & \ding{51} & \ding{51} & \ding{51} & \ding{51} & \ding{51}\\ \hline
    
    Real-time application support & \ding{55} & \ding{55} & \ding{51} & \ding{51} & \ding{55} & \ding{55} & \ding{51} & \ding{51} & \ding{51}\\ \hline
    
    Large-scale application support \hspace{3px} & \ding{51} & \ding{55} & \ding{51} & \ding{51} & \ding{55} & \ding{55} & \ding{51} & \ding{55} & \ding{51}\\ \hline
    
    Standardized & \ding{51} & \ding{51} & \ding{51} & \ding{51} & \ding{55} & \ding{55} & \ding{51} & \ding{55} & \ding{55}\\ \hline
    
    Multiple IoT Applications & \ding{51} & \ding{55} & \ding{51} & \ding{55} & \ding{55} & \ding{55} & \ding{55} & \ding{51} & \ding{51}\\ \hline
    
    Virtualization support & \ding{51} & \ding{55} & \ding{51} & \ding{55} & \ding{55} & \ding{55} & \ding{51} & \ding{51} & \ding{55}\\ \hline
    
    \end{tabularx}
\end{table*}

\subsubsection{Edge Cloud}
When we talk about cloud computing, we mainly talk about ``core'' or ``distant'' clouds, which are far from the user or devices. Core clouds are further from connected things and are responsible for heavy computation. In contrast, ``edge'' clouds are smaller scale compared to core clouds and are closer to the devices. The concept of edge cloud \cite{chang2014bringing} is similar to edge computing. The edge cloud extends cloud capabilities at the edge by leveraging user or operator-contributed compute nodes at the edge of the network. Similar to fog, in edge clouds the ability to run an application in a coordinated manner in both edge and the distant cloud is envisaged. Edge clouds are nodes at the edge, such as micro data centers, cloudlets, and MEC. \cite{225}. 

Researchers have begun studying federation of both edge clouds and core clouds, and proposed the ``osmotic computing'' paradigm \cite{27, 230}. Osmotic computing implies ``the dynamic management of services and micro-services across cloud and edge data centers, addressing issues related to deployment, networking, and security'' \cite{27}. Osmotic computing utilizes both edge and cloud resources, each contained in two separate layers. Application delivery follows an osmotic behavior where virtualized micro-services are deployed opportunistically either in the cloud or edge layers. The ability to control how micro-services can be balanced between edge and cloud is a significant advantage of osmotic computing.

\subsection{Concluding Remarks}
The previous discussion about fog computing and related paradigms demonstrate the importance of understanding the characteristics of these platforms in the changing IT landscape. As demonstrated by the strength and weaknesses attributed to these computing paradigms, some paradigms may be better suited for a particular use case than others. Even so, fog computing is suited for a large number of use cases in the current landscape of IoT and connected devices. The versatility of fog computing makes it suitable for many cases of data-driven computing and low-latency applications, even though it may not be suitable for a few extreme applications, such as disaster zones or sparse network topologies where ad hoc computing (e.g., MACC) or extreme edge clouds (e.g., mist, CoT) may be a better fit. Nonetheless, fog computing is considered a more general form of computing when compared to other similar paradigms (e.g., EC, MEC, cloudlet), because of its comprehensive definition scope, generality, and extensive presence along the thing-to-cloud continuum. Tables \ref{tab:comparison-table-1} and \ref{tab:comparison-table-2} summarize these characteristics. Fog computing offers a bright future for an open-standards environment of connected devices, as it is evident by IEEE Standard's adoption of OpenFog Reference Architecture \cite{fog-ieee-standard}. 

\begin{figure*}[ht]
\centering
\includegraphics[width=0.85\textwidth]{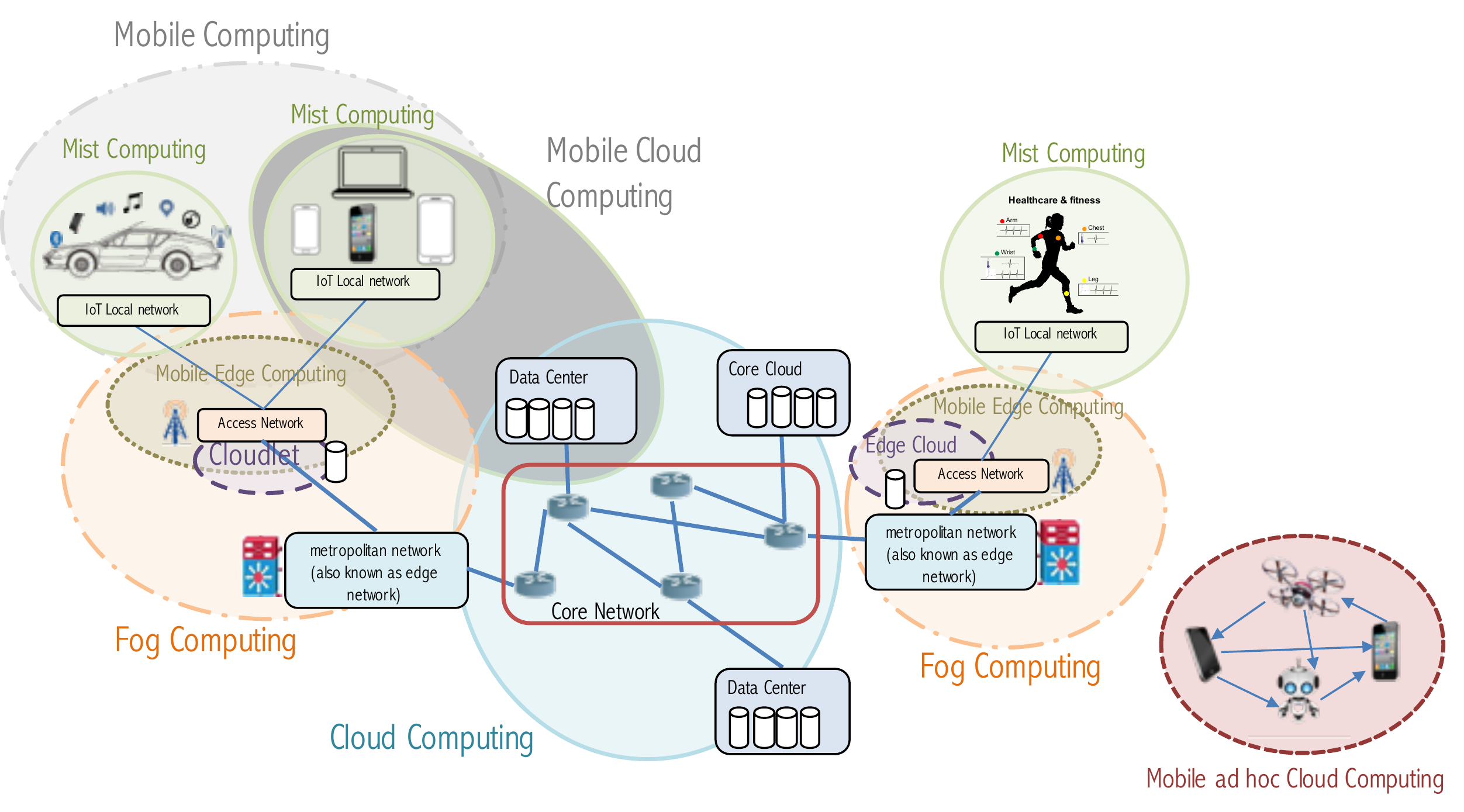}
\caption{Comparison of the infrastructure of fog computing and its related computing paradigms from the networking perspective.}
\label{fig:paradigms}
\end{figure*}

There does not yet exist a globally unanimous distinction between fog computing and related computing paradigms, such as edge computing, mist computing, and cloudlets across researchers and industries, as shown in the previous sections of this paper. We attempt in this survey paper to clarify the distinctions between fog computing and the related computing paradigms. A comparison of the underlying infrastructure of fog computing and its related computing paradigms from the networking perspective is shown in Fig. \ref{fig:paradigms}. In the rest of this paper, we will mainly survey and discuss the recent literature on fog computing, but mention the studies on other related computing paradigms that could be easily extended or directly applied in fog.

\section{Taxonomy of Fog Computing Through a Complete Survey} \label{taxonomy}
\begin{figure*}[ht]
\centering
\includegraphics[width=0.6\textwidth]{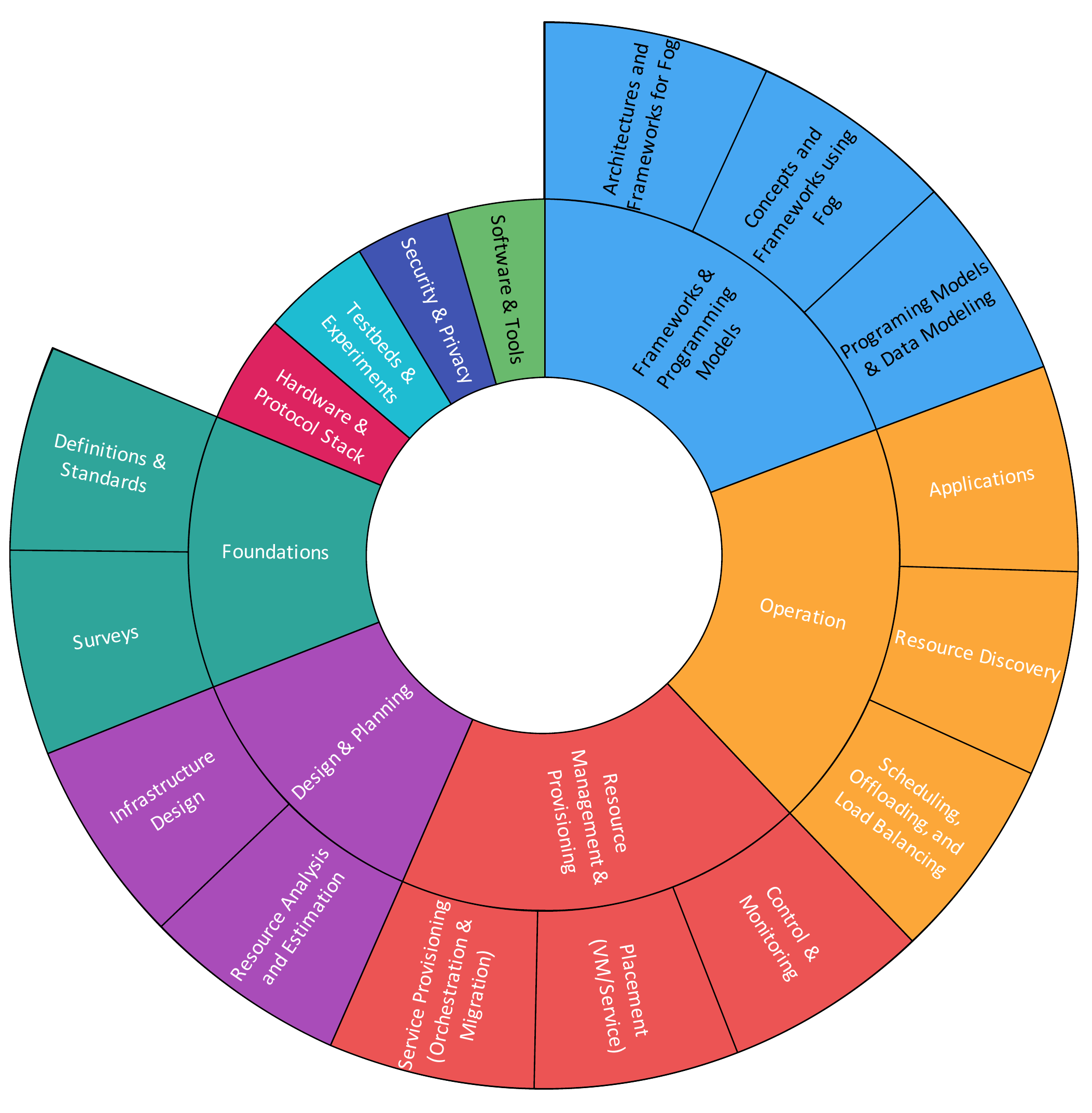}
\caption{Taxonomy of fog computing that is used in this survey.}
\label{fig:fog-taxonomy}
\end{figure*}

In this section, we will introduce a taxonomy of the research in fog computing that is the basis of this survey\footnote{The data (categories and features/objectives of the papers) of this survey are available in the form of several datasets at {\footnotesize https://github.com/ashkan-software/fog-survey-data}}. This taxonomy categorizes the research articles that focus primarily on fog computing from the networking perspective. We have also included research articles from other similar computing paradigms, such as edge computing, if the article is relevant and general enough that it could be easily extended to fog computing. The taxonomy is shown in Fig. \ref{fig:fog-taxonomy}, and the papers on fog computing, edge computing, cloudlet, etc. that are referenced in this survey are categorized into different categories in Table \ref{tab:category-paper}. We will discuss the literature on fog computing in this section, and we categorize the research papers according to this taxonomy. Moreover, we have rigorously checked the objectives of papers (e.g., QoS improvement) and the features they provide (e.g., scalability), and we summarize them in Appendix, Table \ref{tab:features-objectives}. The explanation of the features and objectives along with several examples for each objective/feature are included in Table \ref{tab:objective-explanation}. We extract these objectives/features such that they are comprehensive and useful for a fog system design, and are also closely in compliance with the pillars of OpenFog architecture \cite{29}.

The ``Foundation'' category consists of the research papers that either survey the fog computing area, or try to define and standardize the field of fog computing. The ``Frameworks and Programming Model'' category is where the reader can find research articles that introduce frameworks, architectures, and programming models for fog, or that use fog computing to introduce a new concept (such a vehicular fog computing). The next category is ``Design and Planning,'' which includes the papers that discuss the design and planning of the network and computing infrastructure. The ``Resource Management and Provisioning'' category consists of the research papers that study the management and provisioning of the resources (e.g., service provisioning, VM placement, control and monitoring). The category ``Operation'' includes the papers that discuss operational aspects of fog computing systems (e.g., task scheduling, load balancing, and resource discovery). Each of the mentioned categories has subcategories, and we will discuss each subcategory in the following subsections. The category ``Software and Tools'' will list papers that focus on software, simulators, and tools for fog computing. Likewise, the ``Hardware and Protocol Stack'' category showcases articles that propose a protocol stack or introduce particular hardware for fog computing. Papers that focus on developing testbeds or doing extensive experiments for fog are summarized under the ``Testbeds and Experiments'' category. Finally, papers discussing security and privacy aspects of fog computing are included under the ``Security and Privacy'' category. 

Note that the three categories Design and Planning, Resource Management and Provisioning, and Operation reflect conventional steps in design and operation of distributed computing systems and computer networks. First, in the Design and Planning step, network designers estimate and analyze the required resources for a given network, design the infrastructure and topology of the network, and determine the hardware and resources that must be placed in a particular design. Next, in the Resource Management and Provisioning step, network operators try to manage and provision the resources for better utility and efficiency. For instance, service orchestration and migration methods are used to intelligently allocate and provision the available resources of the network nodes; monitoring techniques are used to monitor the resource usage of the nodes, for instance for placement decisions (e.g., VM or container placement). Finally, in the Operation step, the final improvements for resource usage and efficiency are performed, such as task offloading and scheduling, load balancing, and efficient resource discovery. 


\subsection{Foundations: Definition and Standards}
In this subsection, we survey the articles that are concerned with defining and standardizing fog computing and concepts related to fog computing. The very definition of fog computing and fog nodes is a topic of ongoing discussion, and there is no common consensus on what a fog node is \cite{260}. There are some early efforts to define fog and fog nodes \cite{260, 29}. OpenFog Consortium is one of the pioneers in standardizing and defining fog computing. The OpenFog architecture is established to provide a nonproprietary fog architecture and standard to support current cloud computing in addition to diverse IoT and edge-oriented ecosystems. The white paper introduces security, scalability, openness, agility, among other ``pillars'' of an open fog architecture \cite{29}. Later, IEEE Standards Association adopted OpenFog Consortium's reference architecture as a standard for fog computing through IEEE 1934 \cite{fog-ieee-standard}. 

Vaquero et al. \cite{3} take into account mobile device ubiquity, network management, fog network connectivity, and privacy to propose their definition for fog: {\em a large amount of heterogeneous, ubiquitous, and decentralized devices that can cooperate to form a network for storage and processing without third-party intervention}. The authors in \cite{16} focused on the theoretical modeling and performance metrics of the fog computing architecture. They propose a mathematical formulation for fog computing by defining its components for a generic fog architecture. 

Current communication technologies and standards that could be used in fog networks are presented in \cite{42}. In the paper, a classification of layers and technology settings related to IoT and fog computing is described. On the other hand, the authors in \cite{158} define ``class of service'' for fog applications, a classification of fog services according to their QoS requirements. They also introduce a mapping between certain classes of services and the corresponding processing layers of the fog computing reference architecture. 

With emerging availability of IoT devices and their large volume of data that they produce, timely and reliable transfer of large data streams to a centralized location is a requirement of deep learning models. The authors in \cite{230} introduce ``deep osmosis'' and analyze the research challenges involved with developing edge-cloud-based holistic distributed deep learning algorithms and their corresponding resource models and architecture for cloud and fog computing. 

The study in \cite{235} defines the concept of ``content island'' for fog computing, which interconnects groups of devices to interchange data and processing among themselves and with other content islands. The islands are based on the integration of a publish/subscribe system with disruption-tolerant network (DTN) techniques to provide higher flexibility with respect to data and computation sharing. Another definition in fog computing area is the ``reliability factor'' of a node, which defines the probability of a node being online or available, and is defined in \cite{241}. In \cite{249} ten economic aspects of fog computing (referred to as fogonomics) are introduced. 
 
\subsection{Foundations: Surveys}
In this section, we discuss the previous work in the fog computing domain that are of survey or tutorial nature. Several comprehensive survey papers are already discussed in Section \ref{related}; in this subsection, we will discuss some other survey papers that are not as comprehensive as the ones discussed in Section \ref{related}. Klas from the Vodafone Group presents a clear picture of edge computing and its benefits \cite{60}. This paper highlights relevant edge computing applications and similar research areas (e.g., fog and cloudlet), surveys the latest industrial efforts and available edge computing technologies, and discusses edge computing's potential to improve the telecommunications industry.

Historically, the field of computing has seen cycles between decentralization and centralization \cite{6}. The authors of \cite{6}, in a survey study, advocate for edge-centric computing, a more decentralized paradigm that utilizes peer to peer (P2P) networking at the edge of the network while maintaining access to the cloud. 


The authors in \cite{146} present a comprehensive study of edge computing, starting with the factors leading up to its development, the advantages of edge computing, requirements for successful implementation, application use-cases, and challenges. Several use cases, such as, gaming, real-time image processing, smart grid, smart transportation, are used to emphasize the range of problems that edge computing can help alleviate. Weisong et al. define edge computing and survey its use cases, research issues, and future research directions in \cite{25}. A recent survey on MEC can be found in \cite{284}.


Stojmenovic and Wen in \cite{12} highlight privacy and security concerns of fog computing gateways as major issues. These issues include man-in-the-middle attacks and lack of encryption in gateways that serve as fog nodes. The study in \cite{15} surveys the main features of fog computing, describes its architecture and design goals, and discusses some potential issues of fog computing in 5G. One of the early surveys in fog architecture and taxonomy is the work of Zhang and Chiang \cite{26}. The authors further describe the IoT challenges for which fog computing can provide solutions. The survey in \cite{246} distinguishes and explains edge computing, fog computing, and cloud computing. The study reviews various system architectures, application characteristics, and platform abstractions of fog, edge, and cloud.

In the paper \cite{271}, the authors overview fog computing model architecture, key technologies, and applications. They present the hierarchical architecture of fog computing and its characteristics and compare it with cloud computing and edge computing. Then, the key technologies for fog are introduced to see how they support fog computing deployments. 

\subsection{Frameworks and Programming Models: Architectures and Frameworks for Fog}
Many researchers have independently proposed various architectures and frameworks for fog computing. In this section, we summarize the previous work that have proposed general architectures or frameworks for fog computing. 
\subsubsection{General Architecture for Fog Computing}
\begin{figure*}[ht]
\centering
\includegraphics[width=0.7\textwidth]{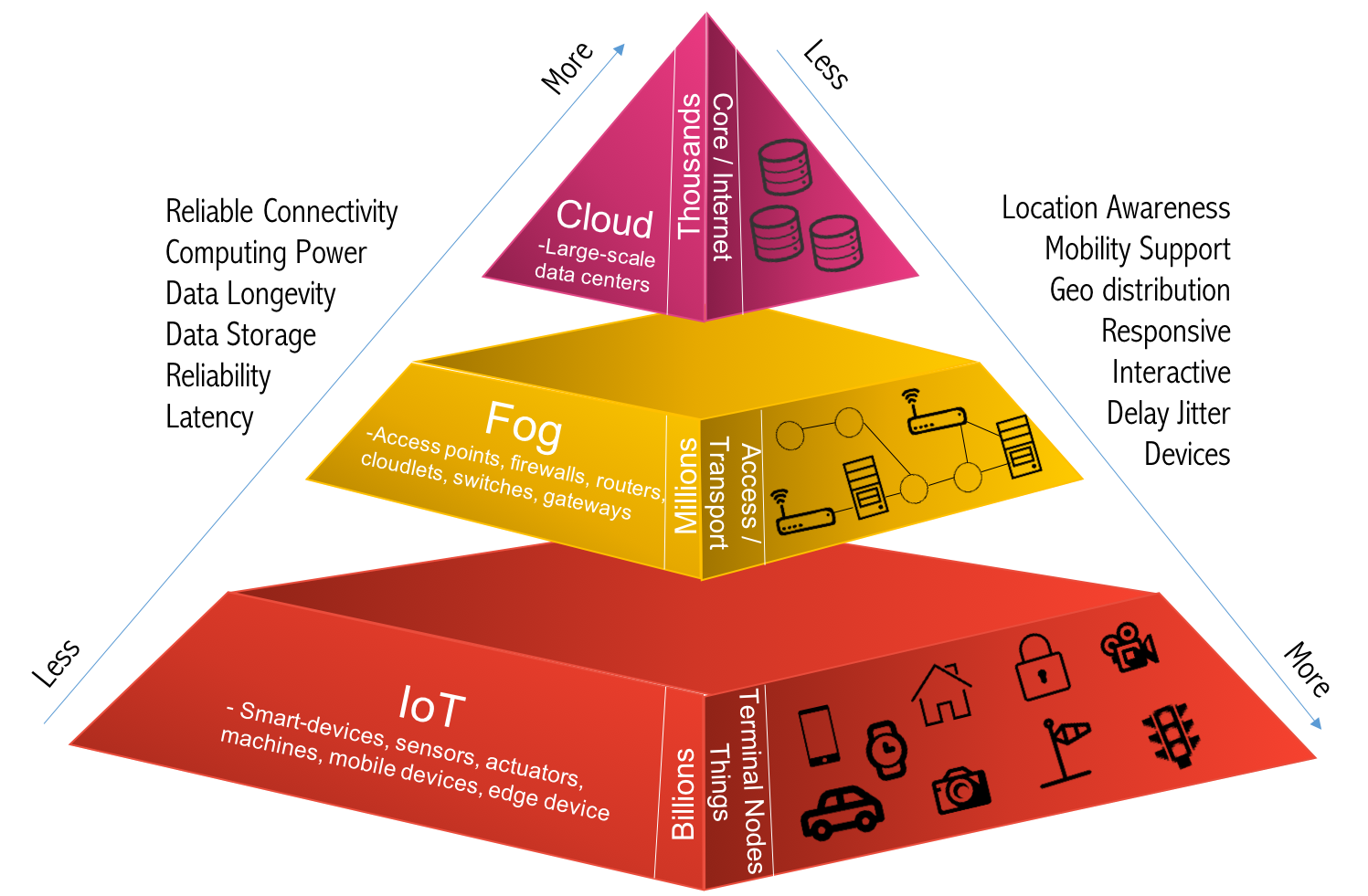}
\caption{Our three-layer architecture for IoT-fog-cloud ecosystem.}
\label{fig:pyramid}
\end{figure*}

A recent study suggests an architectural model for combining MEC and fog computing for 5G networks \cite{199}. The authors claim that fog computing and MEC separately have weaknesses and incompleteness; they further claim the need for convergence of the two computing paradigms for overcoming such limitations. A three-layer general logical architecture for fog computing is introduced in \cite{194} and \cite{18}. The layers are IoT, fog, and cloud, where each layer is partitioned into domains. Similarly, a three-layer architecture including the cloud, MEC, and IoT is proposed in \cite{174}. In the three-layer architecture proposed, the user plane consists of mobile users and IoT devices, the edge computing plane consists of edge servers in close proximity to the users, and the cloud computing plane is the core of the network and contains multiple cloud servers and data centers. Comparably, we propose our three-layer architecture for fog computing in Fig. \ref{fig:pyramid}. 

The authors in \cite{156} propose a fog-to-cloud architecture, consisting of a layered management structure that can bring together different heterogeneous cloud/fog layers into a hierarchical architecture. The paper \cite{118} designs a hierarchical edge cloud architecture, to efficiently utilize the cloud resources for serving the peak loads from mobile users. The proposed architecture consists of servers at the edge, which directly receive workloads from mobile devices via wireless links. These edge servers are connected to higher tiers of edge cloud servers and remote data centers through the Internet backbone. Different from directly using a flat collection of edge cloud servers, the proposed architecture aggregates the peak loads that exceed the capacities of lower tiers of edge cloud servers to other servers at higher tiers in the edge cloud hierarchy.

\subsubsection{Fog Computing Resource Model}
One challenge in fog computing is defining who the fog resource providers are. Is it the case that fog service providers must provide fog resources? Can end users can bring their devices and share their resources? Do network providers offer their edge resources for renting? The articles \cite{147} and \cite{148} present a unified computing, caching, and communication (3C) solution for 5G that allow service, content, and function providers to deploy their services/content/functions near the end users. The solution also allows for the exploitation of extreme edge resources by enabling their owners to form virtual fogs (vFogs) cooperatively; that is, end users will have the ability to become 3C resource providers to the 5G ecosystem. The authors also propose their architecture for fog computing, which consists of vFogs, hyper fogs (constellations of vFog networks to facilitate processing and data exchange that requires resources from more than one vFog), super extreme edge node, regular extreme edge node, and orchestrator. 

One of the many efforts to design a reference framework and infrastructure for the fog-based IoT that considers resource sharing of consumers is Indie Fog \cite{80}. The Indie Fog infrastructure utilizes consumers' equipment (e.g., WiFi access points) to provide fog services for IoT devices. The authors suggest that network infrastructure providers or cloud service providers can make use of the consumer premises equipment to provide their fog-based services. Under this model, they claim that consumers will be willing to share their equipment with the providers for offering their services. Indie Fog uses the general fog architecture proposed by the OpenFog Consortium \cite{29} and adds the Indie Fog services to it, which are interconnected via virtual connections to private and public fog networks. 


\subsubsection{Fog Architecture Design Decisions}
In an article from Cisco \cite{57}, the author proposes a general high-level architecture for fog networks, fog software, and fog nodes. This paper is an early attempt to review and characterize the design/decision parameters of fog networks. The author names these decision criteria ``fog architectural imperatives,'' and discusses them in detail. The fog architectural imperatives are decisions related to design requirements that are difficult to implement on networks with sole reliance on cloud or IoT devices, and that can only be satisfied by using fog resources \cite{57}.

\begin{table*}
\small
\centering
\renewcommand\tabularxcolumn[1]{m{#1}} 
\renewcommand{\arraystretch}{1.5} 
\setlength\tabcolsep{5pt} 
\caption{Overall categories of the papers cited in this survey}
\label{tab:category-paper}
\begin{tabularx}{\textwidth}{|c|c|X|}
    \hline
    
    \textbf{Category} & \textbf{Subcategory} & \multicolumn{1}{c|}{\textbf{Papers}} \\ \hline
    
    \multirow{2}{*}{Foundations} & Surveys & \cite{3,6,12,13,15,25,26,27,38,60,61,78,86,138,146,167,183,246,257,270,271,283,284,290} \\ \cline{2-3}
    
    & Definitions \& Standards & \cite{3,16,22,28,29,35,36,42,158,179,188,196,199,225,230,231,235,241,246,249,260,314,319,327,355,359,360,362,375,381,384} \\ \hline
    
    \multirow{3}{*}{\makecell{Frameworks and \\ Programming Models}} & \makecell{Architectures and Frameworks \\ for Fog} & \cite{1,3,15,16,18,20,21,22,23,29,33,36,41,45,46,51,57,65,69,80,97,98,100,103,105,111,118,130,140,144,147,148,149,150,156,161,163,164,174,175,176,179,194,196,198,199,208,210,216,218,221,222,235,238,249,251,260,265,281,292,295,304,317,319,325,328,333,349,366,368} \\ \cline{2-3}
    
    & \makecell{Concepts and Frameworks \\ using Fog} &
    \vspace{5pt} \cite{39,66,71,95,101,102,121,129,145,162,214,219,252,254,256,262,280,309,320,321,329,340,354,359,376,382,384} \vspace{5pt} \\ \cline{2-3}
    
    & \makecell{Programming Models \\ and Data Modeling} & \vspace{5pt} \cite{28,30,32,38,51,53,54,72,83,97,98,99,107,149,226,262,275,282,313,323,336,341,378} \vspace{5pt} \\ \hline
    
    \multirow{2}{*}{\makecell{Design and Planning}} & Infrastructure Design & \cite{33,41,45,69,88,94,97,155,160,171,195,209,287,297,303,370,375} \\ \cline{2-3}
    
    & \makecell{Resource Analysis \\ and Estimation} & \vspace{5pt} \cite{7,67,68,94,122,173,182,209,212,231,241,277,318,355,360,364,373,375} \vspace{5pt} \\ \hline
    
    \multirow{3}{*}{\makecell{Resource Management \\ and Provisioning}} & \makecell{Service Provisioning \\ (Orchestration \& Migration)} & \cite{9,21,51,54,59,62,82,92,96,98,104,105,106,108,110,111,115,117,128,141,143,161,162,169,180,222,234,250,261,266,279,287,295,334,352,366,379} \\ \cline{2-3}
    
    & \makecell{Placement \\ (VM/Service)} & \cite{8,11,19,24,37,52,74,101,109,116,118,119,133,135,168,172,176,178,179,181,192,205,207,221,238,253,262,287,322,328,331,333,337,349,356} \\ \cline{2-3}
    
    & Control and Monitoring & \cite{14,22,23,37,41,45,77,78,79,97,98,103,108,110,127,169,208,281,296,298,300,301,315,371} \\ \hline
    
    \multirow{3}{*}{Operation} & \makecell{Scheduling, Offloading, \\and Load Balancing} & \cite{1,2,8,10,17,43,46,48,49,55,56,75,76,84,88,89,90,91,93,96,109,120,130,132,134,142,155,159,161,166,174,188,194,215,223,243,245,248,252,269,281,297,298,303,304,305,307,309,310,315,317,338,347,357,358,373} \\ \cline{2-3}
    
    & \makecell{Resource Discovery}  & \cite{7,17,53,54,95,105,110,149,177,185,186,193,223,282,288,294,327,344,363,384} \\ \cline{2-3}
    
    & Applications & \cite{31,40,58,63,64,85,87,95,101,102,116,123,124,125,131,136,139,142,152,167,185,187,192,197,203,211,218,220,228,232,233,234,236,237,239,240,242,244,247,258,259,264,265,268,274,285,286,289,293,302,316,326,330,331,339,340,342,343,346,350,353,354,357,358,367,371,372,380,383} \\ \hline
    
    \multicolumn{2}{|c|}{Software \& Tools} & \cite{32,34,70,73,81,115,126,177,187,198,200,202,206,213,225,227,240,273,312,324,335,336,345,351,377} \\ \hline
    
    \multicolumn{2}{|c|}{Testbeds \& Experiments} & \cite{5,71,111,112,114,137,144,184,189,190,201,202,204,217,220,224,229,255,263,276,278,280,291,306,311,332,345,365,366,367,374,382,384} \\ \hline
    
    \multicolumn{2}{|c|}{Security \& Privacy} & \cite{12,35,44,47,50,113,123,125,127,132,136,138,145,151,153,154,157,170,191,225,227,228,248,257,267,283,286,288,290,296,299,304,308,314,325,326,342,346,361,369} \\ \hline
    
    \multicolumn{2}{|c|}{Hardware \& Protocol Stack} & \cite{33,39,41,45,46,69,99,121,126,135,155,162,163,165,171,177,210,229,235,251,255,303,317,327,348,368,381} \\ \hline
    
\end{tabularx}
\end{table*}

\subsubsection{Fog Architectures for 5G and IoV}
Fog computing is seen as a promising enabler for some of the emerging paradigms, such as 5G, autonomous cars, and Internet of Vehicles (IoV). In their article \cite{251}, the authors propose an SDN-based framework for cloud-fog interoperation in 5G wireless networks. Vilalta et al. propose TelcoFog - a fog computing architecture that is deployed at the network edge for telecom operators to provide cost-effective 5G services for low latency and scalability \cite{98}. TelcoFog consists of three main types of components: scalable TelcoFog nodes, TelcoFog controller, and TelcoFog services. The paper \cite{161} introduces the challenges of handling big data in the IoV environments. The authors emphasize on the role of fog servers and describe a regional cooperative fog computing (CFC) architecture to support IoV applications. The proposed CFC-IoV architecture consists of two layers - the fog layer and edge layer. The fog layer is a federation of geographically distributed fog servers, a coordinator server, and the cloud servers, whereas the edge layer includes the vehicular ad hoc network (VANET), IoT applications, and mobile cellular networks. Other effort suggesting fog architectures for 5G or IoV are \cite{15,199,147,33,97}.

\subsubsection{ICN-based Fog Architecture}
The study in \cite{149} brings together fog computing and information-centric networking (ICN), which enables flexible and efficient data distribution at the network layer. In the introduced ICN-Fog architecture, at the lowest layer are heterogeneous end devices that connect to fog nodes, which run ICN-specific protocols to communicate with other fog nodes. Apart from connecting to other fog nodes, each fog node is also connected to the cloud. The authors note that ICN-Fog relies on the principles of ICN for building smart, horizontal fog-to-fog data communication that leads to reducing the application's dependency on the cloud and distributed processing in the fog. Similarly, the authors of \cite{167} explore the idea of combining Information-Centric Networking(ICN) with MEC to address mobility related issues in the MEC approach that relies heavily on the underlying host-centric networking model. 

\subsubsection{Resource Allocation Frameworks}
Sun and Ansari introduced EdgeIoT, a hierarchical architecture that aims to allocate resources through the use of VMs while maintaining user privacy \cite{22}. Sun and Nirwan use OpenFlow SDN switches to provide network management for aggregated data from IoT devices. The authors of \cite{179} propose a hierarchical MEC architecture for resource allocation in MEC. The architecture introduces the notion of field, shallow, and deep cloudlets, where the field cloudlets are collocated with the base stations, the shallow cloudlets are at aggregation points, and the deep cloudlet is at the mobile backhaul. The architecture can handle peak loads efficiently by utilizing the shallow and deep computing facilities at higher levels when the computing capacity of a field cloudlet is not enough to handle the loads from its corresponding mobile users. 

Lingen et al. \cite{97} focus on a unified approach for computing in fog and cloud computing. They argue that fog computing and cloud computing should not be complementary paradigms, but instead should be fused together. As a result, through the authors' architecture, compute nodes in the fog and cloud have the same architecture, and resources are managed in a unified way. The architecture is extended from the European Telecommunications Standards Institute (ETSI)'s standardized reference architecture for NVF management and orchestration (MANO).

\subsection{Frameworks and Programming Models: Concepts and Frameworks using Fog}

Several studies utilized the concept of fog computing to propose new concepts, ideas, and frameworks based on fog computing. 
\subsubsection{Vehicular Fog Computing}
The authors in \cite{66, 121} proposed the idea of vehicular fog computing (VFC) by utilizing vehicles as the infrastructures for communication and computation. VFC takes advantage of a dynamic group of vehicles to help increase computational power and decrease latency issues. Different from the vehicular cloud computing, the proposed VFC supports geo-distribution, local decision making, and real-time load-balancing. Moreover, VFC depends on the collaboration of near-located vehicles, instead of relying on the remote servers, which reduces the deployment costs and delay.

An architecture for VFC is presented in \cite {121} and is comprised of three layers: the application and services layer, the policy management layer, and the abstraction layer. The application and services layer offers a variety of real-time applications as well as new services to users, whereas the policy management layer is responsible for allocating resources to the tasks and handling basic issues such as monitoring system state dynamically. The abstraction layer is responsible for managing, provisioning, and interfacing with the physical resources and for the security and privacy of the VFC architecture. The benefits, architecture, use cases, and potential issues of VFC are presented in \cite{145}. The authors proposed a high-level architecture of vehicular fog computing, which comprises of three types of entities, namely smart vehicles as the data generation layer, roadside units/fog nodes as the fog layer, and cloud servers as the cloud layer. 

Similar to VFC, unmanned aerial vehicles (UAVs) have been considered as means to provide computing capabilities \cite{309}. In this model, UAVs act as fog nodes and provide computing capabilities with enhanced coverage for IoT nodes. Similarly, the concept of vehicular micro clouds based on map information is introduced, and, by a simulation study, investigated in \cite{256}. Vehicular micro clouds are virtual edge servers and are essentially clusters of cars that help to aggregate and preprocess data that is transferred to the cloud. 

The study in \cite{129} argues that cloud and fog computing using the current mobile networks may not be ideally suited to provide the desired levels of QoS for moving electric vehicles in vehicle-to-grid (V2G) services. They propose a hybrid computing model called ``Foud,'' in which the cloud and fog come together and are made available to the V2G systems. In the proposed model, the cloud allows virtualized computing, storage, and network resources to be available to the V2G system entities, whereas the fog temporarily integrates the stationary and mobile computing resources located at the edge of V2G networks to expand the computing capacity. 

\subsubsection{Beyond Conventional Fog Nodes}
Prazeres et al. \cite{71} proposed a new paradigm called fog of things (FoT) which uses fog computing platforms for the IoT. The authors note that, in the proposed FoT, IoT services are defined at the edge of the network and are distributed through message and service-oriented middleware. Additionally, fog of things is self-organized, consists of FoT devices, FoT gateways, and FoT servers, and can deliver IoT services in a distributed manner. With the described FoT paradigm, the authors further propose a platform for the actual implementation of the FoT paradigm. The authors in \cite{280} propose human-driven edge computing (HEC) as a new model to ease the provisioning and to extend the coverage of traditional fixed MEC solutions by utilizing devices that humans carry. 

The study in \cite{102} looks at the latency issues that may be experienced by delay-sensitive IoT applications due to the geographical distances between the physical IoT devices and the data centers. The authors consider the mobile IoT federation as a service (MIFaaS) paradigm that leverages the pool of devices managed by individual cloud providers as a whole in order to help support delay-sensitive applications. The network model considered is a cellular IoT environment with multiple LTE femtocells as the network edge nodes that supports the MIFaaS paradigm. 

\subsubsection{Fog for Transparent Computing}
The paper \cite{162} examines the question of how to leverage transparent computing to build scalable IoT platforms and proposes a tailored, transparent computing architecture for IoT applications. Transparent computing eliminates the dependency of hardware and software and allows the provisioning of cross-platform and on-demand services on lightweight IoT devices. The proposed architecture consists of several layers - end user layer, edge server layer, core network layer, cloud layer, and the management and interface layer. The end user layer consists of a variety of IoT devices, and the edge server layer is responsible for distributing computing, control, and storage functions to end user devices at the edge of the network. The core network layer is the core of the Internet, and the cloud layer is composed of a cluster of servers with massive computing and storage resources.

\subsubsection{Volunteer Edge Computing}
Researchers from the University of Minnesota studied the possibility of using volunteer resources near the edge for both computing and storage, and proposed Nebula. Some existing systems that exploit volunteer edge computing and data sharing are Grid and peer-to-peer systems such as BitTorrent, BOINC \cite{boinc}, and SETI@home \cite{seti}. While these volunteer platforms either are for compute-intensive applications (such as BOINC and SETI), or file-sharing systems (e.g., BitTorrent), Nebula supports distributed data-intensive applications through a close interaction between compute and storage resources \cite{254}. Nebula utilizes edge resources for in-situ data-intensive computing, through location-aware data and computation placement, replication, and recovery. 

\subsubsection{Path Computing}
Path computing is paradigm based on the fog computing paradigm, where a multi-tier cloud architecture is deployed over the geographic span of the network. Path computing provides storage and compute resources along a succession of data centers of increasing size, located between the IoT devices and the cloud data centers, and enables the deployment of a multi-level hierarchy of data centers along the path that traffic follows \cite{262}. The authors in \cite{262} propose path computing and CloudPath (a platform for path computing). CloudPath enables dynamic installation of light-weight stateless functions, and a distributed eventual consistent storage system. CloudPath also automatically migrates application data across data centers to minimize latency and bandwidth usage. 

\subsubsection{Fog Nodes as IoT Hubs}
A proposed application of fog nodes to allow for the interoperability of heterogeneous IoT devices is presented in \cite{39}. These specialized fog nodes are referred to as ``IoT Hubs.''  IoT devices have different computational power and different energy requirements; thus they are restricted to different communication protocols. The IoT Hub serves as a bridge between all the different physical networks and then merges them all using an all-IP network. 

\subsection{Frameworks and Programming Models: Programming Models and Data Modeling}

In this subsection, we survey the studies that introduce and propose programming models or data modeling tools for fog computing. 
\subsubsection{Distributed Data Modeling and Frameworks}

With the rise of IoT, the demand for distributed big data analytics follows; however, data is rarely shared between stakeholders due to resource, security, and privacy concerns. Zhang et al. \cite{30} proposed a new programming model, called Firework, that takes advantage of fog computing to enable safe and reliable distributed data sharing between stakeholders. Firework merges geographically distributed data through the creation of virtually shared data views that end users can access via interfaces defined by the data owners. An example use case of Firework proposed by Zhang et al. is the shared data from security cameras. Security cameras from different sources in a common geographical location can be a part of a single Firework instance. Police departments can then request access to a specific footage at a particular time from the datasets in the Firework instance to track a person or object of interest. 

One fundamental question in fog computing is the distribution of tasks between fog nodes and the cloud. How should the workload be balanced? Should the user be able to choose the balance? If yes, how so? The authors in \cite{38} propose the design of WM-FOG, a flexible software architecture that allows developers to customize policies regarding the workflow. WM-FOG allows the developer to define the synchronization policy, to choose how much data is sent to the back-end cloud and how much data is sent to the fog node. 

The study in \cite{72} focuses on the distributed deployment of process-aware IoT applications. The authors propose a mechanism to design a distributed IoT application in one place and annotate the different components of the application which are supposed to be deployed at different edge nodes with their location information. After defining deployment locations, the model identifies application fragments which need to be deployed in different locations, and annotates each fragment with the deployment location. It then decomposes the process-aware IoT application into a set of fragments based on the annotated locations. Crystal \cite{313} provides an abstraction for fault-tolerant and distributed fog application development. A fog application using Crystal can take full advantage of location transparency, self-healing, auto-scaling, and mobility support.

\subsubsection{Fog Programming Models}
The paper \cite{28} studied QoS-aware application deployment in fog infrastructure from the programming perspective. They present a prototype to support application deployment in the IoT-fog-cloud scenario\footnote{Available at {\scriptsize https://github.com/di-unipi-socc/FogTorch}}. As a future direction, one could consider other QoS attributes and include cost information to get a richer classification of eligible fog deployments. Moreover, one could account for multiple and multi-tenant deployments on the same fog infrastructure. Renart et al. proposed an edge-based programming framework that enables users to define data-driven reactive behaviors based on the content and source of data streams \cite{226}. Users define how data stream must be processed on the edge based on their content and the location of the data source. 

Saurez et al. propose Foglets, a programming model that facilitates distributed programming across fog nodes \cite{54}. Foglets provides APIs for spatio-temporal data abstraction for storing and retrieving application-generated data on the local nodes. Through the Foglets API, Foglets processes are set for a certain geospatial region and Foglets manages the application components on the Fog nodes. Foglets is implemented through container-based visualization. The Foglets API takes into account QoS and load balancing when migrating persistent (stateful) data between fog nodes. 

\subsubsection{Fog Software Stack for Android}
The authors in \cite{32} have developed a software stack for Android to implement fog computing by using smartphones as fog nodes. To do this, several challenges need to be considered. For example, mobile operating systems typically do not have priorities when dealing with memory allocation, it is hard to track the resources used by each app, and it is difficult to ensure adaptability with current software. The proposed software stack is split into four components, each handling the above challenges. 

\subsubsection{PaaS for Fog}
Current platform as a service (PaaS) models are intended for traditional cloud applications that are not latency-sensitive nor large-scale. However, using fog computing can be difficult due to the orchestration of highly dynamic and heterogeneous fog resources. Hong et al. \cite{83} propose a PaaS model called Mobile Fog that provides a programming abstraction and allows applications to easily use fog resources while supporting dynamic scaling at runtime. The Mobile Fog communication API is composed of event handlers that must be implemented and standard functions that can be called by the application. The same code can be run on a different device like smartphones, vehicles, or cameras; the developer only needs to write the code once.

Similarly, the authors in \cite{51} have proposed a PaaS architecture for automating application provisioning in hybrid fog/cloud environment. To extend existing PaaS, the authors used Cloud Foundry droplets (an open-source PaaS service), Docker containers, and REST to provide interactions between PaaS and the fog. This enabled their architecture to provide development, deployment, and management phases of hybrid cloud/fog applications. In the proposed architecture, the controller exposes an API for PaaS users to allocate resources for their applications in the deployment phase. The Cloud Foundry droplets are responsible for running the applications in the management phase. 

\subsubsection{Service Modeling}
Lingen et al. \cite{97} used the YANG modeling language to model fog services and devices, which allows for better resource orchestration and system design. In addition, modeling services and devices using YANG allows the authors to transform service intention into service instantiation in their architecture. The authors then break down the complexity of service modeling for reuse and building higher-level services for better overall service management. Through the YANG modeling language, the authors in \cite{98} are able to effectively model IoT services description for their proposed controller to handle a large volume of data. By using the YANG modeling language, the controller can optimally allocate resources across multiple networks.

\subsection{Design and Planning: Infrastructure Design}

Network planning and design of fog networks is an important research topic, and yet not many studies are performed in this area. This is due to lack of standardized protocols and definitions for fog in general. Nevertheless, there are some early efforts, and we discuss them in this subsection. The authors of \cite{171} propose a framework for cloudlet-based network design and planning. The goal of the study is to design a network based on time division multiplexed passive optical networks (TDM-PON) to optimize the network infrastructure cost while satisfying latency constraints. Fan et al. address the problem of placing the cloudlets to minimize the deployment cost of cloudlet providers and to minimize the end-to-end delay of user requests \cite{88}. Since the total deployment cost of a cloudlet provider depends on the location of cloudlets and the amount of resources for cloudlets, cloudlet providers must take into account both the end-to-end delay of user requests and the deployment cost. 


\subsubsection{Virtualization-based Infrastructure}
The paper \cite{160} addresses the tradeoff between computing and communication and propose an architecture based on F-RAN. The F-RAN-based architecture consists of radio access equipment, F-RAN nodes that provide the computing resources, end devices, and an F-RAN controller that is in charge of receiving service requests and distributing tasks to fog nodes. The authors observe that this architecture can meet the demands of ultra-low latency applications by relying on front-haul wireless communications and distributing computation tasks to multiple F-RAN nodes near the end users. 

Fog networks can be implemented by SDN and virtualization to reduce the management costs, and to improve the scalability and resource utilization. The authors in \cite{45} propose an integrated network architecture for software-defined and virtualized radio access networks with fog computing. The proposed architecture is hierarchical and has two control levels: the higher level that is the SDN controller and the lower layer that is the local controller, which could be collocated with fog nodes. The SDN controller instructs the local controllers how to process specific applications or requests. The authors further introduce a SaaS called OpenPine that enables virtualization at the network level and user control of network operation.


\subsubsection{Capacity Planning}
The authors of \cite{209} address the questions of where the edge data centers must be located and how much compute capacity needs to be allocated to each DC for cost-effectiveness while also satisfying the bandwidth and performance requirements of applications. They conclude that adding edge layer data centers results in high-cost savings for network-intensive applications while adding an intermediate data center close to the root data center is beneficial for medium to low demand compute-intensive applications. Noreikis et al. improve edge resource utilization by taking advantage of network resources such as GPUs in addition to CPUs \cite{94}. Their study includes initial capacity planning in edge nodes to meet QoS requirements.

\subsection{Design and Planning: Resource Analysis and Estimation}
As many publications on fog computing highlight the importance of this technology for ubiquitous access, service provisioning, and service discovery, several studies also focus on the resource pricing, estimation, and analysis of fog computing. In this subsection, we describe these studies. 

\subsubsection{Fog Resource Pricing}
In fog computing, fog service providers aim to offer their services to their customers as close as possible to the customers' locations. On the other hand, infrastructure providers try to maximize their infrastructure utility by renting their edge resources to the cloud/fog service providers. An attempt to model the edge resource pricing and auction between service and content providers and edge infrastructure providers is described in \cite{231}. Aazam and Huh in \cite{7} formulated an estimation for pricing of fog services based on CPU, memory, storage, and bandwidth. Their pricing model also includes incentivized pricing for active customers and takes into account fog device mobility and customer history to determine a more fair price. 

\subsubsection{Fog Energy Estimation}
Centralized data centers consume a significant amount of energy compared to small distributed servers, or nano data centers. Jalali et al. identify applications that are more energy-efficient when implemented on nano data centers than centralized data centers \cite{67}. In their models, Jalali et al. use a ``flow-based'' energy consumption model for equipment shared by many users and services, while for network equipment close to end-users, the authors use a ``time-based'' energy consumption model based on the amount of time the equipment needs to provide access to services. The authors conclude the best energy savings with nano data centers is in applications that generate and distribute a large quantity of data near end-users that is not frequently accessed (e.g., local video surveillance in homes). 

Similar to the above study, the study in \cite{68} focuses on the energy consumption of IoT applications using both fog and cloud. To increase the power efficiency of IoT applications, Jalali et al. propose the use of both fog and microgrids \cite{68}. In their paper, the authors find that if a centralized grid provides power, fog computing is more energy efficient if there is little computation to be done with the data. However, cloud computing is more efficient if high data processing is required. 


\subsubsection{Fog Resource Estimation}
The allocated resources to fog nodes must be elastic, scalable, and dynamic, because heterogeneous IoT nodes forming IoT networks are highly dynamic, both spatially and temporally. To address this challenge, the authors in \cite{122} propose an analytical model that analyzes the required fog resources for some offered IoT workload and scales fog nodes according to incoming IoT workloads. Dynamic scaling of fog nodes is scaling up or down allocated resources according to the incoming IoT workload. Ideally, when the workload is high, more fog nodes (or more resources on the fog nodes) are provisioned, whereas when the workload is low, it may be possible to free up some fog nodes or release the allocated fog resources.

Similarly, the authors in \cite{173} take a close look at server utilization. Specifically, the paper studies the relationship between the extent to which MEC is deployed and the resulting average server utilization and latency. The metric for latency is defined as the distance between the base stations and the servers processing the traffic, and the efficiency metric for server deployment and utilization is defined as the ratio between the average and peak traffic processed by the servers. 

\subsubsection{Estimating Load and Response Time}
Fog computing can reduce the web response times of modern websites by enabling the generation of dynamic web contents at edge nodes close to end users. In \cite{182}, the authors proposed a simple formula to estimate the lower bound of this reduction of response time. Moreover, they measured the response times of about 1000 popular web pages from 12 locations across the world to evaluate whether edge computing is suitable for offering dynamic web content. Based on their experiments, edge computing can reduce web responses across the world except in North America and Europe where the round-trip times are considerably short. In \cite{212}, a location-aware load prediction for edge data centers is proposed. For each edge data center, the load is predicted using its historical load time series and those of its neighboring data centers.

\subsection{Resource Management and Provisioning: Service Provisioning (Orchestration and Migration)}

Due to the limited storage capacity of fog nodes, proper resource utilization of fog nodes has a significant impact on their performance. 
\subsubsection{Service Provisioning}
The study in \cite{250} focuses on dynamic service provisioning in edge clouds from a theoretical perspective. The model in the study captures the limited capacity of fog nodes, the unknown arrival process of requests, the cost of forwarding requests to the remote cloud, and the cost of onloading a new service on a fog node. In a simulation study, the authors of \cite{21} suggest a conceptual framework for fog resource provisioning. They introduce the concept of ``fog cell,'' which is a software component running on fog devices that controls and monitors a particular group of IoT devices. Using this and other related concepts, they model orchestration of IoT devices using a hierarchical cloud/fog resource control and provide a suitable resource provisioning solution for distributing tasks among them. Recently, researchers in \cite{366} present the {\em Red Wedding Problem}, a real-world scenario motivating the need for stateful computations at the edge. They implement a prototype database for operation at the edge that addresses the issues presented in the Red Wedding Problem.

\subsubsection{Service Migration}
The authors of \cite{59} study the service migration problem in edge clouds, in response to user movement and network performance. The solution is based on based on Markov Decision Process (MDP) that considers network state and server response time in making migration decisions. In \cite{261}, it is suggested to use multi-path TCP for live migration of VMs across edge nodes to improve VM migration time and network transparency of applications. 

Farris et al. define two integer linear programming (ILP) optimization schemes to minimize QoE degradation and cost of replica deployment in service replication for MEC \cite{92}. They distinguish classic reactive service migration from proactive migration: reactive service migration is dependent on user movement and accommodates this movement by locating the most suitable target edge and then starting the process for migration; however, proactive service migration deploys multiple replicas of the user service to neighboring nodes. As a future direction, one could study path-oriented proactive migration and trade-offs between the probability of reactive migration and the cost of service replications.

\subsubsection{Orchestration Frameworks}
To align deployed applications in distributed systems, Wen et al. underscore the importance of fog orchestration \cite{62}. The authors develop methodologies for studying fog orchestration systems with a focus on challenges in reliability, scalability, and security for fog orchestration. Another effort for service orchestration is the work of the authors in \cite{82} called ``Foggy.'' Foggy is a framework for continuous and automated application deployment in the fog. It facilitates dynamic resource provisioning and automated IoT application deployment in fog architectures. Foggy assumes that IoT nodes can host Docker containers. Developers push their containerized application package and their requirement specification to an orchestration center, which is a central authority that is in charge of monitoring each IoT node's resources and deploying services on IoT nodes. 

The authors in \cite{110} propose a service-oriented middleware that aims to distribute services over fog nodes for scalability, and with the help of SDN, performs QoS-aware orchestration by scheduling flows between services. The architecture proposed mainly consists of two components - the service-oriented middleware and the distributed service orchestration engine. The service-oriented middleware abstracts device functionalities, allowing all the nodes to act as service hosting platforms, whereas the distributed service orchestration performs orchestration. Similarly, the authors in \cite{51} proposed a Platform as a service (PaaS) architecture for automating application provisioning and orchestration in hybrid fog/cloud environments.

\subsubsection{Virtualization Technologies for Fog Computing}
The study in \cite{104} explores container-oriented operational frameworks for IoT from the perspective of how lightweight virtualization can help exploit the resources offered by IoT devices. The authors propose to use Docker-based service provisioning in wireless resource-constrained IoT environments. To analyze and identify the impact of Docker management, the authors consider two use cases, based on which they propose two distinct approaches to container-based IoT service provisioning.

In their proposed framework called FADES, researchers from TUM advertise the use of unikernels (single-purpose standalone kernels) to isolate and embed application logic into bootable images \cite{169}. FADES is a modular orchestration architecture for handling heavy computation from IoT devices to edge nodes. Each FADES unit supervises a subset of IoT devices. When needed, FADES pulls the necessary application/service from the cloud to handle the requests of IoT devices. Similar to MCC, IoT devices resort to computational resources of edge nodes to handle heavy computations. In this sense, FADES is an orchestration framework for pulling required services for IoT devices from the cloud.

\subsubsection{Provisioning of Resource-limited IoT Devices}
To satisfy QoS requirements in an edge cloud built by single-board devices (e.g., Raspberry Pi), the University of Cambridge researchers in \cite{115} present a platform, PiCasso\footnote{Available at {\scriptsize https://github.com/AdL1398/PiCasso}}, for service deployment based on the service specifications and the status of the resources at the hosting devices. The architecture of PiCasso consists of edge nodes and a service orchestrator. Similar to PiCasso, an edge node resource management framework with provisioning and deployment capability is introduced in \cite{128}. The framework integrates a fog node with a cloud server and supports auto-scaling to manage edge resources dynamically. The framework is developed for resource-limited environments and hence is shown to be simple. The authors then validate the proposed framework on a location-aware and latency-sensitive online game use case (PokeMon Go). 

\subsubsection{Handover}
Handover issues of mobile IoT devices between access points should be considered in designing orchestration or migration frameworks. In a recent study, it is shown that VM handoff in the edge across cloudlets is more than an order of magnitude faster than live migration methods currently used in data centers, for typical WAN bandwidths \cite{266}. The authors in \cite{141} observe that traditional mobile network handover mechanisms cannot handle the demands of fog computation resources and the low-latency requirements of mobile IoT applications. The authors propose {\em Follow Me Fog} framework to guarantee service continuity and reduce latency during handovers. The key idea proposed is to continuously monitor the received signal strength of the fog nodes at the mobile IoT device, and to trigger pre-migration of computation jobs before disconnecting the IoT device from the existing fog node. Comparably, the authors in \cite{143} use Docker container migration between edge nodes for service handoff. 

\subsection{Resource Management and Provisioning: Placement (VM and Service)}

In this subsection, we survey research articles that address placement problems (e.g., service placement, VM placement, content placement, and caching) in fog networks. 

\subsubsection{Application (Service) Placement}
Gu et al. developed a linear programming-based two-phase heuristic algorithm to compute an optimal solution for service placement in fog computing medical cyber-physical systems \cite{8}. The authors formulate service placement problem based on user association, task distribution, VM placement, and QoS constraints. Souza et al. also formulate service allocation in a combined fog-cloud ecosystem in \cite{11}. The authors in \cite{253} address the problem of multi-component application placement on fog nodes. Each application could be modeled as a graph, where each node is a component of the application, and the edges indicate the communication between them. 

The researchers in \cite{24} investigate the problem of optimal resource provisioning and fog service placement while taking into account their QoS requirements. The authors define the fog service placement problem and formulate it as an integer linear programming problem with the objective of maximizing the utilization of the fog landscape. As a research direction, we motivate the reader to consider other constraints such as the availability of resources, the reliability of services, and the cost of resources. Similarly, an uncoordinated strategy for service placement in edge-clouds is studied in \cite{207}.

Virtual network functions (VNFs) are network services that provide some network functionality and provide flexible ways to deploy network services \cite{kong2017guaranteed}. The study in \cite{19} focuses on the QoS-aware VNF placement and provisioning problem over an edge-cloud infrastructure. They propose a strategy to determine the required resources and placement of VNFs in the two-tier carrier cloud infrastructure while considering SLA requirements. They formulate the VNF placement problem as Mixed Integer Linear Program (MILP).

The paper \cite{52} addresses the problem of allocating computing resources (specifically, containers) in edge networks. The authors introduce a contract model between cloud service providers and edge infrastructure providers for resource sharing. Based on the contract, service providers can provision their services on the micro data centers owned by edge infrastructure provides. This essentially decouples the management of the edge infrastructures with that of the service placement performed by service providers. 

\subsubsection{VM Placement}
The paper \cite{74} analyzes the VM placement problem in fog radio access networks (F-RANs) with the objective to minimize the overall back-haul traffic. The back-haul traffic is incurred in two ways: the VM replication and data transmission to the cloud. When a user connects to a fog node and requests an application service, there is no back-haul bandwidth consumption if the fog node has the application VM. Otherwise, the VM has to be replicated on the fog node, or the request is forwarded to the cloud. They formulate the replica-based VM placement problem by considering the computing and storage of fog nodes, the user service constraint, as well as the edge bandwidth constraint. Similarly, the study \cite{133} addresses the issues of launching VM replicas and migrating them for latency-sensitive, computation and memory intensive applications in a MEC environment. 

Guaranteeing availability in a fog network needs a careful design, as fog nodes are presumed to be less reliable than always-provisioned cloud data centers. The authors in \cite{172} study VM placement in MEC with respect to availability. The goal is to find placement strategies for different types of multi-access edge applications with low cost while satisfying the availability requirements of an application. Comparably, the traffic-aware VM placement problem in the cloudlet mesh with the objective of minimizing the total inter-cloudlet communication traffic is studied in \cite{205}. The study in \cite{119} proposes a messaging method with low overhead that notifies fog nodes about nearby replica nodes so that the replica nodes can be used for handling requests instead of depending on cloud storage. 


\subsubsection{Caching}
Analogous to service placement, caching data on fog nodes can also considerably reduces the data retrieval delay compared to solely relying on a central data store. In \cite{168}, the authors propose a cache grouping mechanism for managing content in edge clouds. They also present a cache coherency mechanism to ensure data consistency within a cache group. Another study on cache placement is discussed in \cite{176}, where cache placement is done via the MEC paradigm in wireless networks to satisfy the demands of automated driving services. The service requests submitted by autonomous vehicles are first processed by the edge server to determine if they can be processed locally or need to be handed to the remote cloud. 

In their recent study, the authors of \cite{178} study the cache placement problem in F-RANs and consider flexible physical-layer transmission schemes. They develop centralized and distributed cache placement strategies to minimize users' average download delay while meeting the fog storage capacity constraints. Similar to cache placement, the cloudlet placement problem is of prominent importance. \cite{181} introduces the concept of movable cloudlets and explores the problem of how to cost-effectively deploy these movable cloudlets to enhance cloud services for dynamic context-aware mobile applications. To this end, the authors propose an adaptive cloudlet placement method for mobile applications. Another use of edge caching may be for 360-degree vedeo streaming. Streaming high-quality 360-degree video is challenging due to high bandwidth and low latency requirements \cite{nasrabadi2017adaptive}. In \cite{anahitaFoV} a probabilistic model of common field of view (FoV) for 360-degree video based on previous users' viewing histories to improve caching performance at the edge is introduced.

\subsection{Resource Management and Provisioning: Control and Monitoring} \label{control-and-monitoring}

\subsubsection{Control and Monitoring of Fog-based Networks}
The traditional network architecture was not designed with high levels of scalability of IoT devices in mind. Tomovic et al. \cite{23} and Xu et al. \cite{79} propose a control and monitoring framework for the fog-based IoT network that utilizes SDN. Using SDN allows a system to know the requirements of the network and all of the resources it has, giving it the ability to handle large waves of data \cite{mirkhanzadeh2018twolayer}. In the proposed model, a global view of the network allows for easy fog orchestration. The SDN controller controls fog orchestration as it can see all the available fog nodes and their resources such as RAM, storage, and software applications. The authors of \cite{79} use the message queuing telemetry transport (MQTT) protocol in addition to SDN to enable effective and reliable delivery of IoT data. The SDN controller is a fog node that is the broker for MQTT clients. 

The study \cite{103} explores the idea of implementing a wide area SDN controller for fog and cloud while also taking into consideration the problem of minimizing the overall carbon footprint of data centers. The paper \cite{77} proposed an SDN-enabled wireless fog architecture by combining Openflow and distributed wireless protocols. The architecture deploys a hybrid SDN control plane to address the limitations of both the centralized and distributed SDN controllers. In case of a controller failure or overhead, additional controllers are added at runtime as required to balance the network performance. Likewise, the study \cite{78} motivates how edge computing can benefit from SDN, proposes a collaboration model between SDN and edge computing through practical architectures, and shows that SDN can feasibly operate within the edge computing infrastructures.

\subsubsection{Virtualization for Control and Monitoring}
Some studies suggest bringing virtualization of network services to the edge devices. The authors in \cite{108} argue that the heavy footprint of today's virtualization platforms is responsible for preventing them from being utilized at the network edge. They present Glasgow Network Functions (GNF), a container-based NFV platform that runs and orchestrates lightweight VNFs. They show that the presented framework has low VNF instantiation time and memory requirements as compared with other existing virtualization technologies, making it suited to run on a variety of edge devices. On the other hand, the TelcoFog controller \cite{98} extends the functionalities of an NFV orchestrator for dynamic deployment of virtualized functions. The main component in the TelcoFog controller is the Resource Orchestration, which defines and enforces orchestration logic based on fog node status provided by the Resource monitoring module. 

\subsubsection{Control and Monitoring Other Networks via Fog}
Vehicular ad hoc networks (VANETs) face many issues such as unreliable connectivity, delay constraints, and poor scalability. The authors of \cite{14} suggest that using principles of fog computing along with SDN could solve many of the current problems with VANETs. In their architecture, the vehicles, which act as end-users, are SDN wireless nodes. These wireless nodes send their data to Road Side Units (RSUs) which are installed alongside road systems. Once the data is sent to an RSU, it is then sent to an RSU controller (RSUC) which is a cluster of RSUs connected by broadband. The RSUC is capable of data storage and processing. Finally, the RSUCs communicate with the SDN controller, which has global knowledge of the VANET system. 

\subsection{Operation: Scheduling, offloading, and Load Balancing}

In this subsection, we survey the studies that address job scheduling, job dispatching, task offloading, and load balancing in fog networks. 
\subsubsection{Offline vs. Online}
It is worth mentioning that we discuss several candidate papers among all of the papers in this area since there are a large number of papers in this area and most of them study ``centralized'' (referred to as {\em offline}) scheduling, dispatching, offloading, and load balancing. In centralized settings, either full information about the tasks, network, or nodes is known; or a centralized entity decides where each task could be sent. Even though there is a large body of research in this area, a more challenging setting is the ``distributed'' version of the problem (referred to as {\em online}) where there is no central authority knowledgeable of the tasks, nor is the full information about the tasks, network, or nodes known. Note that, in the online scenario, information on each job is unknown before its release.


\subsubsection{Cooperative Offloading and Load Sharing}
The authors in \cite{194} propose a delay-minimizing collaboration and offloading policy for fog-capable devices that aims to reduce the service delay for IoT applications. They also develop an analytical model to evaluate the policy and show how the proposed framework helps to reduce IoT service delay. In \cite{223}, the authors proposed a locality-aware workload sharing scheme for mobile edge computing environments. In the proposed scheme, each node is aware of its neighboring nodes and their current workloads and utilizes such information for workload sharing in case of high workload. Another distributed latency-aware task processing and offloading model is proposed in the study \cite{159}. In this model, gateways, which act as fog nodes, dynamically exchange their processing and storage capability information. Based on this information, fog nodes probabilistically forward their task to their neighboring fog nodes or the cloud, when there is a limit in local processing or storage. 

The authors in \cite{258} argue that offloading computation-intensive tasks, such as those of image recognition, to fog nodes is not always a right decision. The last-mile network latency due to the wireless communication may not be tolerable for some applications; also, the fog nodes themselves may become the bottleneck of processing delay, if many tasks are offloaded to them. Hence, they propose to use the available resources on edge devices and propose Precog, a collaborative scheme between edge devices and fog nodes to prefetch parts of the trained models of image recognition onto the device. Comparably, the authors in \cite{248} introduce the notion of edge computing coalition, which is a collaborative edge-based resource pool of small cell base stations with cloud computing capabilities to serve computation requests. The collaborative edge computing scheme accommodates more computation workloads by offloading workload from overloaded nodes to lightly-loaded nodes. The coalition framework is based on coalition game theory, which follows a payment-based incentive mechanism to form stable groups, and which builds a social trust network for managing risks among edge nodes.

The authors in \cite{215} propose and formulate a cooperative offloading policy between two edge data centers for load balancing. The model is based on a simple rule: if a service request arrives at one data center when its buffer is full, the request is offloaded to the other cooperating data center and served by that data center. On the other hand, the study in \cite{245} analyzes an offloading policy between multiple fog data centers installed at the edge of the network in a ring topology. The study also quantitatively models and estimates the gain achieved via cooperation between neighboring fog data centers in a ring topology. 

\subsubsection{Offloading in Dynamic/Uncertain Fog Networks}
Lee et al. study the problem of fog network formation and task distribution in a hybrid cloud-fog architecture \cite{75}. Their framework differs from other studies on task allocation for fog nodes by accounting for the dynamic formation of a fog network. Since the locations fog node neighbors is an uncertainty, the authors use an online approach for quickly obtaining information of the fog network and minimizing computational latency accordingly. Their online k-secretary algorithm allows a given fog node to observe its unknown environment and determine how to offload computational tasks. A recent study that investigates the computation offloading in an uncertain wireless environment is the work of authors in \cite{166}. The authors study the computation offloading when the mobile users' behavior is subjective and may deviate from rationality. In this framework, which is modeled as a non-cooperative game, users compete for limited communication resources.

\subsubsection{Fog Offloading for Robotics}
The study \cite{48} proposes a scheme to estimate the processing time of robotic tasks in fog networks. They use OS profiling tools to roughly measure the execution times of runtime task on heterogeneous hardware. Based on the estimated computation time, the robot can decide (using a basic offloading condition) whether to do the computation task locally or offload to a nearby fog node. The authors also profile conventional robotic runtime algorithms to estimate computational times more accurately. To do so, they estimate processing time of multiple image, video, and map processing algorithms using OpenCV. In their other study, the authors investigate the problem of computation offloading in fog computing \cite{49}. An optimization formulation is presented conserving the computation, energy, and communication overheads when making an offloading decision. Networked robotics is proposed as the motivating example, where the objective is to perform a computationally intensive task over the collected data by robots, to actuate in a short time. 




\subsubsection{Privacy-aware Offloading and Scheduling}
Privacy protection is also a challenge when offloading tasks in fog networks. The authors in \cite{132} consider a system in which the mobile device generates tasks and the objective of the mobile device is to find a scheduling policy that minimizes expected long-term cost. The authors observe that in their formulated problem, the privacy issues of location and usage pattern are ignored. To this end, they propose a heuristic privacy metric that jointly quantifies location and usage pattern. 

\begin{table*}
\centering
\small
\newcolumntype{Y}{>{\raggedright\arraybackslash}X} 
\renewcommand\tabularxcolumn[1]{m{#1}} 
\renewcommand{\arraystretch}{1.5} 
\setlength\tabcolsep{3pt} 
\caption{Objectives of the papers and their corresponding explanation.}
\label{tab:objective-explanation}
\begin{tabularx}{\textwidth}{|l|Y|Y|}
    \hline
    
    \textbf{Objective} & \textbf{Explanation} & \textbf{Example} \\ \hline
    
    QoS & The proposed scheme deals with improving the Quality of Service or quality of experience (e.g., by minimizing or controlling latency or success rate) in the fog. & Algorithms for enabling real-time applications, migration engines, task offloading, dynamic fog service provisioning. \\ \hline
    
    Cost & The work considers cost parameters, such as operational cost or capital cost, in the proposed scheme. & Cost-aware replica placement, cost estimation or capacity planning for designing a fog network. \\ \hline
    
    Energy & The authors analyze the energy consumption or power in the paper. & Energy-aware computation offloading, energy-aware mobility management, federation of constrained devices. \\ \hline
    
    Bandwidth & (efficiency) The paper discusses and proposes algorithms that affect bandwidth and throughput using fog computing. & Fog resource sharing, in-network processing, edge analytics, capacity planning for designing a fog network. \\ \hline
    
    Security & Security and privacy aspects of fog computing are considered in these studies. These include vulnerabilities, security mechanisms, privacy issues, and security protocols (e.g.authentication). & Anomaly detection using fog, location privacy, authentication schemes for fog nodes. \\ \hline
    
    Foundation & The fundamental and foundations of fog computing are discussed in these papers. & Surveys, standards, reference architectures, reference frameworks, new concepts based on fog. \\ \hline
    
    RAS & Reliability, Availability, Survivability. The proposed scheme improves the reliability, availability or survivability of the fog, or uses fog to provide/improve the reliability, availability or survivability, in the event of a network/node failure. & Survivable replica placement, availability analysis of fog services, availability-aware VM placement. \\ \hline
    
    Mobility & The paper considers the mobility of IoT devices or fog nodes. & Mobility-aware fog node placement, service migration based on mobility. mobility-aware service placement. \\ \hline
    
    Scalability & The proposed scheme can efficiently scale to the large magnitude of IoT networks. & Edge analytics, scalable IoT node management, computation offloading and task assignment (not per task). \\ \hline
    
    Heterogeneity & The paper discusses heterogeneity or proposes frameworks that handle heterogeneity of devices. The algorithms and frameworks in the paper do not assume any particular type of node or network. & Cloud of things, computation offloading among fog nodes, federation of fog nodes or IoT nodes. \\ \hline
    
    Management & The paper proposes management, monitoring, or federation schemes, where fog nodes, (or IoT nodes) are managed, monitored or federated using some method of management (e.g., SDN). & SDN-enabled control of fog nodes, fog operating system, orchestration of IoT services on fog nodes, orchestration of fog nodes. \\ \hline
    
    Programmability & The proposed framework is a programming language, programming framework or data modeling for fog computing. & Fog YANG models, distributed data flow for fog, data modeling and labeling.\\ \hline
    
\end{tabularx}
\end{table*}

\subsubsection{Energy-aware Offloading and Scheduling}
Deng et al. focused on investigating power consumption and network delay tradeoff in cloud-fog services by developing a computation-based framework and workload scheduling \cite{1}. Similarly, Xiao et al. study the workload offloading problem in fog networks \cite{56}. In their scheme, fog nodes in their scheme can process or offload to other fog nodes part of the workload that was initially sent to the cloud. Fog nodes decide to either offload the workload to neighbors or locally process it, under a given power constraint. The authors study the tradeoff between quality of service and power efficiency when considering fog offloading. They propose a distributed optimization algorithm that solves the problem of optimal workload allocation to maximize QoS in terms of response time. 

\subsubsection{Quality of Results in Offloading}
The authors of \cite{188} propose a systematic optimization framework with the key idea that relaxing Quality of Results (QoR) in applications where a perfect result is not always necessary. Relaxing QoR alleviates the required computation workload and enables a significant reduction of response time and energy consumption. For the proposed framework, the authors consider a mobile edge environment where the computing tasks can be divided, offloaded, and processed in parallel by distributed edge nodes. Thus, the goal of the framework is to minimize both response time and energy consumption by jointly optimizing the selection of edge nodes' QoR levels and task assignments to all edges. 

\subsubsection{Scalability Issues in Task Scheduling and Dispatching}
The problem of task scheduling and dispatching in edge clouds has been studied extensively by several researchers in the area \cite{120,2,9,109,243}. One example is the work of Tan et al. \cite{120}, where they formulate the problem of online job scheduling and dispatching in edge clouds. According to this study, the job scheduling problem is to determine which task should be served first, and the job dispatching problem is to determine where to send the job, based on latency, required resources, etc. As a future direction for task scheduling and dispatching, it is suggested to investigate the scalability of the proposed frameworks. Often, the framework fails to scale to the large magnitude of edge cloud networks, since, for example, when an IoT device generates a job, it has to be sent to all edge cloud nodes, or all edge cloud nodes need to calculate some function and inform the IoT device about the suitable edge cloud. Often, this has to be done for every job of every IoT device; hence the scalability and communication overhead are the two major issues associated with the proposed algorithms. 

\subsection{Operation: Resource Discovery}

In this subsection, we summarize the studies that have investigated the problem of resource discovery or selection in fog computing. By ``resources,'' we mean resources in the fog networks such as IoT nodes, fog nodes, nearby devices, fog services, etc. 

\subsubsection{IoT Resource Discovery and Selection}
The goal in IoT resource discovery and selection is to provide applications with global discovery and access of IoT resources irrespective of their location \cite{294}. In order to preserve the distributed nature of the federation of IoT devices, in \cite{294} the service is realized by IoT gateways (fog nodes) through a P2P overlay. The service is implemented using a distributed hash table (DHT), where information about all available IoT resources is stored for global lookup store. 

ACACIA \cite{177} uses context awareness and employs LTE-direct for service discovery, which is a proximity service discovery technique using D2D communication in IoT. In \cite{95}, edge computing nodes are used as cloud agents near the edge to discover, virtualize, and form a cloud network of IoT devices, named Cloud of Things. This network is a geographically distributed infrastructure, in which cloud agents continuously discover resources of IoT devices and pool them as cloud resources. Similarly, in FocusStack \cite{105}, IoT devices can be selected and orchestrated using their geolocation information. Comparably, in \cite{moeini2017routing} a semantic-based and space-efficient routing protocol for IoT service discovery is proposed. 

\subsubsection{Fog Resource Discovery and Selection}
The paper \cite{193} examines the problem of discovering surrogates, which are micro-clouds, fog nodes, or cloudlets, used by client devices to offload computation tasks in a fog computing environment. In order to enable the discovery and selection of available surrogates, the authors propose a brokering mechanism in which available surrogates advertise themselves to the broker. The broker receives client requests and considers a number of attributes such as network information, hardware capabilities, and distance to find the best available surrogate for the client. The proposed mechanism is implemented on off-the-shelf home routers. The authors in \cite{53} discuss a comprehensive architecture for resource (container) selection in fog nano data centers. They introduced a 5-layered framework for task scheduling over containers, which selects a container based on energy-efficiency to meet the users' SLA requirements. Container selection and task scheduling occur through a cooperative game between special middle entities called brokers. For container scheduling and migration, Docker is used, which also helps schedule tasks over containers. 

\subsection{Operation: Applications}
In this subsection, we survey the papers related to fog computing that have used the concept of fog to develop and propose new applications in other domains. 

\subsubsection{Data Stream Processing}
Cloud-based data stream processing applications are not able to keep up with geo-distributed IoT systems. Cheng et al. designed GeeLytics, an edge analytics platform, to process real-time data streams from network edges and in the cloud \cite{87}. To process IoT data streams, the authors design their platform to account for unstructured stream data that is constantly generated, mobility and colocation of sensors, low latency, heterogeneity, and ubiquity. 
Jayaraman et al. propose a context-aware real-time data analytics platform, CARDAP, to enable energy-efficient data delivery strategies in mobile crowdsourcing applications \cite{63}. 

Microsoft Research in their paper \cite{85} has developed a real-time video analytics system which relies on edge computing. The proposed system processes live camera feeds from the traffic intersections in the city of Bellevue, Washington, and raises alerts on anomalous traffic patterns. The system could be expanded to operate in other cities and can identify dangerous traffic patterns to reduce traffic casualties eventually. Similarly, Yang in the article \cite{58} describes general model and architecture of fog data stream processing and analytics. The framework in \cite{40} enables fog nodes to support query evaluations, specifically with weather data, in a federated environment in which high data volume is expected. The authors in their study \cite{268} propose the use of vehicles as fog nodes, to process live video streams from the in-vehicle dashboard-mounted cameras; unlike smartphones that are constrained in compute resources, vehicles can support efficient computing platforms.

\subsubsection{Bandwidth Savings}
The authors of \cite{228} propose a fog-based crowdsourcing framework for precise task allocation and secure data deduplication. Fog nodes in their scheme can detect and erase the repeated data in crowdsourced reports without deducing any information. Fog helps in data size reduction by erasing duplicate data. The researchers from Georgia Institute of Technology built AppFlux, a novel mobile app streaming system based on edge-clouds, for fast and efficient delivery of mobile apps and their updates \cite{236}. This approach also relieves users from having to deal with app updates and could potentially save bytes on their mobile plans. EdgeCourier proposed in \cite{247} uses edge computing to address the bandwidth issues caused by cloud-based office applications. In a measurement study, they show that contemporary cloud storage office services (e.g., word editor, or spreadsheet) consume unnecessary bandwidth since they transmit the whole file when an update happens. They design an effective office-document-aware incremental sync approach, and EdgeCourier, that uses edge-hosted unikernels for low-bandwidth mobile document synchronization in cloud storage services.

\subsubsection{Data Analytics}
Another study to make use of edge computing is the work in \cite{239} that introduces a Wi-Fi-based in-bus monitoring and tracking system that observes mobile devices and provides analytics about people both within and outside the vehicle. The system can further use the data that is collected by the vehicle-mounted wireless device to track passenger movements, detect pedestrian flows, and evaluate how external factors impact human mobility, which provides useful analytics to transit operators. Another use of resources near the edge and edge computing is in vehicular applications \cite{242}, shown by the researchers of the University of Michigan. Edge computing is used to do some analysis on the user's interactions with the car's application, to determine what priority the current interaction should have and how much of the driver's attention should be demanded. 

\subsubsection{Healthcare}
Several studies have considered the use of fog in healthcare \cite{259,131}. In \cite{259} the authors introduce processing ECG features using fog nodes, which results in low-bandwidth and low-latency data processing. In \cite{131}, the authors present a hierarchical fog-assisted computing architecture for remote IoT-based patient monitoring systems. The hierarchical computing scheme enables partitioning of analytics and decision making between the fog and the cloud and deploys. The idea is based on mapping the heavy training procedures in the cloud while outsourcing the trained hypothesis to the fog nodes periodically, and exploiting the knowledge at the edge. 

Through the utilization of the MQTT publish/subscribe protocol and fog computing paradigm, Zao et al. implemented a pervasive neuroimaging system to demonstrate the benefits of fog computing \cite{64}. They used mobile devices to act as an interface, fog servers for brain state classification, and cloud servers for further brain state analysis and archiving. Thus, the authors were able to take advantage of fog computing to classify brain states in real-time. Other researchers introduced the use of fog computing in measuring ultraviolet (UV) radiation via smartphones \cite{244}. Due to the sensitivity of CMOS sensors in mobile phone cameras to UV, the researchers have found that mobile phones have potential to measure UV radiation. Through fog servers, UV measurement results can be gathered and improved to provide more accurate UV measurements.

\subsubsection{Video and Game Analytics}
Video analytics are either computation-intensive or bandwidth hungry. Even with mobile cloud computing, there are still issues of unpredictable latency, unexpected service outage, and limited bandwidth. To address this, the authors in \cite{142} present an edge computing platform called Latency-Aware Video Edge Analytics (LAVEA) to provide low-latency video analytics at places closer to the users. Cloud gaming comes with disadvantages such as long response latency, user coverage, QoS, and bandwidth cost. The authors in \cite{185} explore approaches to deal with the challenges of thin-client massively multiplayer online gaming and propose a lightweight fog-based system called CloudFog. Fog is formed by idle machines that are close to the end-users and connect to the cloud. In CloudFog, the intensive computation of the new game state of the virtual world is conducted in the cloud which then sends update messages to fog nodes. The fog nodes update the virtual world, render game videos, and stream videos to the players. 

\subsubsection{Image and Face Recognition}
The authors of \cite{192} focus on image recognition based mobile applications that are latency sensitive and are soft real-time in nature. They present the idea of using an edge server as a cache with computing resources. The authors show that using an edge server as a typical web cache does not reduce latencies much, and therefore propose Cachier. Cachier is an image recognition cache that leverages the spatiotemporal locality of requests by storing appropriate requests locally and minimizes expected latency by dynamically adjusting its cache size. 

Traditionally, face recognition includes face identification and resolution, and requires performing computation-intensive tasks in the cloud and transmitting raw facial images to the cloud, which is bandwidth intensive. The authors in \cite{139} observe that migrating part of the resolution tasks to fog nodes and transmitting only the feature value to the cloud can significantly reduce network traffic. To this end, they propose fog-based face identification and face resolution frameworks. 

\subsubsection{Artificial Intelligence and Machine Learning}
\cite{197} presents edge stochastic gradient descent (EdgeSGD), a decentralized SGD algorithm for solving linear regression problem with the objective of estimating the feature vector on the edge node. EdgeSGD is used to predict subsurface seismic anomaly via real-time imaging. The edge nodes form a mesh network, and the algorithm obtains the image by collaboratively optimizing the objective function over the edge network. The proposed algorithm is entirely decentralized and does not require synchronization. Comparably, an edge-assisted adaptive deep learning framework for mobile object recognition is introduced in \cite{232}.


\subsection{Software and Tools}

In this subsection, we describe the software and tools that are developed for fog computing. 

\subsubsection{Simulation and Emulation}
Gupta et al. focused on implementing resource management techniques in fog computing to measure latency and throughput, and implemented iFogSim\footnote{Available at {\scriptsize https://github.com/cloudslab/ifogsim}}, a Java-based tool for simulation of fog networks \cite{34}. iFogSim relies on CloudSim, a cloud simulation tool that enables modeling and simulation of cloud systems and application provisioning environments. iFogSim supports cloud-only placement and edge-ward placement to demonstrate the scalability of fog-based applications. An extension\footnote{Available at {\scriptsize http://www.lrc.ic.unicamp.br/fogcomputing/}} of iFogSim to support mobility through migration of VMs between cloudlets is implemented in \cite{200}. Another extension is proposed in \cite{273} to simulate scenarios with strategies aiming to optimize data placement\footnote{Available at {\scriptsize https://github.com/medislam/iFogSimWithDataPlacement}}.

Similarly, the authors in \cite{213} propose another edge computing simulation environment, EdgeCloudSim\footnote{Available at {\scriptsize https://github.com/CagataySonmez/EdgeCloudSim}}, that considers both network and computational resources and covers all aspects of edge computing simulation modeling, including network and computational modeling. Similar to iFogSim, EdgeCloudSim relies on CloudSim as well. Additionally, EdgeCloudSim provides a modular architecture to provide support for a variety of critical functionality and supports simulating multi-tier scenarios where multiple edge servers are running in coordination with upper layer cloud solutions. 

Nevertheless, compared to simulation, emulation supports both repeatable and controllable experiments with real applications. The authors in \cite{324} propose their software implementation of EmuFog\footnote{Available at {\scriptsize https://github.com/emufog/emufog}}, an extensible emulation framework for fog computing. EmuFog enables the user to design network topologies, embed fog nodes in the topology, and run Docker-based applications on those nodes connected by an emulated network. The modules of EmuFog are easily extensible, although EmuFog provides a default implementation for them \cite{324}. Another emulation environment for fog computing is FogBed\footnote{Available at {\scriptsize https://github.com/fogbed/fogbed}} that is based on Mininet and Docker \cite{351}.  

\subsubsection{Edge Computing Middleware}
One of the software implementations for fog is developed by the researchers of the University of Wisconsin, named ParaDrop \cite{81}. ParaDrop is an edge computing platform that runs on WiFi access points to enable edge computing at the extreme edge of the network. Developers can use this edge computing platform to deploy services, which should be based on Docker containers, on the WiFi access points. ParaDrop is available as an open source project\footnote{Available at {\scriptsize https://paradrop.org}}, along with the documents and tutorials. ParaDrop has three components: ParaDrop access points, the ParaDrop controller, and the ParaDrop API. Using ParaDrop API, cloud services could be dynamically deployed on access points using the ParaDrop controller, which is a back-end controller that developers interact with to develop their desired services.

Analogous to offloading computation and task to either cloud or fog, a group of researchers from Georgia Institute of Technology argue the utility of ``onloading'' cloud services to the edge of the network to address the bandwidth and latency challenges of IoT networks \cite{225}. They view the cyber-foraging (e.g., code/task offloading) research domain as client-based methodology, while their proposed approach is ``backend-based'' cyber foraging or onloading. They define onloading specific services (e.g., caching, or aggregating traffic) near end-users, with a goal of minimizing user-perceived latency. They propose AirBox, a software platform based on containers for fast, scalable and secure onloading of edge services.

Bruneo et al. designed Stack4Things\footnote{Available at {\scriptsize http://stack4things.unime.it}}, a framework based on OpenStack IaaS middleware that adopts a cloud-oriented model for IoT resource provisioning \cite{70}. Their framework allows injected code at runtime through the cloud, which they define as ``contextualization.'' Similarly, the authors in \cite{198} revise OpenStack Nova service for compatibility with fog/edge systems, by leveraging a distributed key/value store instead of the centralized SQL backend\footnote{More information and code available at {\scriptsize http://beyondtheclouds.github.io}}. 

One commercial edge computing platform for IoT gateways is the {\em Everyware Software Framework} recently developed by Eurotech\footnote{Available at {\scriptsize https://esf.eurotech.com}}. Similarly, EdgeX Foundry\footnote{Available at {\scriptsize https://www.edgexfoundry.org}} is a vendor-neutral open source project building a common open framework for edge computing. UC Berkeley researchers also implemented a generic and platform-agnostic open source edge computing framework\footnote{Available at {\scriptsize https://github.com/marckoerner/oci}} called Open Carrier Interface (OCI) \cite{335}.

\subsubsection{Edge-based Data Analytics}
Xu et al. proposed edge analytics as a service (EAaaS) to promote a lightweight, scalable, and low-latency service model for IoT devices \cite{73}. The primary motivations for EAaaS are the lack of desired real-time responsiveness, a pricy pay-as-you-go model, and data privacy concerns associated with cloud-centered IoT analytic services. EAaaS provides RESTful interfaces for outside applications, an edge analytics agent on the gateway side, and an edge analytics SDK to allow users to develop methods for integrating with devices. EAaaS is provided as a part of IBM Watson IoT platform on IBM Bluemix Cloud. In their work, the authors plan to provide software upgrade capabilities as part of the existing RESTful services and utilize machine learning for existing analytic models. Comparably, SpanEdge\footnote{Available at {\scriptsize https://github.com/Telolets/StormOnEdge}} provides a programming environment that allows programmers to specify parts of their applications that need to be close to the data source, without knowledge of the number of data sources and their geographical distributions \cite{240}. 

\subsubsection{Distributed Deep Learning}
Harvard University researchers recently proposed a framework and its software implementation\footnote{Available at {\scriptsize https://github.com/kunglab/ddnn}} for distributed deep neural networks (DDNN), that can span over cloud, fog, and end devices \cite{187}. The framework map sections of a deep neural network (DNN) onto a distributed computing graph. While the resulting DDNN allows for deep inferences in the cloud, it accommodates fast and localized inference via some shallow layers of the DNN at the edge and end devices. DDNN inherently enhances sensor fusion, data privacy, and fault tolerance for DNN applications. According to this study, DDNN can reduce the communication overhead by a factor of 20x, compared with the traditional method of sending raw sensor data to the cloud.

\subsubsection{Trust Establishment}
A traditional solution for establishing trust between two entities is to create and share credentials in advance, and then to use a third-party to validate the credentials of the nodes. Nevertheless, the characteristics of some environments (e.g., environments with intermittent or no Internet connectivity), do not consistently provide access to third-party authority. A team of researchers from CMU designed and implemented a system for establishing a trusted identity solution based on cloudlets \cite{227}. The authors discuss a software implementation\footnote{Available at {\scriptsize https://github.com/SEI-TTG/pycloud/wiki}} of an in-the-field solution for establishing trusted identities in disconnected edge environments. 


\subsection{Testbeds and Experiments}

Several studies in the fog computing research area implement testbeds or conduct experiments to verify the concepts and ideas experimentally. In this subsection, we look at such studies. 

\subsubsection{Where is The Edge?}
One might ask ``Where is the edge of the network?'' Edge computing or MEC as general concepts do not restrict the specific location of the edge compute and storage nodes, and most researchers assume they are ``at the edge of the network'' or ``in the users' proximity.'' A recent measurement study by Bell Labs researchers \cite{255} shed light on actual values of delay and provided a realistic picture of LTE deployments for edge computing. They found that the first hop (UE to base station) imposes latency of 10-12 ms and adds a sawtooth pattern with an amplitude of about 40 ms. They also found that in some cases the latency of the first aggregation stage dominates the end-to-end latency.

\subsubsection{Experiments with Single-board Computers as Enablers of Fog Computing}
In \cite{184}, the authors proposed a container-based architecture for edge-based PaaS, in which applications are orchestrated between nodes at the edge. To show that the proposed architecture can meet the requirements of edge computing, such as cost-efficiency, and low power consumption, the researchers implemented their solution on a cluster of Raspberry Pi (RPi) devices, which are resource-limited devices. Researchers in \cite{229} have looked at the problem of bringing cloud services to rural and remote areas in developing countries, where building large and expensive data centers may not be feasible. They propose a hardware platform based on a cluster of single-board computer  (e.g., RPi) for making cloudlet nodes in rural and remote areas that offer cloud services. 

Elkhatib et al. \cite{112} propose the use of micro-clouds, which are small deployable computational infrastructures, to deploy fog networks. A device as small as an RPi can be used to create a micro-cloud. The authors note that, unlike mini data centers, micro-clouds are portable, easy to set up, low cost, and can be deployed in rougher environments. They run tests to compare the traditional cloud architecture to the micro-clouds and concluded that micro-clouds are more suitable for scenarios where the cloud server is far away, suffering from high latency. Besides, the boot time of micro-clouds is significantly faster than cloud. New RPi models are capable of booting up 40\% faster than an Amazon EC2.  

\subsubsection{Experiments with Lightweight Virtualization Technologies as Enablers of Fog Computing}
In \cite{190}, the authors evaluated three container orchestration tools, namely Google Kubernetes, Docker Swarm and Apache Marathon, to study their applicability for IoT networks. To do so, they defined three requirements that an effective container orchestration solution for such an environment must meet. First, adding/removing a new fog node to a cluster must be seamless, requiring a minimal software package installation on that node. Second, scheduling an application to a specific fog node must be possible. Third, all available hardware resources on a fog node must be accessible by the container on top of it. Therefore, they proposed a new container orchestration framework based on Docker Swarm for fog computing environment that can meet all of the requirements above. 

Similarly, The paper \cite{201} presents an evaluation of Docker as a container to host applications at the edge for enabling edge computing. The evaluation is based on four fundamental requirements for edge computing, namely, deployment and termination, resource and service management, fault tolerance, and caching. To evaluate and examine Docker as a candidate for edge computing, a testbed with a data center and three edge sites is simulated, and the four requirements of edge are evaluated in this testbed. Authors in \cite{lightweight-virtualization} study how lightweight virtualization technologies, such as containers and unikernels, can be integrated with edge architectures and be suitable for IoT pervasive environments. They present three IoT use cases, in which lightweight virtualization solutions can bring benefits and desirable design flexibility.

\subsubsection{Experiments with High Computation Applications}
Interactive wearable cognitive assistance application is often considered a killer app for edge computing and cloudlets \cite{263}. In an empirical study on latency \cite{263}, the performance of several such applications is evaluated. The authors in \cite{217} proposed a new serverless architecture for MEC environments, in which mobile app computation is offloaded among edge nodes to reach high throughput while keeping the latency low. To evaluate their architecture, they studied a mobile augmented reality application, in which captured frames from camera must be analyzed to detect the point of interest of objects. The authors in their recent study \cite{291} investigate the benefits game developers may obtain by exploring emerging edge cloud technology, specifically use of Google's Edge Network. They demonstrate how massively multiplayer online games benefit from this new system through simulations. Another recent experimental study is the work in \cite{332} where authors compare the performance of machine learning packages on the edges, including TensorFlow, Caffe2, MXNet, PyTorch, and TensorFlow Lite.


\subsection{Hardware and Protocol Stack}

In this subsection, we summarize the literature that introduced specific hardware for implementation of fog computing, cloudlet, F-RAN, etc., or that proposed a protocol stack for fog. 

\subsubsection{Hardware}
Intel recently has released a documentation for Fog Reference Unit, a reference design in a self-contained enclosed chassis for testing and demonstration of fog use cases \cite{fog-intel-node}. Intel's Fog Reference Unit can be seen as a generic fog node. The authors of \cite{155} propose to use optical fiber networks with MEC and present a generic fiber-wireless (FiWi) architecture for IoT. The network architecture consists of two parts: the backbone network for connecting to centralized cloud servers, and the FiWi network for providing MEC services for IoT devices. The authors in \cite{171} propose the use of time division multiplexed passive optical networks (TDM-PON) and optical network units (ONUs) for designing a cloudlet-based network. The architecture for fog computing in \cite{33} is based on the general-purpose processor (GPP) platform, which allows the architecture for processing general and shared resources. 

In \cite{165}, the authors focus on the requirements of an industrial IoT (IIoT) gateway with respect to heterogeneous network communication, management, big data, and other services. They describe the advantages of using a multi-Microcontroller architecture. Their architecture incorporates a high-speed parallel bridge controller using a reconfigurable field programmable gate array (FPGA) to overcome the serial communication bottlenecks in a traditional MCU architecture. The proposed architecture of the IIoT multi-Microcontroller gateway consists of three major modules: a master controller for the cloud, a high-speed bridge controller for data and instruction exchange, and slave controllers for IoT management and database operations. The high-speed bridge controller module is the core of the IIoT gateway and is responsible for packaging tasks and controlling communication with the other modules. 

The paper \cite{135} proposed a Cloud-Fog Radio Access Network (CF-RAN) by jointly considering the fog computing paradigm over a Time and Wavelength-Division Multiplexing Passive Optical Networks (TWDM-PON). The proposed CF-RAN can place services provided by the cloud onto the fog nodes by adopting NFV to process the baseband signals sent from the RRHs. Similarly, \cite{163} focuses on interference mitigation, resource optimization, and mobility management in F-RAN. The authors first present the system architecture that illustrates how the various components in F-RAN, such as macro RRHs (MRRHs), small RRHs (SRRHs), fog computing access points (F-APs), and smart user equipment (F-UE), work together for the successful implementation of F-RAN. The MRRHs, SRRHs and the F-APs connect to the BBU pool which supports resource optimization and provides centralized storage and communications in F-RAN. 

\subsubsection{Protocol Stack}
The IoT hub proposed in \cite{39} introduces a protocol stack for an IoT gateway (e.g., fog node) that interacts with the smart object devices using a variety of network protocols such as IEEE 802.15.4, IEEE 802.11, or Bluetooth Low-Energy. The job of the application layer includes service and resource discovery, maintaining a resource directory, acting as an origin server, and providing a CoAP-to-CoAP/HTTP-to-CoAP Proxy and cache. The network and physical/link layer both work together to provide border router functionality, allowing the IoT Hub to act as a gateway/bridge between multiple constrained networks.

Another protocol stack for fog is introduced as FogOS \cite{99}. The authors view the entire IoT ecosystem as a computer and utilize operating system concepts to create FogOS to manage this abstract computer. FogOS is composed of four main layers: service abstraction layer, application manager layer, resource manager layer, and device abstraction layer. The role of the service and device abstraction component is to provide a service APIs and device data model, respectively. Resource management then pools or slices resources of the fog and devices as needed. The application manager manages IoT applications, finds the proper edge resource for a service request, and resources of currently running applications. 


\begin{table*}
\centering
\footnotesize
\newcolumntype{B}{>{\raggedright\arraybackslash}X} 
\newcolumntype{s}{>{\hsize=.28\hsize\raggedright\arraybackslash}X} 
\newcolumntype{S}{>{\hsize=.5\hsize\raggedright\arraybackslash}X}  
\newcolumntype{b}{>{\hsize=.7\hsize\raggedright\arraybackslash}X}  
\renewcommand\tabularxcolumn[1]{m{#1}} 
\renewcommand{\arraystretch}{1.5} 
\setlength\tabcolsep{3pt} 
\caption{Summary of challenges and future research directions}
\label{tab:future}
\begin{tabularx}{\textwidth}{sbBsS}
    \hline
    
    \textbf{Challenge} & \textbf{Current Limitation} & \textbf{Research Direction or Potential Solution} & \textbf{Related Features or Objectives} & \textbf{Related Categories} \\ \hline
    
    Fog system SLA & SLAs are not defined for fog systems. Current SLAs are defined for cloud services or network infrastructure. & $\bullet$ Define new and compatible SLA for fog systems. ~~$\bullet$ Design SLA management techniques and framework for fog computing. $\bullet$ Support for multi-vendor or provider SLA for fog systems & QoS, Cost & Architectures and Frameworks for Fog; Control and Monitoring  \\ \hline
    
   Multi-objective fog system design & Many schemes (e.g., offloading, load balancing) consider few objectives and ignore other objectives. & $\bullet$ Propose schemes that consider multiple objectives (e.g., latency, bandwidth, energy) simultaneously (e.g., an efficient task offloading scheme that considers bandwidth, waiting time, availability, security, and energy). & QoS, Cost, Energy, Bandwidth & Resource Analysis and Estimation; Scheduling, offloading, and Load Balancing; Testbeds and Experiments \\ \hline
    
   Bandwidth-aware fog system design & Few works consider bandwidth savings through the use of fog computing, even though one of the promising features of fog computing is to reduce bandwidth usage of the core. & $\bullet$ Need more studies on bandwidth savings through the use of fog computing. $\bullet$ Perform measurement studies to capture the actual bandwidth usage in the presence of fog. & Bandwidth & Testbeds and Experiments; Scheduling, offloading, and Load Balancing; Control and Monitoring; Resource Analysis and Estimation; Infrastructure Design \\ \hline
    
   Scalable design of fog schemes & Many of the existing schemes and algorithms for fog do not scale to the magnitude of IoT networks. & $\bullet$ Design scalable algorithms and schemes for fog systems, e.g., online task offloading scheme that does not consider individual IoT nodes for decision making $\bullet$ Verify scalability of the algorithm and schemes by actual implementation. & Scalability & Service Provisioning; Placement; Scheduling, offloading, and Load Balancing; Applications \\ \hline
    
   Mobile fog computing & Most of the existing literature assumes fog nodes are fixed, or focus on the mobility of IoT devices. If fog nodes are mobile, resource availability, offloading, and resources provisioning will be more challenging. & $\bullet$ Propose mobile fog computing, where fog nodes can move. $\bullet$ Scheme for management or federation of mobile fog nodes. $\bullet$ Provisioning method for mobile fog services to keep the service always-available for IoT nodes. $\bullet$ Design of mobility-aware task offloading and scheduling schemes when fog nodes are mobile. & Mobility, Management & Resource Discovery; Concepts and Frameworks using Fog; Programming Models and Data Modeling; Service Provisioning; Security and Privacy; Scheduling, offloading, and Load Balancing \\ \hline
    
   Fog resource monitoring & Few studies address monitoring of fog resources. Monitoring is more challenging if multiple operators use a fog node. & $\bullet$ Multi-operator fog resource monitoring techniques. $\bullet$ SDN-based monitoring software for resource monitoring and resource advertisement. & Management, programmability & Control and Monitoring; Software and Tools \\ \hline
   
   Green fog computing & Improving the overall energy consumption of fog has not been well studied (literature considered energy-aware computation offloading, energy-aware mobility management, and federation of IoT devices to improve energy consumption). & $\bullet$ Use of energy harvesters and battery storages for IoT devices and sensors. $\bullet$ Energy-aware fog node placement, e.g., close to renewable energy resources (solar, wind, or vibration) & Energy & Infrastructure Design; Resource Analysis and Estimation \\ \hline
    
   Support of high-speed users & Current communication protocols do not support high-speed users. & $\bullet$ Develop fast or stateless handshake protocols for high-speed users, e.g., users in vehicles or for automotive communication. $\bullet$ Develop machine-learning-based mobility prediction algorithms. & Mobility, RAS & Architectures and Frameworks for Fog; Service Provisioning; Resource Discovery \\ \hline
    
   Fog node security & Fog nodes normally are located close to users, e.g., at the base stations, routers, or even at extreme network edge such as WiFi access points. This makes their security challenging. & $\bullet$ Design of physically secure fog nodes against site attacks $\bullet$ Design secure hardware, safe against physical damage, jamming, etc. $\bullet$ Design strong access-control policies for fog nodes. & Security, Heterogeneity & Security and Privacy; Infrastructure Design; Hardware and Protocol Stack \\ \hline
    
    SDN support for fog & SDN does not have native support for fog computing. & $\bullet$ Enhancing and standardizing SDN (e.g., OpenFlow northbound, southbound, east-west bound interface) for fog use cases. & Foundation, programmability & Definition and Standards; Software and Tools \\ \hline
       
   Fog node site selection & Few studies address the fog node site selection problem, which is a design problem for finding appropriate locations for deploying for nodes. & $\bullet$ Developing fog node site selection strategies that considers communication, storage, and computing (a communication hotspot may not be a storage or computing hotspot). $\bullet$ Considering cost in fog node site selection strategies (e.g., deploying fog nodes in Manhattan may be a good decision concerning latency and bandwidth, but may not be a good decision with respect to rental costs). & Cost, RAS, QoS, Energy & Infrastructure Design; Resource Analysis and Estimation \\ \hline
    
   Resilient fog system design & Current fog networks do not consider failure or fault in the network. Also, denial of service (DoS) attacks are more possible on fog nodes, since they are more resource-constrained than cloud servers. & $\bullet$ Fault detection, fault prevention, and fault recovery in fog-based networks $\bullet$ DoS-resilient fog system design $\bullet$ Design a coordinated protection mechanism that considers fog and cloud to guarantee availability. & RAS, Security & Control and Monitoring; Infrastructure Design; Service Provisioning; Security and Privacy \\ \hline

	 Fog federation & There is no fog federation framework or software similar to that of hybrid cloud federation schemes. & $\bullet$ Design new schemes for the federation of fog nodes, across different operating domains. $\bullet$ Design resource sharing models for fog nodes from different vendors/operators $\bullet$ Define new pricing models for federated fog resource sharing schemes & Management, programmability & Placement; Software and Tools; Resource Discovery. Service Provisioning \\ \hline
      
\end{tabularx}
\end{table*}

\begin{table*}
\centering
\footnotesize
\newcolumntype{B}{>{\raggedright\arraybackslash}X} 
\newcolumntype{s}{>{\hsize=.28\hsize\raggedright\arraybackslash}X} 
\newcolumntype{S}{>{\hsize=.5\hsize\raggedright\arraybackslash}X}  
\newcolumntype{b}{>{\hsize=.7\hsize\raggedright\arraybackslash}X}  
\renewcommand{\arraystretch}{1.5} 
\setlength\tabcolsep{3pt} 
\renewcommand\cellalign{tl} 
\begin{tabularx}{\textwidth}{sbBsS}
    \hline
    
    \textbf{Challenge} & \textbf{Current Limitation} & \textbf{Research Direction or Potential Solution} & \textbf{Related Features or Objectives} & \textbf{Related Categories} \\ \hline
           
   P2P fog computing & Current fog computing resource models assume simple client-server model. & $\bullet$ Design a complete P2P resource framework for fog computing that handles heterogeneity and mobility of fog nodes. $\bullet$ P2P fog computing is beneficial when there is no connectivity to the cloud (e.g. in disasters such as flooding, earthquake, etc.) ~~~$\bullet$ P2P fog resource pricing, where blockchain could be used for recording fog resource transactions. & Foundation, Mobility, Heterogeneity & Concepts and Frameworks using Fog; Resource Discovery; Service Provisioning; Security and Privacy \\ \hline
    
   Trust and authentication in heterogeneous fog systems & Heterogeneity of fog nodes and IoT nodes makes the conventional trust and authentication protocols unsuitable for fog systems. & $\bullet$ Design new authentication and trust mechanisms that could cope with heterogeneity of fog nodes and IoT nodes. $\bullet$ Design authentication protocols for fog nodes of different vendors/operators. & Heterogeneity, Security & Definition and Standards; Security and Privacy; Hardware and Protocol Stack \\ \hline
    
   Secure fog offloading & Offloading tasks to fog nodes might incur some security and privacy risks. & $\bullet$ Design secure and private offloading and load balancing schemes. $\bullet$ A mechanism for receivers to verify the correctness and integrity of the offloaded task. & Security, QoS & Scheduling, offloading, and Load Balancing; Security and Privacy \\ \hline
    
   PaaS for fog computing & Lack of a PaaS for fog systems, where developers can easily develop software across fog, IoT, and cloud. & $\bullet$ Developing a PaaS for fog computing, which is transparent to users and supports different communication- and application-level protocols and APIs. $\bullet$ Developing plugins for PaaS for different fog computing applications & programmability, Management & Software and Tools; Service Provisioning; Programming Models and Data Modeling \\ \hline
    
   Standardizing fog computing & Many independent definitions for fog (and fog-related computing paradigms) are being proposed. & $\bullet$ Unanimous and universally-agreed on the definition of fog computing. & Foundation & Definition and Standards \\ \hline
    
   Hardware technologies for fog & Most studies do not use new available hardware or communication technologies. & $\bullet$ Use of new hardware and communication technologies, such as non-volatile storage technologies, optical networks, and FPGAs. & Scalability & Hardware and Protocol Stack \\ \hline
    
    \end{tabularx}
\end{table*}

\subsection{Security and Privacy}

In this subsection, we discuss the studies that consider and address security and privacy issues in fog computing or utilize fog computing to improve current security systems and protocols. In \cite{138}, the authors discussed various privacy and security challenges in fog computing and surveyed the existing literature addressing these problems. Alrawis et al. \cite{113} explain security concerns with current IoT environments. They proposed a scheme that utilizes fog computing to address the security issues faced in IoT environments, specifically the distribution of certificate revocation information. Security and privacy issues of vehicular crowdsensing (VFCS) are discussed in \cite{257}. 

\subsubsection{Location Privacy}
Ting et al. \cite{191} proposed several strategies that can be utilized to prevent cyber eavesdroppers from tracking a user's location by observing the user's service migrations across edge clouds. The underlying idea is to use chaff services, or fake services, in addition to the real services and move them between the edge clouds to confuse eavesdroppers. The authors showed that by carefully moving the chaff services in conjunction with the real ones, tracking accuracy could be close to zero when the entropy of the user movement between edge clouds is sufficiently high. 

To protect the privacy of users using applications with location-based services, several private proximity detection algorithms exist. However, these algorithms fall short in the vicinity range and come with high communication and computation costs. \cite{125} presents a secure homomorphic protocol for private proximity detection in applications of location-based services that addresses the drawbacks of existing approaches. To achieve privacy in the data transmission process, the authors propose to use a homomorphic encryption scheme that provides fast encryption and decryption of data. 

\subsubsection{Data Privacy}
The authors in \cite{50} propose a privacy-preserving protocol for vehicular road surface condition monitoring. They propose a certificate-less signcryption scheme and a data transmission protocol for road surface condition monitoring that provides confidentiality, mutual authenticity, integrity, privacy, and anonymity. In \cite{154}, the authors proposed a new privacy-preserving scheme based on differential privacy technology for fog computing. The primary goal in this scheme is to prevent colluded nodes in a fog network from learning the data shared by the IoT devices, thus preserving their privacy. The underlying idea is to introduce artificial noise to the data before outsourcing, such that the colluded node cannot infer the original data by conducting statistical analysis.

A lightweight privacy-preserving data aggregation scheme for IoT networks is proposed in \cite{153}. The scheme enables a control center to compute the average and variance of data collected by various types of IoT sensors while preserving their privacy; fog nodes and control centers cannot learn the original data collected by devices. In this scheme, IoT devices are authenticated by the fog nodes. Also, the scheme is fault tolerant in the sense that if some IoT devices malfunction, the control center still can calculate the correct average and variance of the remaining devices without breaching their privacy. Similarly, for information sharing, a lightweight privacy-preserving fog-assisted information sharing scheme for health data collected by medical IoT devices is proposed in \cite{136}. 

\subsubsection{Intrusion Detection}
Although utilizing SDN switches as fog nodes seems promising (as discussed in Section \ref{control-and-monitoring}), it can increase the security risks in SDN networks. The attacker who compromises a fog node can also attempt to take advantage of the SDN switch to control the network. In \cite{127}, the authors described a man-in-the-middle attack on SDN networks relying only on the TLS protocol to secure their control channels between SDN switches and controllers. They also proposed a lightweight countermeasure utilizing Bloom Filters to detect such an attempt. Fog can be used to improve the current security system. In \cite{157}, the authors propose a new fog-based intrusion system. 

\subsubsection{Secure Protocols and Secure Data Transfer}
A recent study in \cite{170} investigates secure data transmission between IoT nodes and fog nodes. The study analytically models a threat model and some possible attacks; it then provides the necessary proofs to show that the framework is secure against the discussed attacks. The authors of \cite{47} propose constructing leakage-resilient functional encryption schemes for the fog that provide privacy, fine-grained access control, and security against side channel attacks. 

Identification and resolution of human subjects are crucial in cyber-physical systems where humans are involved. Hu et al. \cite{123} proposed a new face identification and resolution framework that utilizes fog computing to address bandwidth issues that arise in conventional cloud-based schemes. In \cite{123}, they provided the details of the security protocols that they devised to address the security and privacy concerns that arise in their system, such as confidentiality and integrity.


\section{Challenges and Future Research Directions} \label{future-section}

\begin{figure}[!t]
\centering
\includegraphics[width=\linewidth]{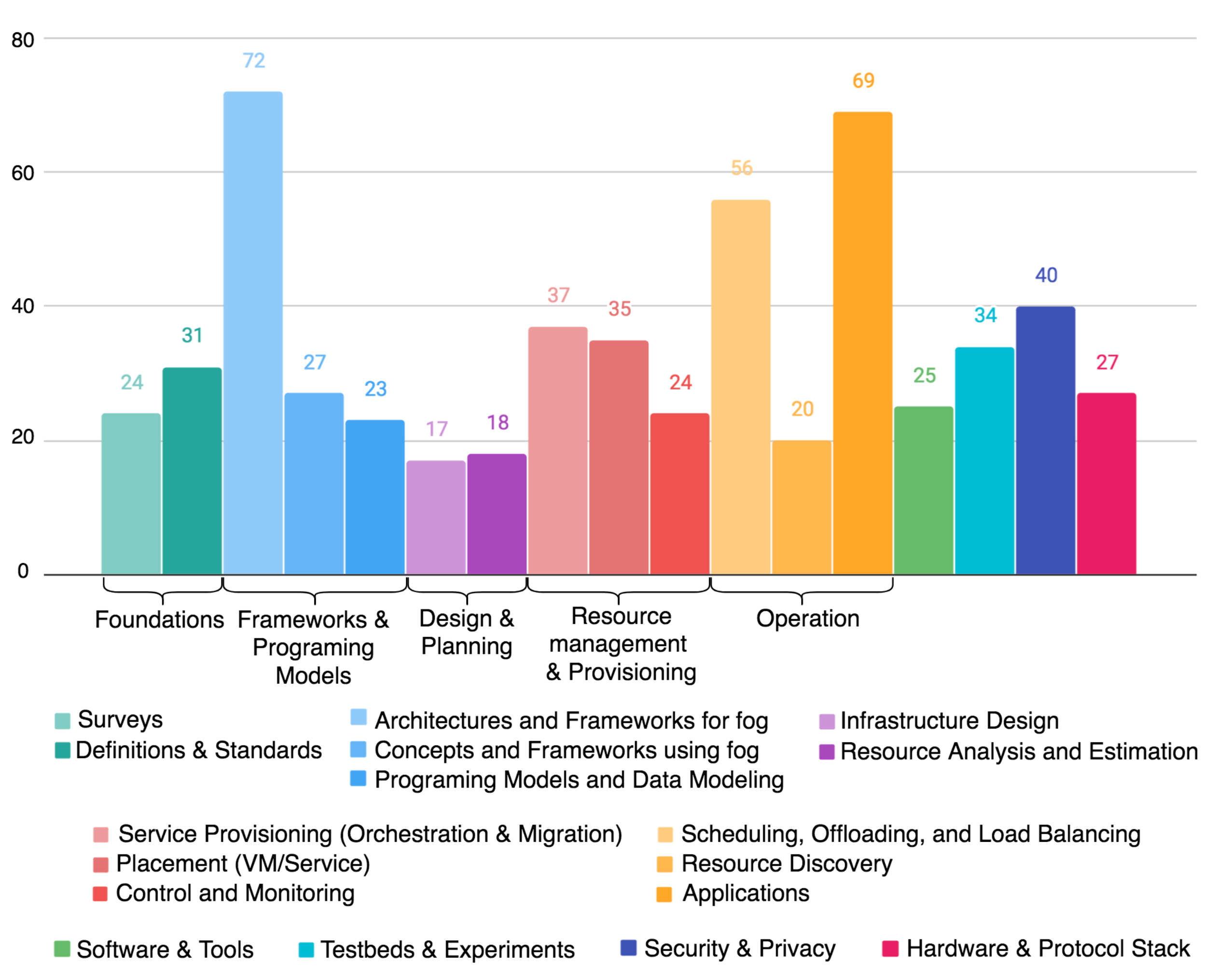}
\caption{A bar chart showing the number of research articles in this survey paper under each category of our proposed taxonomy.}
\label{fig:numbers-categories}
\end{figure}

\begin{figure}[!t]
\centering
\includegraphics[width=\linewidth]{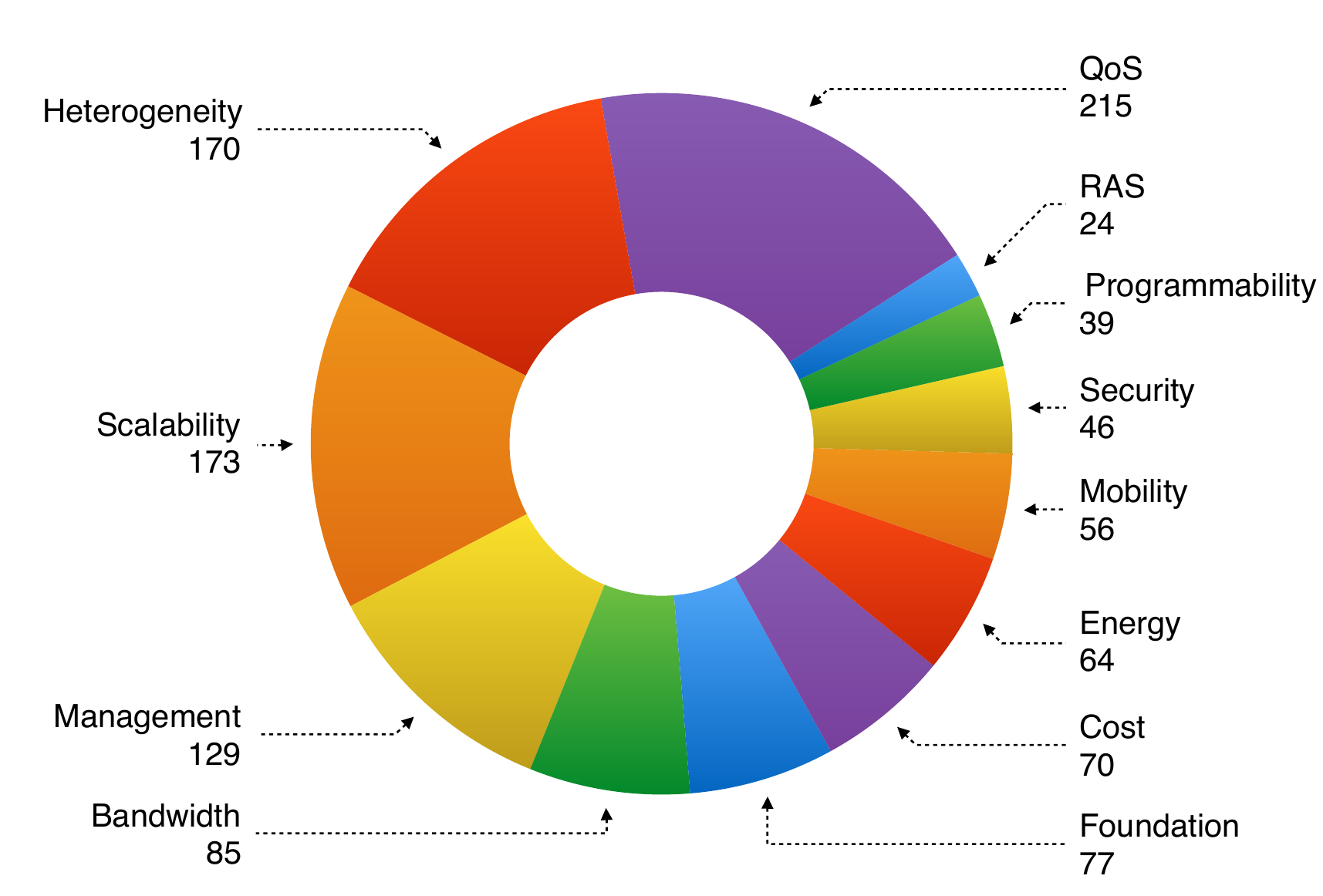}
\caption{A pie chart showing number of research articles in this survey paper that address/support a particular objective/feature.}
\label{fig:numbers-features}
\end{figure}

In this section, we discuss the current challenges and limitations of the research in the fog computing area, and we provide future directions and potential starting points for those challenges. The summary of the challenges and future research directions are included in Table \ref{tab:future}. Fig. \ref{fig:numbers-categories} shows the number of research articles in this survey that fall under each category of our taxonomy. Fig. \ref{fig:numbers-features} illustrates the number of research articles in this survey paper that address/support a particular objective/feature. The number of research articles may be an indicator of potential future research directions.


\subsubsection{Fog System SLA} 
Service level agreements (SLAs) are not currently defined for fog systems. Current SLAs that are used for fog systems are defined for cloud services (e.g., 99.99\% availability guarantee for cloud services) or network infrastructure. Moreover, a fog system may have multiple providers/operators and span across multiple operating domains. A potential research direction is defining new and compatible SLA for fog systems (e.g., guaranteeing latency and bandwidth). Additionally, designing SLA management techniques and framework for fog computing that supports for multi-vendor or provider is another potential direction. 

\subsubsection{Multi-objective Fog System Design}
Most of the existing schemes that are proposed for fog systems, such as offloading, load balancing, or service provisioning, only consider few objectives (e.g., QoS, cost) and assume other objectives do not affect the problem (e.g., \cite{56,19,110,28,134}). A new research direction will be to design schemes that consider many objectives (e.g., QoS, bandwidth, energy, cost) simultaneously. For instance, developing an efficient task offloading scheme that considers bandwidth, waiting time, availability, security, and energy at the same time is a promising direction. 

\subsubsection{Bandwidth-aware Fog System Design}
Few studies consider bandwidth savings through the use of fog computing (e.g., \cite{247}), even though one of the promising features of fog computing is to reduce bandwidth usage in the core of the Internet. There is a need for more research on bandwidth savings through the use of fog computing. These studies could be measurement studies that capture the actual bandwidth usage in the presence of fog computing.
 
\subsubsection{Scalable Design of Fog Schemes}
Many of the existing schemes and algorithms for fog do not scale to the magnitude of IoT networks, since the authors neglect scalability in their fundamental design. We believe scalability is critical in designing fog systems; fog systems should be scalable so that they could be implemented in IoT networks. For instance, a scalable algorithm for fog offloading is an online offloading scheme that does not need information of individual IoT nodes for decision making (e.g., \cite{194,245,248}). We encourage researchers in the fog computing area to verify the scalability of their proposed algorithms and schemes (e.g., by an actual implementation).

\subsubsection{Mobile Fog Computing}
Most of the existing literature assumes that the fog nodes are fixed, or only considers the mobility of IoT devices (e.g., \cite{43,163}). Less attention has been paid to mobile fog computing and how the mobile fog nodes can improve the QoS, cost, and energy consumption. When fog nodes are mobile, fog resource availability, resource discovery, task offloading, and resources provisioning will be more challenging. Mobile fog computing, where fog nodes can move and form new networks, is an interesting and challenging research direction. Moreover, designing a scheme for management or federation of mobile fog nodes is another possible direction. Along with mobile fog computing, there needs to exist new provisioning methods for mobile fog services, such that fog services are available for IoT nodes and users. Similarly, one could design a task offloading and scheduling scheme when fog nodes are mobile. An early effort for mobile fog nodes can be found in \cite{281}.

\subsubsection{Fog Resource Monitoring}
Few studies in the literature propose monitoring schemes for fog resources \cite{208}. Monitoring is useful when multiple operators use a fog node, or when a fog node is located in a location where many users use the fog node. A possible direction is developing fog resource monitoring techniques that support multi-operator access. Use of SDN-based monitoring software for fog resource monitoring and fog resource advertisement is also a promising approach.

\subsubsection{Green Fog Computing}
Few studies in the reviewed literature have addressed the energy criterion in their system design (e.g. \cite{53,55,56,67,126,63}). Most of the studies on energy are about energy-aware computation offloading, energy-aware mobility management, and federation of IoT devices to improve the energy consumption of fog systems. However, improving the overall energy consumption of fog has not been well studied. Energy consumption of a fog network includes three major portions: (1) energy consumption of IoT devices sending data to the fog, (2) energy consumption of the network interconnecting IoT devices and the fog nodes, and (3) energy consumption of the fog nodes. 

To reduce the energy consumption of IoT devices, use of energy harvesters and battery storage for IoT devices and sensors are potential research directions. Energy harvesters can improve energy consumptions while bringing new challenges to the system, such as uncertainty and unpredictability. In order to reduce the energy consumption of the network interconnecting IoT devices and the fog nodes, one of the potential research directions is to identify where to put fog nodes and how close they should be to the end users. Mobile fog nodes is also a compelling use case for energy consumption. To reduce the energy consumption of fog nodes, one potential research direction is to reduce the distance between fog servers and local renewable energy sources (such as solar, wind, or vibration). This problem can be addressed in different ways: the traffic from IoT devices can be rerouted to the nearest fog node that is powered by renewable energy. The other way is for telecommunication companies to identify the location of fog nodes that need the high amount of power to serve the traffic, and to encourage people to use their local renewable energy for their local micro-grid to power up their local fog nodes. 

\subsubsection{SDN Support for Fog}
SDN software does not have native support for fog computing. SDN is mostly commercially viable inside large data centers or campus networks \cite{78}. Enhancing and standardizing SDN software (e.g., OpenFlow northbound, southbound, east-west bound interface) for fog use cases is a direction we believe will ease the development of fog computing software. Moreover, with multiple vendors/operators in fog systems, there will be a need for new SDN architectures with multiple domains and hierarchies of SDN controllers. 

\subsubsection{Support of High-speed Users}
Current communication protocols that are proposed for fog computing environments do not support high-speed users, such as users in cars, users on trains, and vehicular computing. A research direction is to develop quick or stateless handshake and authentication protocols for high-speed users and automotive communication. Note that there have been already some early efforts in this field, such as the articles \cite{66,161,167}; nevertheless, we are still far from having a working and resilient communication protocol for high-speed users and automotive communication for fog computing. One can use machine-learning-based mobility prediction algorithms in the design of the handshake and authentication protocols, to predict the location of the high-speed users and analyze their mobility patterns for fog computing.

In addition to handshake and authentication protocols for high-speed users, fog service provisioning for IoT applications is required to be dynamic and proactive due to the rapid changes (such as connectivity, bandwidth fluctuations, or failure) in mobile and high-speed IoT environments. To address dynamic and proactive fog service provisioning, predicting the behavior and location of IoT devices and high-speed users based on historical data or machine learning methods is another potential solution that requires further investigation. 

\subsubsection{Fog Node Security}
Fog nodes are going to be placed near users, in locations such as on the base stations, or routers, or even at the extreme network edge on WiFi access points. This makes it challenging to provide security for fog nodes. Site attacks are more possible on the fog nodes than cloud data centers. A research direction may be designing secure fog node sites, safe against physical damage, jamming, etc. Also, another direction may be to designing strong access-control policies for fog nodes, so that they are secure in the presence of malicious users in the vicinity. A potential starting point for access control in fog computing can be found in \cite{47,283}.

\subsubsection{Fog Node Site Selection}
Few studies address the fog node site selection problem, which is a design problem for finding appropriate locations for deploying nodes. Fog node site selection strategies should consider communication, storage, and computing at the same time for finding an appropriate location (a communication hotspot may not necessarily be a storage or computing hotspot). Moreover, cost should also be a deciding factor in fog node site selection strategies; deploying fog nodes in Manhattan may be a good decision concerning reduction in latency and bandwidth, but may not be a good decision with respect to rental costs. Furthermore, fog node security considerations that are discussed in ``Fog Node Security'' previously could also affect the fog node site selection decision. We refer the interested reader to the recent article \cite{openfog-sec} that describes the security requirements and approaches of an open fog architecture.

\subsubsection{Resilient Fog System Design}
From the reliability and availability perspective, fog services and fog networks bring new challenges to the current network and service provisioning methods. To guarantee the availability and reliability of the fog services, a coordinated service provisioning mechanism that considers both fog and cloud computing is needed. For example, if a fog service needs some functions to process a stream of data, providing extra replicas of those functions can improve the availability of the service. On the other hand, due to the limited computing resources of the fog nodes compared to the cloud data centers, allocation of the function replicas to provide availability and reliability is not a straightforward decision. As a future direction, availability may be considered in addition to constraints, such as latency, throughput, and security when designing provisioning methods for fog services. 

Most of the articles in the literature about fog computing do not consider failure or fault in the fog network. Another research direction is to provide different protection and restoration mechanism across different layers. In addition, failure detection, prevention, and recovery are efficient ways to improve the availability of the fog services. Additionally, fog nodes are more prone to denial of service (DoS) attacks, since they are more resource-constrained than adequately-secure cloud data centers; also, recently compromising IoT nodes and embedded systems are becoming new sources of distributed DoS attacks \cite{darki16don}. Novel classes of proactive defense techniques based on moving target defense paradigm (sometimes referred to as {\em address mutation/randomization}) could be used to thwart DoS attacks \cite{ashkan-privacy, amir-moving-target}. Researchers of UC Berkeley recently have proposed a resilient edge computing framework \cite{327}, which is a good starting point in the direction of resilient fog system design.

\subsubsection{Fog Federation}
Currently, there is no fog federation framework or software (similar to that of hybrid cloud federation schemes), which controls and federates fog resources across multiple operating domains. There is a need for new schemes for the federation of fog nodes, especially when they belong to different operating domains. The federation scheme should account for resource sharing models for fog nodes from different vendors/operators. Similarly, one can define new pricing models for federated fog resources. Finally, one can propose policies for new fog resource sharing schemes (e.g., P2P fog computing resource sharing model) under the federation framework. The recent article in \cite{318} might be a good start for potential research about fog federation. 

\subsubsection{P2P Fog Computing}Current fog computing resource models assume a simple client-server model, where IoT devices (clients) use cloud computing or fog computing (server) to process their requests, while fog nodes may also offload computation among themselves or to the cloud. We argue that design of a complete peer-to-peer (P2P) resource framework for fog computing is a promising direction. Under the P2P fog computing model, fog nodes share resources in a P2P manner, where individual fog nodes share their resources (e.g., compute or storage), without an intermediary third-party. The P2P fog computing resource model could further handle heterogeneity and mobility of fog nodes (e.g., when a fog node leaves the P2P network). P2P fog computing is the most beneficial when there is no connectivity to the cloud, for instance when there are disasters such as flooding or earthquake and the connection to the cloud resources is gone. P2P fog resource pricing is another research direction, where different pricing and incentive methods could be proposed for P2P resource sharing model. Blockchain technology could be a choice for keeping track of fog resource transactions when designing a P2P fog resource pricing model. 

\subsubsection{Trust and Authentication in Heterogenous Fog Systems}
In addition to mobility that is discussed in ``Support of High-speed Users,'' heterogeneity of fog nodes and IoT devices makes the conventional trust and authentication protocols unsuitable for fog systems. Also, fog nodes may belong to different vendors and operators. Hence, we need to design new authentication and trust mechanisms for new fog systems that could cope with the heterogeneity of fog nodes and IoT devices. Some researchers have started to address heterogeneity in their fog system design, such as the articles \cite{39,186}.

\subsubsection{Secure Fog Offloading}
Offloading tasks among fog nodes might incur some security and privacy risks. The risk is when tasks containing security- and privacy-critical information is offloaded. Also, a security risk might be when a fog node becomes over-loaded (e.g., by the requests sent by a malicious user) and starts offloading security- and privacy-critical information to other fog nodes (which may be accessible to the malicious user). A research direction, hence, is to design and implement secure offloading and load balancing schemes. An early effort for privacy-aware offloading in MEC is included in \cite{132}. Furthermore, designing a lightweight and efficient mechanism for IoT receivers to verify the correctness and integrity of the offloaded tasks.

\subsubsection{PaaS for Fog Computing}
There are not any solid implementations of a PaaS for fog systems, where developers can easily develop software across fog, IoT, and cloud. \cite{51,184} are some of the very few efforts in this direction. Developing a PaaS for fog computing can ease the general development and acceptability of fog computing. The future PaaS for fog computing should hide the fog configuration specification (e.g., location of fog nodes, their interconnection, their capacity) for users, provision applications, and services proactively and automatically with minimal effort from developers, and support different communication- and application-level protocols and APIs. Once this PaaS is available, its modular design could also be taken into account so that various plugins for different fog computing applications and services could be easily integrated with the PaaS. 

\subsubsection{Standardizing Fog Computing}
Different research teams are proposing many independent definitions of fog (and fog-related computing paradigms, such as edge computing). In Section \ref{comparison} we saw that the definitions of fog computing and its related computing paradigms are not completely standardized. We believe there is a research gap in the definitions and standards for fog computing and other fog-related computing paradigms that needs to be filled by standards and universally-agreed definitions. Once the definitions are agreed upon, researchers become more clear when defining problems, and there will be more agreement among researchers and industry about these paradigms. Organizations such as OpenFog Consortium and OpenEdge Computing are already developing standards and definitions for fog computing and edge computing.


\subsubsection{Hardware Technologies for Fog}
Most studies in the area of fog computing or edge computing do not make use of new hardware or communication technologies, such as non-volatile storage technologies, optical networks, fiber-wireless (FiWi), or FPGAs. Use of new hardware and communication technologies for the design of fog networks, (e.g., fog-to-cloud interconnection) is a direction worth exploring.

\section{Conclusion} \label{conclusion}

The Internet of Things accelerates digital transformation and provides benefits to many industries, including manufacturing, energy, transportation, smart cities, education, retail, healthcare, and government. Due to IoT's fundamental benefits, the number of connected devices and the IoT networks is on the rise, as individuals and companies deploy more and more IoT devices. IoT is expected to connect billions of devices and humans to bring promising advantages for us. Fog computing is one of the promising solutions for handling the big data that is being produced by the IoT, which is often security-critical and time-sensitive. In this paper, we provided a tutorial on what fog computing is and how it relates to or differs from other computing paradigms, such as cloudlets, MEC, and edge computing. Next, we provided a taxonomy of research topics in fog computing and summarized the relevant papers on fog computing and its related computing paradigms. Finally, we provided challenges and future directions for research in fog computing.

\footnotesize
\bibliography{references}

\begin{table*}
\bibliographystyle{unsrtnat}
\tiny
\centering
\newcolumntype{Y}{>{\raggedright\arraybackslash}X} 
\renewcommand\tabularxcolumn[1]{m{#1}} 
\renewcommand{\arraystretch}{1.5} 
\setlength\tabcolsep{3pt} 
\caption{(Appendix 1) Features and objectives of the cited papers in this survey}
\label{tab:features-objectives}

\begin{minipage}{0.48\linewidth}
\begin{tabularx}{\linewidth}{|Y|c|c|c|c|c|c|c|c|c|c|c|c|}
    \hline
    
    \center{Paper} & \rotatebox{90}{QoS} & \rotatebox{90}{Cost} & \rotatebox{90}{Energy} & \rotatebox{90}{Bandwidth} & \rotatebox{90}{Security} & \rotatebox{90}{Foundation} & \rotatebox{90}{RAS} & \rotatebox{90}{Mobility} & \rotatebox{90}{Scalability} & \rotatebox{90}{Heterogeneity} & \rotatebox{90}{Management} & \rotatebox{90}{Programmability\hspace{3pt}} \\ \hline
    
    \citet{1} & \ding{51} & & \ding{51} & & & & & & \ding{51} & \ding{51} & & \\ \hline
    
    \citet{2} & \ding{51} & & & \ding{51} & & & & & & & & \\ \hline
    
    \citet{3} & & & & & & \ding{51} & & & & & \ding{51} & \\ \hline
    
    \citet{4} & & & & & & \ding{51} & & & & & & \\ \hline
    
    \citet{5} & \ding{51} & & \ding{51} & & & & & & \ding{51} & & & \\ \hline
    
    \citet{6} & & & & & & \ding{51} & & & & & & \\ \hline
    
    \citet{7} & & \ding{51} & & & & & & & & & \ding{51} & \\ \hline
    
    \citet{8} & \ding{51} & \ding{51} & & & & & & & & & \ding{51} & \\ \hline
    
    \citet{9} & \ding{51} & \ding{51} & & & & & & \ding{51} & & & & \\ \hline
    
    \citet{10} & \ding{51} & \ding{51} & & & & & & & & & & \\ \hline
    
    \citet{11} & \ding{51} & & & & & & & & & \ding{51} & & \\ \hline
    
    \citet{12} & & & \ding{51} & & & \ding{51} & & & & & & \\ \hline
    
    \citet{13} & \ding{51} & & & & & \ding{51} & & \ding{51} & \ding{51} & & & \\ \hline
    
    \citet{14} & & & & & & & & \ding{51} & & & \ding{51} & \\ \hline
    
    \citet{15} & & & & & & & & & & \ding{51} & \ding{51} & \\ \hline
    
    \citet{16} & \ding{51} & & \ding{51} & & & & & & & \ding{51} & & \\ \hline
    
    \citet{17} & \ding{51} & \ding{51} & & \ding{51} & & & & & & & & \\ \hline
    
    \citet{18} & \ding{51} & \ding{51} & \ding{51} & \ding{51} & & & & & & \ding{51} & & \\ \hline
    
    \citet{19} & \ding{51} & \ding{51} & & & & & & & & & \ding{51} & \\ \hline
    
    \citet{20} & & & & & & \ding{51} & & & & & & \\ \hline
    
    \citet{21} & \ding{51} & \ding{51} & & & & \ding{51} & &  & \ding{51} & \ding{51} & \ding{51} & \\ \hline
    
    \citet{22} & & & & & & \ding{51} & & & & \ding{51} & \ding{51} & \\ \hline
    
    \citet{23} & & & & & & & & & \ding{51} & \ding{51} & \ding{51} & \\ \hline
    
    \citet{24} & \ding{51} & \ding{51} & & & & & & & & \ding{51} & \ding{51} & \\ \hline
    
    \citet{25} & & & & & & \ding{51} & & & & & & \\ \hline
    
    \citet{26} & & & & & & \ding{51} & & & & & & \\ \hline
    
    \citet{27} & & & & & & \ding{51} & & & & & & \\ \hline
    
    \citet{28} & \ding{51} & & & & & & & & & \ding{51} & & \ding{51} \\ \hline
    
    \citet{29} & & & & & & \ding{51} & & & & & & \\ \hline
    
    \citet{30} & & & & & & & & & & & & \ding{51} \\ \hline
    
    \citet{31} & & & & \ding{51} & & & & & & & & \\ \hline
    
    \citet{32} & & & & & & & \ding{51} & & & & & \\ \hline
    
    \citet{33} & \ding{51} & & & & & & & & & \ding{51} & & \\ \hline
    
    \citet{34} & \ding{51} & & \ding{51} & & & & & & \ding{51} & \ding{51} & & \ding{51} \\ \hline
    
    \citet{35} & & & & & \ding{51} & & & & & & \ding{51} & \\ \hline
    
    \citet{36} & & & & & & \ding{51} & & & & & & \\ \hline
    
    \citet{37} & \ding{51} & & & & & & & & & \ding{51} & \ding{51} & \\ \hline
    
    \citet{38} & \ding{51} & & & & & \ding{51} & & & & & & \ding{51} \\ \hline
    
    \citet{39} & & & & & & & & & \ding{51} & \ding{51} & \ding{51} & \\ \hline
    
    \citet{40} & & & & \ding{51} & & & & & \ding{51} & & & \\ \hline
    
    \citet{41} & & & & \ding{51} & & & & & \ding{51} & \ding{51} & \ding{51} & \\ \hline
    
    \citet{42} & & & & & & \ding{51} & & & & & & \\ \hline
    
    \citet{43} & \ding{51} & & & & & & & \ding{51} & & & & \\ \hline
    
    \citet{44} & & & & & \ding{51} & \ding{51} & & & & & & \\ \hline
    
    \citet{45} & \ding{51} & & & \ding{51} & & & & \ding{51} & \ding{51} & & \ding{51} & \\ \hline
    
    \citet{46} & \ding{51} & & & & & & & & & & & \\ \hline
    
    \citet{47} & & & & & \ding{51} & & & & & & & \\ \hline
    
    \citet{48} & \ding{51} & & & & & & & & & \ding{51} & & \\ \hline
    
    \citet{49} & \ding{51} & & \ding{51} & & & & & & & \ding{51} & & \\ \hline
    
    \citet{50} & & & & & \ding{51} & & & & & & & \\ \hline
    
    \citet{51} & & & & & & & & & \ding{51} & \ding{51} & \ding{51} & \\ \hline
    
    \citet{52} & \ding{51} & \ding{51} & & & & & & & & & \ding{51} & \\ \hline
    
    \citet{53} & \ding{51} & & \ding{51} & & & & & & & & \ding{51} &  \\ \hline
    
    \citet{54} & \ding{51} & & & & & & & \ding{51} & & & \ding{51} & \ding{51} \\ \hline
    
    \citet{55} & \ding{51} & \ding{51} & \ding{51} & & & & & & & & & \\ \hline
    
    \citet{56} & \ding{51} & & \ding{51} & & & & & & & & & \\ \hline
    
    \citet{57} & & & & & & \ding{51} & & & & & & \\ \hline
    
    \citet{58} & & & \ding{51} & & & \ding{51} & & & & & & \\ \hline
    
    \citet{59} & \ding{51} & & & \ding{51} & & & & \ding{51} & \ding{51} & \ding{51} & \ding{51} & \\ \hline
    
    \citet{60} & & & & & & \ding{51} & & & & & & \\ \hline
    
     \citet{61} & & & & & & \ding{51} & & & & & & \\ \hline

\end{tabularx}
\end{minipage}
\hspace{10pt}
\begin{minipage}{0.48\linewidth}
\begin{tabularx}{\linewidth}{|Y|c|c|c|c|c|c|c|c|c|c|c|c|}
    \hline
    
    \center{Paper} & \rotatebox{90}{QoS} & \rotatebox{90}{Cost} & \rotatebox{90}{Energy} & \rotatebox{90}{Bandwidth} & \rotatebox{90}{Security} & \rotatebox{90}{Foundation} & \rotatebox{90}{RAS} & \rotatebox{90}{Mobility} & \rotatebox{90}{Scalability} & \rotatebox{90}{Heterogeneity} & \rotatebox{90}{Management} & \rotatebox{90}{Programmability\hspace{3pt}} \\ \hline
    
    \citet{62} & \ding{51} & & & & & & & & & & \ding{51} & \\ \hline
    
    \citet{63} & & \ding{51} & \ding{51} & & & & & & \ding{51} & & \ding{51} & \\ \hline
    
    \citet{64} & & & & & & & & & & & & \ding{51} \\ \hline
    
    \citet{65} & & & & \ding{51} & & & & & & & & \\ \hline
    
    \citet{66} & & & & & & \ding{51} & & \ding{51} & & & \ding{51} & \\ \hline
    
    \citet{67} & & & \ding{51} & \ding{51} & & & & & & & & \\ \hline
    
    \citet{68} & & & \ding{51} & & & & & & & & & \\ \hline
    
    \citet{69} & & & \ding{51} & & & & & \ding{51} & & & & \\ \hline
    
    \citet{70} & & & & & & & & \ding{51} & & & & \\ \hline
    
    \citet{71} & & & & & & & & & \ding{51} & \ding{51} & \ding{51} & \\ \hline
    
    \citet{72} & & & & & & & & & & & & \ding{51} \\ \hline
    
    \citet{73} & & & & & & & & & \ding{51} & & \ding{51} & \\ \hline
    
    \citet{74} & & & & \ding{51} & & & & \ding{51} & & & & \\ \hline
    
    \citet{75} & \ding{51} & & & & & & & & & \ding{51} & & \\ \hline
    
    \citet{76} & \ding{51} & & \ding{51} & & & & & & & & & \\ \hline
    
    \citet{77} & & & & \ding{51} & & & & & & & \ding{51} & \\ \hline
    
    \citet{78} & & & & & & \ding{51} & & & & & \ding{51} & \\ \hline
    
    \citet{79} & & & & & & & & & & & \ding{51} & \\ \hline
    
    \citet{80} & & & & & & & & & & \ding{51} & \ding{51} & \\ \hline
    
    \citet{81} & \ding{51} & & & \ding{51} & & & & & \ding{51} & & \ding{51} & \ding{51} \\ \hline
    
    \citet{82} & & & & & & & & & & \ding{51} & \ding{51} & \ding{51} \\ \hline
    
    \citet{83} & \ding{51} & & & & & & & & \ding{51} & & & \ding{51} \\ \hline
    
    \citet{84} & & \ding{51} & & \ding{51} & & & & & \ding{51} & & \ding{51} & \\ \hline
    
    \citet{85} & \ding{51} & & & & & & & & \ding{51} & & & \\ \hline
    
    \citet{86} & & & & & \ding{51}& \ding{51}& & & & & & \\ \hline 
    
    \citet{87} & & & & & & \ding{51} & & & & & & \\ \hline
    
    \citet{88} & \ding{51} & \ding{51} & & & & & & \ding{51} & & & & \\ \hline
    
    \citet{89} & \ding{51} & & & & & & & & & & & \\
    \hline
    \citet{90} & & \ding{51} & \ding{51} & & & & & \ding{51} & & & & \\
    \hline
    \citet{91} & \ding{51} & & \ding{51} & \ding{51} & & & & & & & & \\
    \hline
    \citet{92} & & & & & & & & \ding{51} & & & & \\
    \hline
    \citet{93} & \ding{51} & \ding{51} & & & & & & & & & & \\
    \hline
    \citet{94} & \ding{51} & \ding{51} & & \ding{51} & & & & & & & & \\
    \hline
    \citet{95} & & & & & & \ding{51} & & & \ding{51} & \ding{51} & & \\
    \hline
    \citet{96} & \ding{51} & & \ding{51} & \ding{51} & & & & & \ding{51} & \ding{51} & & \\
    \hline
    \citet{97} & & & & & & \ding{51} & & & \ding{51} & \ding{51} & \ding{51} & \ding{51} \\
    \hline
    \citet{98} & \ding{51} & & & & & \ding{51} & & & & \ding{51} & \ding{51} & \ding{51} \\
    \hline
    \citet{99} & & & & & & & & & \ding{51} & \ding{51} & \ding{51} & \\
    \hline
    \citet{100} & & & & & & \ding{51} & & & & & & \\
    \hline
    \citet{101} & \ding{51} & \ding{51} & & & & & & & & \ding{51} & & \\
    \hline
    \citet{102} & \ding{51} & \ding{51} & & & & & & \ding{51} & \ding{51} & \ding{51} & & \\
    \hline
    \citet{103} & \ding{51} & & \ding{51} & & & & & & & \ding{51} & \ding{51} & \\ \hline
    
    \citet{104} & & & & & & & & & & \ding{51} & \ding{51} & \\ \hline
    
    \citet{105} & & & \ding{51} & & & & & & \ding{51} & \ding{51} & \ding{51} & \\ \hline
    
    \citet{106} & \ding{51} & \ding{51} & & & & & & \ding{51} & & & & \\ \hline
    
    \citet{107} & & & & & & & & \ding{51} & \ding{51} & \ding{51} & & \ding{51} \\ \hline
    
    \citet{108} & \ding{51} & & & & \ding{51} & & & \ding{51} & & & \ding{51} & \\ \hline
    
    \citet{109} & \ding{51} & & & & & & \ding{51} & & & & & \\ \hline
    
    \citet{110} & \ding{51} & & & & & & & & \ding{51} & \ding{51} & \ding{51} & \\ \hline
    
    \citet{111} & & & & & & \ding{51} & & & \ding{51} & \ding{51} & \ding{51} & \\ \hline
    
    \citet{112} & \ding{51} & & & & & & & & \ding{51} & & & \\ \hline
    
    \citet{113} & & & & & \ding{51} & & & & & & & \\ \hline
    
    \citet{114} & & & & & & & & & & & \ding{51} & \\ \hline
    
    \citet{115} & \ding{51} & & & & & & & & & & \ding{51} & \ding{51} \\ \hline
    
    \citet{116} & \ding{51} & & & \ding{51} & & & & & & \ding{51} & \ding{51} & \\ \hline
    
    \citet{117} & \ding{51} & & & \ding{51} & & & & & & & & \\ \hline
    
    \citet{118} & \ding{51} & & & & & \ding{51} & & & \ding{51} & & \ding{51} & \\ \hline
    
    \citet{119} & \ding{51} & \ding{51} & & \ding{51} & & & & & \ding{51} & \ding{51} & & \\ \hline
    
    \citet{120} & \ding{51} & & & & & & & & & \ding{51} & & \\ \hline
    
    \citet{121} & & & & & & \ding{51} & & & \ding{51} & \ding{51} & & \\ \hline
    
    \citet{122} & \ding{51} & & & & & & & & \ding{51} & \ding{51} & \ding{51} & \\ \hline

\end{tabularx}
\vspace*{0pt} 
\end{minipage} 

\end{table*}

\begin{table*}
\tiny
\centering
\newcolumntype{Y}{>{\raggedright\arraybackslash}X} 
\renewcommand\tabularxcolumn[1]{m{#1}} 
\renewcommand{\arraystretch}{1.5} 
\setlength\tabcolsep{3pt} 

\begin{minipage}{0.48\linewidth}
\begin{tabularx}{\linewidth}{|Y|c|c|c|c|c|c|c|c|c|c|c|c|}
    \hline
    
    \center{Paper} & \rotatebox{90}{QoS} & \rotatebox{90}{Cost} & \rotatebox{90}{Energy} & \rotatebox{90}{Bandwidth} & \rotatebox{90}{Security} & \rotatebox{90}{Foundation} & \rotatebox{90}{RAS} & \rotatebox{90}{Mobility} & \rotatebox{90}{Scalability} & \rotatebox{90}{Heterogeneity} & \rotatebox{90}{Management} & \rotatebox{90}{Programmability\hspace{3pt}} \\ \hline
    
    \citet{123} & & & & & \ding{51} & & & & & \ding{51} & & \\ \hline
    
    \citet{124} & \ding{51} & & \ding{51} & & & & & & \ding{51} & & & \\ \hline
    
    \citet{125} & & & & & \ding{51} & & & & & \ding{51} & & \\ \hline
    
    \citet{126} & & & \ding{51} & & & & & & \ding{51} & \ding{51} & & \\ \hline
    
    \citet{127} & & & & & \ding{51} & & \ding{51} & & & & \ding{51} & \\ \hline
    
    \citet{128} & \ding{51} & & & & & & & & \ding{51} & \ding{51} & \ding{51} & \\ \hline
    
    \citet{129} & \ding{51} & & \ding{51} & & & & & \ding{51} & \ding{51} & & & \\ \hline
    
    \citet{130} & \ding{51} & \ding{51} & & & & & & & & \ding{51} & & \\ \hline
    
    \citet{131} & \ding{51} & & & & & & \ding{51} & & \ding{51} & & \ding{51} & \\ \hline
    
    \citet{132} & \ding{51} & \ding{51} & \ding{51} & & \ding{51} & & & & & & & \\ \hline
    
    \citet{133} & & \ding{51} & & \ding{51} & & & & & \ding{51} & & \ding{51} & \\ \hline
    
    \citet{134} & \ding{51} & & \ding{51} & & & & & & & & & \\ \hline
    
    \citet{135} & & \ding{51}& \ding{51}& \ding{51}& & & & & & &\ding{51} & \\ \hline 
    
    \citet{136} & & \ding{51} & & & \ding{51} & & & & \ding{51} & & & \\ \hline
    
    \citet{137} & & & & \ding{51} & & & & & \ding{51} & \ding{51} & & \\ \hline
    
    \citet{138} & & & & & \ding{51} & & & & & & & \\ \hline
    
    \citet{139} & \ding{51} & & & & & & & & \ding{51} & \ding{51} & & \\ \hline
    
    \citet{140} & & & & & & \ding{51} & & & & & & \\ \hline
    
    \citet{141} & \ding{51} & & & & & & & \ding{51} & & \ding{51} & & \\ \hline
    
    \citet{142} & \ding{51} & & & \ding{51} & & & & & & \ding{51} & & \\ \hline
    
    \citet{143} & \ding{51} & & & \ding{51} & & & & \ding{51} & \ding{51} & \ding{51} & \ding{51} & \\ \hline
    
    \citet{144} & & & & & & & \ding{51} & & \ding{51} & \ding{51} & & \\ \hline
    
    \citet{145} & & & & & \ding{51} & \ding{51} & & & & & & \\ \hline
    
    \citet{146} & & & & & & \ding{51} & & & & & & \\ \hline
    
    \citet{147} & \ding{51} & & & & & \ding{51} & & & \ding{51} & \ding{51} & \ding{51} & \\ \hline
    
    \citet{148} & & & \ding{51} & & & \ding{51} & & & \ding{51} & \ding{51} & \ding{51} & \\ \hline
    
    \citet{149} & & & & & & \ding{51} & & \ding{51} & \ding{51} & \ding{51} & & \\ \hline
    
    \citet{150} & & & & \ding{51} & & & & & & \ding{51} & & \\ \hline
    
    \citet{151} & & \ding{51} & & & \ding{51} & & & & \ding{51} & & & \\ \hline
    
    \citet{152} & \ding{51} & & & & & & & & \ding{51} & & & \\ \hline
    
     \citet{153} & & \ding{51} & & \ding{51} & \ding{51} & & & & & \ding{51} & & \\ \hline
    
    \citet{154} & & & & & \ding{51} & & & & & \ding{51} & & \\ \hline
    
    \citet{155} & \ding{51} & & \ding{51} & & & & & & & & & \\ \hline
    
    \citet{156} & \ding{51} & & & & & \ding{51} & & & \ding{51} & \ding{51} & \ding{51} & \\ \hline
    
    \citet{157} & & & & & \ding{51} & & & & & \ding{51} & & \\ \hline
    
    \citet{158} & & & & & & \ding{51} & & & & & & \\ \hline
    
    \citet{159} & \ding{51} & & & & & & & & \ding{51} & \ding{51} & & \\ \hline
    
    \citet{160} & \ding{51} & \ding{51} & & \ding{51} & &     \ding{51} & & & \ding{51} & & & \\ \hline
    
    \citet{161} & \ding{51}& &\ding{51} & & &\ding{51} & &\ding{51} &\ding{51} & &\ding{51} & \\ \hline 
    
    \citet{162} &\ding{51} & &\ding{51} & & & & & &\ding{51} &\ding{51} & & \\ \hline 
    
    \citet{163} & & & & & &\ding{51} & &\ding{51} &\ding{51} & & & \\ \hline 
     
     \citet{164} & \ding{51} & & \ding{51} & & & & \ding{51} & & \ding{51} & \ding{51} & & \\ \hline
    
    \citet{165} & & & & & & & & & & \ding{51} & & \\ \hline
    
    \citet{166} & \ding{51} & & \ding{51} & & & & & & & & & \\ \hline
    
    \citet{167} & \ding{51} & & & & & \ding{51} & & & & & & \\ \hline
    
    \citet{168} & \ding{51} & & & & & & & & \ding{51} & \ding{51} & & \\ \hline
    
    \citet{169} & \ding{51} & & & & & & & & \ding{51} & \ding{51} & \ding{51} & \\ \hline
    
    \citet{170} & & & & & \ding{51} & & & & & & \ding{51} & \\ \hline

    \citet{171} & \ding{51} & \ding{51} & \ding{51} & \ding{51} & & & & & & \ding{51} & & \\ \hline
    
    \citet{172} & & \ding{51} & & \ding{51} & & & \ding{51} & & \ding{51} & & & \\ \hline
    
    \citet{173} & \ding{51} & & & & & \ding{51} & & & & & & \\ \hline
    
    \citet{174} & \ding{51} & \ding{51} & \ding{51} & & & & & & \ding{51} & \ding{51} & & \\ \hline
    
    \citet{175} & & & & & & & & & \ding{51} & & \ding{51} & \\ \hline
    
    \citet{176} & & & & & & & & \ding{51} & \ding{51} & \ding{51} & \ding{51} & \\ \hline
    
    \citet{177} & \ding{51} & & & \ding{51} & & & & \ding{51} & \ding{51} & \ding{51} & \ding{51} & \\ \hline
    
    \citet{178} & \ding{51} & & & & & & & & \ding{51} & \ding{51} & \ding{51} & \\
    \hline
    
    \citet{179} & \ding{51} & \ding{51} & \ding{51} & \ding{51} & & & & & & & \ding{51} & \\
    \hline
    
    \citet{180} & \ding{51} & \ding{51} & & \ding{51} & & & & \ding{51} & \ding{51} & & \ding{51} & \\ \hline
    
    \citet{181} & & & & & & & & \ding{51} & \ding{51} & & & \\ \hline
    
    \citet{182} & \ding{51} & & & & & \ding{51} & & & & & & \\ \hline
    
    \citet{183} & & & & & & \ding{51} & & & & & & \\ \hline
    
    \citet{184} & & & & & & & & & \ding{51} & & \ding{51} & \\ \hline
    
    \citet{185} & \ding{51} & \ding{51} & & \ding{51} & & & & & \ding{51} & \ding{51} & & \\ \hline

\end{tabularx}
\vspace*{0pt}
\end{minipage}
\hspace{10pt}
\begin{minipage}{0.48\linewidth}
\begin{tabularx}{\linewidth}{|Y|c|c|c|c|c|c|c|c|c|c|c|c|}
    \hline
    
    \center{Paper} & \rotatebox{90}{QoS} & \rotatebox{90}{Cost} & \rotatebox{90}{Energy} & \rotatebox{90}{Bandwidth} & \rotatebox{90}{Security} & \rotatebox{90}{Foundation} & \rotatebox{90}{RAS} & \rotatebox{90}{Mobility} & \rotatebox{90}{Scalability} & \rotatebox{90}{Heterogeneity} & \rotatebox{90}{Management} & \rotatebox{90}{Programmability\hspace{3pt}} \\ \hline
    
    \citet{186} & & & & & & & & & \ding{51} & \ding{51} & & \\ \hline
    
    \citet{187} & & & & \ding{51} & & & & & \ding{51} & & & \\ \hline
    
    \citet{188} & \ding{51} & & \ding{51} & & & \ding{51} & & & & \ding{51} & & \\ \hline
    
    \citet{189} & & & & & & & & & & \ding{51} & \ding{51} & \ding{51} \\ \hline
    
    \citet{190} & \ding{51} & & & & & & & & & & & \\ \hline
    
    \citet{191} & & & & & \ding{51} & & & & & & & \\ \hline
    
    \citet{192} & \ding{51} & & & \ding{51} & & & & & \ding{51} & \ding{51} & & \\ \hline
    
    \citet{193} & \ding{51} & & & & & & & \ding{51} & & \ding{51} & & \\ \hline
    
    \citet{194} & \ding{51} & & & & & & & & \ding{51} & \ding{51} & & \\ \hline
    
    \citet{195} & \ding{51} & & \ding{51} & & & & & & & & & \\ \hline
    
    \citet{196} & \ding{51} & & & & & & & \ding{51} & \ding{51} & \ding{51} & & \\ \hline
    
    \citet{197} & & \ding{51} & & & & & & & \ding{51} & & & \\ \hline
    
    \citet{198} & \ding{51} & & & & & & & & \ding{51} & \ding{51} & \ding{51} & \ding{51} \\ \hline
    
    \citet{199} & & & & & & \ding{51} & & & & \ding{51} & & \\ \hline
    
    \citet{200} & \ding{51} & & & & & & & \ding{51} & & \ding{51} & & \ding{51} \\ \hline
    
    \citet{201} & & & & & & & & & & \ding{51} & \ding{51} & \ding{51} \\ \hline
    
    \citet{202} & \ding{51} & & & & & & & \ding{51} & \ding{51} & & & \\ \hline
    
    \citet{203} & \ding{51} & & & & & & & & \ding{51} & \ding{51} & & \\ \hline
    
    \citet{204} & & & & \ding{51} & & & & & & & \ding{51} & \\ \hline
    
    \citet{205} & & & & \ding{51} & & & & & \ding{51} & & \ding{51} & \\ \hline
    
    \citet{206} & & & & & & & & & & & & \ding{51} \\ \hline
    
    \citet{207} & \ding{51} & & & & & & & & \ding{51} & \ding{51} & & \\ \hline
    
    \citet{208} & & & & & & \ding{51} & & & \ding{51} & \ding{51} & \ding{51} & \\ \hline
    
    \citet{209} & & \ding{51} & & \ding{51} & & & & & \ding{51} & \ding{51} & & \\ \hline
    
    \citet{210} & \ding{51} & & \ding{51} & & & \ding{51} & & \ding{51} & & \ding{51} & \ding{51} & \\ \hline
    
    \citet{211} & & & & & & & & & \ding{51} & & & \\ \hline
    
    \citet{212} & & & & \ding{51} & & & & \ding{51} & & \ding{51} & & \\ \hline
    
    \citet{213} & & & & & & & & \ding{51} & & \ding{51} & & \ding{51} \\ \hline
    
    \citet{214} & & & & \ding{51} & & \ding{51} & \ding{51} & & \ding{51} & \ding{51} & & \\ \hline
    
    \citet{215} & \ding{51} & & & & & & & & & \ding{51} & & \\ \hline
    
    \citet{216} & \ding{51} & & \ding{51} & & & & \ding{51} & & & \ding{51} & & \\ \hline
    
    \citet{217} & \ding{51} & & & & & & & & \ding{51} & & & \\ \hline
    
    \citet{218} & & & & & & & & & \ding{51} & \ding{51} & \ding{51} & \\ \hline
    
    \citet{219} & \ding{51} & & & \ding{51} & & & & & & \ding{51} & & \\ \hline
    
    \citet{220} & \ding{51} & & & \ding{51} & & & & & \ding{51} & & & \\ \hline
    
    \citet{221} & \ding{51} & \ding{51} & & & & & & & \ding{51} & \ding{51} & \ding{51} & \\ \hline
    
    \citet{222} & \ding{51} & \ding{51} & & & & & & & & \ding{51} & \ding{51} & \\
    \hline
    
    \citet{223} & \ding{51} & & & & & & \ding{51} & & \ding{51} & \ding{51} & & \\ \hline
    
    \citet{224} & \ding{51} & & & & & \ding{51} & & & \ding{51} & \ding{51} & & \ding{51} \\ \hline
    
    \citet{225} & \ding{51} & & & \ding{51} & \ding{51} & & & & \ding{51} & \ding{51} & \ding{51} & \\ \hline
    
    \citet{226} & \ding{51} & & & & & & & & \ding{51} & \ding{51} & & \ding{51} \\ \hline
    
    \citet{227} & & & & & \ding{51} & & & & & & \ding{51} & \ding{51} \\ \hline
    
    \citet{228} & & & & \ding{51} & \ding{51} & & & & \ding{51} & \ding{51} & & \\ \hline
    
    \citet{229} & & \ding{51} & \ding{51} & \ding{51} & & & \ding{51} & & \ding{51} & & & \\ \hline
    
    \citet{230} & & & & & & \ding{51} & & & & \ding{51} & & \\ \hline
    
    \citet{231} & & \ding{51} & & & & & & & & \ding{51} & & \\ \hline
    
    \citet{232} & \ding{51} & & \ding{51} & & & & & & \ding{51} & & & \\ \hline
    
    \citet{233} & \ding{51} & & & \ding{51} & & & & & \ding{51} & \ding{51} & & \\ \hline
    
    \citet{234} & \ding{51} & \ding{51} & & \ding{51} & & & & \ding{51} & \ding{51} & & \ding{51} & \\ \hline
    
    \citet{235} & \ding{51} & & & & & \ding{51} & & & \ding{51} & \ding{51} & & \\ \hline
    
    \citet{236} & \ding{51} & \ding{51} & & \ding{51} & & & & & \ding{51} & & & \\ \hline
    
    \citet{237} & \ding{51} & & \ding{51} & & & & & & \ding{51} & & & \\ \hline
    
    \citet{238} & \ding{51} & & & & & & & & & \ding{51} & \ding{51} & \\ \hline
    
    \citet{239} & & & & & & & & \ding{51} & \ding{51} & & & \\ \hline
    
    \citet{240} & \ding{51} & & & \ding{51} & & & & & \ding{51} & & & \ding{51} \\ \hline
    
    \citet{241} & & & & \ding{51} & & & \ding{51} & & & \ding{51} & & \\ \hline
    
    \citet{242} & \ding{51} & & & \ding{51} & & & & & \ding{51} & & & \\ \hline
    
    \citet{243} & \ding{51} & \ding{51} & & & \ding{51} & & & & \ding{51} & \ding{51} & & \\ \hline
    
    \citet{244} & & & & & & & & & & \ding{51} & & \\ \hline
    
    \citet{245} & \ding{51} & & & & & & & & & \ding{51} & & \\ \hline
    
    \citet{246} & & & & & & \ding{51} & & & & & & \\ \hline
    
    \citet{247} & \ding{51} & & \ding{51} & \ding{51} & & & & & \ding{51} & & & \\ \hline
    
    \citet{248} & \ding{51} & \ding{51} & \ding{51} & & \ding{51} & & & & \ding{51} & \ding{51} & & \\ \hline

\end{tabularx}
\end{minipage} 

\end{table*}

\begin{table*}
\tiny
\centering
\newcolumntype{Y}{>{\raggedright\arraybackslash}X} 
\renewcommand\tabularxcolumn[1]{m{#1}} 
\renewcommand{\arraystretch}{1.5} 
\setlength\tabcolsep{3pt} 

\begin{minipage}{0.48\linewidth}
\begin{tabularx}{\linewidth}{|Y|c|c|c|c|c|c|c|c|c|c|c|c|}
    \hline
    
    \center{Paper} & \rotatebox{90}{QoS} & \rotatebox{90}{Cost} & \rotatebox{90}{Energy} & \rotatebox{90}{Bandwidth} & \rotatebox{90}{Security} & \rotatebox{90}{Foundation} & \rotatebox{90}{RAS} & \rotatebox{90}{Mobility} & \rotatebox{90}{Scalability} & \rotatebox{90}{Heterogeneity} & \rotatebox{90}{Management} & \rotatebox{90}{Programmability\hspace{3pt}} \\ \hline
    
    \citet{249} & & & & & & \ding{51} & & & & & & \\ \hline
    
    \citet{250} & & \ding{51} & & & & & & & & \ding{51} & \ding{51} & \\ \hline
    
    \citet{251} & \ding{51} & & & & & \ding{51} & & & & \ding{51} & \ding{51} & \\ \hline
    
    \citet{252} & \ding{51} & & & & & & & & \ding{51} & \ding{51} & \ding{51} & \\ \hline
    
    \citet{253} & \ding{51} & \ding{51} & & & & & & \ding{51} & & \ding{51} & \ding{51} & \\ \hline
    
    \citet{254} & \ding{51} & & & \ding{51} & & & \ding{51} & & \ding{51} & \ding{51} & & \\ \hline
    
    \citet{255} & \ding{51} & & & & & & & & & & & \\ \hline
    
    \citet{256} & \ding{51} & & & & & \ding{51} & & \ding{51} & \ding{51} & \ding{51} & & \\ \hline
    
    \citet{257} & & & & & \ding{51} & \ding{51} & & & & & & \\ \hline
    
    \citet{258} & \ding{51} & & \ding{51} & & & & & & \ding{51} & & & \\ \hline
    
    \citet{259} & \ding{51} & & & \ding{51} & & & & & \ding{51} & \ding{51} & & \\ \hline
    
    \citet{260} & & & & & & \ding{51} & & & & \ding{51} & & \\ \hline
    
    \citet{261} & \ding{51} & & & \ding{51} & & & & \ding{51} & \ding{51} & \ding{51} & \ding{51} & \\ \hline
    
    \citet{262} & \ding{51} & & & & & & \ding{51} & & \ding{51} & \ding{51} & & \ding{51} \\ \hline
    
    \citet{263} & \ding{51} & & & \ding{51} & & & & & & & & \\ \hline
    
    \citet{264} & \ding{51} & & & & & & & & & & & \\ \hline
    
    \citet{265} & \ding{51} & & & & & \ding{51} & & & \ding{51} & \ding{51} & & \\ \hline
    
    \citet{266} & \ding{51} & & & \ding{51} & & & & & \ding{51} & & \ding{51} & \\ \hline
    
    \citet{267} & & & & & \ding{51} & & & & & \ding{51} & \ding{51} & \\ \hline
    
    \citet{268} & \ding{51} & & & & & & & & \ding{51} & \ding{51} & & \\ \hline
    
    \citet{269} & & & & \ding{51} & & & & & \ding{51} & \ding{51} & \ding{51} & \\ \hline
    
    \citet{270} & & & & & & \ding{51} & & & & & & \\ \hline
    
    \citet{271} & & & & & & \ding{51} & & & & & & \\ \hline

    \citet{272} & \ding{51} & & & \ding{51} & & \ding{51} & & & & & & \\ \hline

    \citet{273} & \ding{51} & & & & & & & &\ding{51}  & &\ding{51}  & \ding{51} \\ \hline
    
    \citet{274} & \ding{51} & & & & & & & & \ding{51}  & & & \\ \hline
    
    \citet{275} & \ding{51} & & & & & & & & \ding{51} & \ding{51} & &\ding{51}  \\ \hline
    
    \citet{276} & \ding{51}  & & & & & & & & \ding{51} & & & \\ \hline
    
    \citet{277} & & \ding{51} & & & & & & & & \ding{51} & & \\ \hline
    
    \citet{278} & \ding{51} & & & & & & & & & & & \\ \hline
    
    \citet{279} & \ding{51} & & & & & & \ding{51} & & & \ding{51} & \ding{51} & \\ \hline
    
    \citet{280} & & & & & & & & \ding{51} & \ding{51} & & & \\ \hline
    
    \citet{281} & \ding{51} & \ding{51} & & & & & & \ding{51} & \ding{51} & & \ding{51} & \\ \hline
    
    \citet{282} & \ding{51} & & & \ding{51} & & & & & \ding{51} & \ding{51} & & \ding{51} \\ \hline
    
    \citet{283} & & & & & \ding{51} & \ding{51} & & & & & & \\ \hline
    
    \citet{284} & & & & & & \ding{51} & & & & & & \\ \hline
    
    \citet{285} & \ding{51} & & & & & & & & \ding{51} & & & \\ \hline
    
    \citet{286} & \ding{51} & & & & \ding{51} & & & & \ding{51} & & & \\ \hline

    \citet{287} & \ding{51} & \ding{51} & & & & & & \ding{51} & & & \ding{51} & \\ \hline
   
	\citet{288} & & & & & \ding{51} & & & & & \ding{51} & & \\ \hline
    
    \citet{289} & \ding{51} & & \ding{51} & & & & & \ding{51} & & & & \\ \hline

    \citet{290} & & & & & \ding{51} & \ding{51} & & & & & & \\ \hline
   
    \citet{291} & \ding{51} & & & & & & & & \ding{51} & & & \\ \hline
    
    \citet{292} & & & & & & \ding{51} & & & & & & \\ \hline
    
    \citet{293} & \ding{51} & & & \ding{51} & & & & & & & & \\ \hline
    
    \citet{294} & \ding{51} & & & & & & & & \ding{51} & \ding{51} & & \\ \hline
    
    \citet{295} & \ding{51} & \ding{51} & & & & & & & \ding{51} & \ding{51} & \ding{51} & \\ \hline
    
    \citet{296} & & & \ding{51} & \ding{51} & \ding{51} & & & & \ding{51} & & \ding{51} & \\ \hline

    \citet{297} & \ding{51} & & & & & & & & \ding{51} & & \ding{51} & \\ \hline
    
    \citet{298} & \ding{51} & & \ding{51} & \ding{51} & & & & \ding{51} & \ding{51} & & \ding{51} & \\ \hline
    
    \citet{299} & & & & & \ding{51} & & & & & & & \\ \hline

	\citet{300} & \ding{51} & & & \ding{51} & & & & \ding{51} & & & \ding{51} & \\ \hline

	\citet{301} & \ding{51} & \ding{51} & & & & & & \ding{51} & & & \ding{51} & \\ \hline

	\citet{302} & \ding{51} & \ding{51} & \ding{51} & \ding{51} & & & & & & \ding{51} & \ding{51} & \\ \hline

	\citet{303} & \ding{51} & & & \ding{51} & & & \ding{51} & & \ding{51} & & \ding{51} & \\ \hline

	\citet{304} & \ding{51} & & & & \ding{51} & & & & \ding{51} & & \ding{51} & \\ \hline

	\citet{305} & \ding{51} & & \ding{51} & \ding{51} & & & & & & & & \\ \hline

	\citet{306} & & & & \ding{51} & & \ding{51} & & & \ding{51} & & & \\ \hline

	\citet{307} & \ding{51} & \ding{51} & & \ding{51} & & & & & \ding{51} & \ding{51} & & \\ \hline

	\citet{308} & & & & & \ding{51} & & & & & & & \\ \hline

	\citet{309} & \ding{51} & & \ding{51} & & & & & \ding{51} & & & & \\ \hline

	\citet{310} & & \ding{51} & \ding{51} & & & & & & \ding{51} & & \ding{51} & \\ \hline

	\citet{311} & \ding{51} & & \ding{51} & \ding{51} & & & & & & & \ding{51} & \\ \hline

\end{tabularx}
\vspace*{0pt}
\end{minipage}
\hspace{10pt}
\begin{minipage}{0.48\linewidth}
\begin{tabularx}{\linewidth}{|Y|c|c|c|c|c|c|c|c|c|c|c|c|}
    \hline
    
    \center{Paper} & \rotatebox{90}{QoS} & \rotatebox{90}{Cost} & \rotatebox{90}{Energy} & \rotatebox{90}{Bandwidth} & \rotatebox{90}{Security} & \rotatebox{90}{Foundation} & \rotatebox{90}{RAS} & \rotatebox{90}{Mobility} & \rotatebox{90}{Scalability} & \rotatebox{90}{Heterogeneity} & \rotatebox{90}{Management} & \rotatebox{90}{Programmability\hspace{3pt}} \\ \hline

	\citet{312} & & & & & & & & & & \ding{51} & \ding{51} & \ding{51} \\ \hline

	\citet{313} & & & & & & & \ding{51} & \ding{51} & \ding{51} & \ding{51} & & \ding{51} \\ \hline

	\citet{314} & & & & & \ding{51} & \ding{51} & & & & & & \\ \hline

	\citet{315} & \ding{51} & & \ding{51} & & & & & & \ding{51} & & \ding{51} & \\ \hline

    \citet{316} & & & & & & & & & \ding{51} & & & \\ \hline

	\citet{317} & \ding{51} & & & & & & & & \ding{51} & & \ding{51} & \\ \hline

	\citet{318} & & \ding{51} & \ding{51} & & & & & & & & \ding{51} & \\ \hline

	\citet{319} & & & & & & \ding{51} & & & & & & \\ \hline

	\citet{320} & \ding{51} & & & & & & \ding{51} & & & & & \\ \hline

	\citet{321} & & & & & & \ding{51} & & & & & & \\ \hline

	\citet{322} & & & & \ding{51} & & & & & & \ding{51} & & \\ \hline

	\citet{323} & \ding{51} & & & & & & \ding{51} & & \ding{51}& \ding{51} & & \ding{51} \\ \hline

	\citet{324} & & & & & & & & & \ding{51} & & & \ding{51} \\ \hline

	\citet{325} & & & & & \ding{51} & & & & & & & \\ \hline
	
	\citet{326} & \ding{51} & & & & \ding{51} & & & & & & & \\ \hline
    
    \citet{327} & & & & & & & \ding{51} & \ding{51} & & \ding{51} & & \\ \hline
    
    \citet{328} & \ding{51} & \ding{51} & & & & & & & \ding{51} & \ding{51} & \ding{51} & \\ \hline
    
    \citet{329} & & \ding{51} & \ding{51} & & \ding{51} & & & & \ding{51} & \ding{51} & & \\ \hline
    
    \citet{330} & \ding{51} & & & & & & & & \ding{51} & & & \\ \hline
    
    \citet{331} & \ding{51} & & & & & & & & & \ding{51} & \ding{51} & \\ \hline
    
    \citet{332} & \ding{51} & & \ding{51} & & & & & & & & & \\ \hline
    
    \citet{333} & \ding{51} & \ding{51} & & \ding{51} & & & & & & \ding{51} & \ding{51} & \\ \hline
    
    \citet{334} & \ding{51} & \ding{51} & & & & & & & & & \ding{51} & \\ \hline
    
    \citet{335} & \ding{51} & & & & & & & & \ding{51} & \ding{51} & \ding{51} & \ding{51} \\ \hline
    
    \citet{336} & \ding{51} & & & & & & & & \ding{51} & & & \ding{51} \\ \hline
    
    \citet{337} & \ding{51} & \ding{51} & & & & & & \ding{51} & & \ding{51} & \ding{51} & \\ \hline
    
    \citet{338} & \ding{51} & & & & & & & & \ding{51} & \ding{51} & & \\ \hline
    
    \citet{339} & \ding{51} & & \ding{51} & & & & & & & & & \\ \hline
    
    \citet{340} & \ding{51} & & & \ding{51} & & & & & \ding{51} & \ding{51} & & \\ \hline
    
    \citet{341} & & \ding{51} & & & & & & & \ding{51} & & \ding{51} & \ding{51} \\ \hline
    
    \citet{342} & & & & & \ding{51} & & \ding{51} & & & & & \\ \hline
    
    \citet{343} & \ding{51} & & & & & & & & \ding{51} & & & \\ \hline
    
    \citet{344} & \ding{51} & & & & & & & \ding{51} & \ding{51} & & & \\ \hline
    
    \citet{345} & & & & & & & & & & & \ding{51} & \\ \hline
    
    \citet{346} & \ding{51} & & & & \ding{51} & & & & & & & \\ \hline
    
    \citet{347} & \ding{51} & & & & \ding{51} & & & & \ding{51} & & \ding{51} & \\ \hline
    
    \citet{348} & & & \ding{51} & \ding{51} & & & & & \ding{51} & \ding{51} & & \\ \hline
    
    \citet{349} & \ding{51} & \ding{51} & & & & & \ding{51} & & & \ding{51} & \ding{51} & \\ \hline
    
    \citet{350} & \ding{51} & & & \ding{51} & & & & & & \ding{51} & & \\ \hline
    
    \citet{351} & & & & & & & & & & \ding{51} & \ding{51} & \ding{51} \\ \hline
    
    \citet{352} & \ding{51} & & & \ding{51} & & & & \ding{51} & \ding{51} & \ding{51} & \ding{51} & \\ \hline
    
    \citet{353} & \ding{51} & & & \ding{51} & & & & & & & & \\ \hline
    
    \citet{354} & & & & \ding{51} & & & \ding{51} & & \ding{51} & \ding{51} & & \\ \hline
    
    \citet{355} & & \ding{51} & & & & & & & & \ding{51} & & \\ \hline
    
    \citet{356} & \ding{51} & \ding{51} & & & & & & & & \ding{51} & \ding{51} & \\ \hline
    
    \citet{357} & & & & \ding{51} & \ding{51} & & & & \ding{51} & & & \\ \hline
    
    \citet{358} & \ding{51} & & & & & & & & \ding{51} & & & \\ \hline
    
    \citet{359} & & & & & & \ding{51} & & & \ding{51} & & & \\ \hline
    
    \citet{360} & & \ding{51} & & & & \ding{51} & & & & \ding{51} & & \\ \hline
    
    \citet{361} & \ding{51} & \ding{51} & & & \ding{51} & & & & & & & \\ \hline
    
    \citet{362} & & & & & & & & & & & \ding{51} & \\ \hline
    
    \citet{363} & & & \ding{51} & & & & & & & & & \\ \hline
    
    \citet{364} & \ding{51} & \ding{51} & \ding{51} & & & & & & & \ding{51} & \ding{51} & \\ \hline
    
    \citet{365} & \ding{51} & & & \ding{51} & & & & & & \ding{51} & & \\ \hline
    
    \citet{366} & \ding{51} & \ding{51} & & & & & & & \ding{51} & & \ding{51} & \\ \hline
    
    \citet{367} & \ding{51} & & \ding{51} & \ding{51} & & & & & \ding{51} & \ding{51} & & \\ \hline
    
    \citet{368} & \ding{51} & & & & & & & & \ding{51} & & & \\ \hline
    
    \citet{369} & & & & & \ding{51} & & & & & & & \\ \hline
    
    \citet{370} & \ding{51} & & & & & & & & \ding{51} & & & \\ \hline
    
    \citet{371} & & & & & & & & & \ding{51} & & \ding{51} & \\ \hline
    
    \citet{372} & \ding{51} & & & & & & & & & \ding{51} & \ding{51} & \\ \hline
    
    \citet{373} & \ding{51} & \ding{51} & \ding{51} & & & & & & & \ding{51} & & \\ \hline
    
    \citet{374} & \ding{51} & & & & & & & & \ding{51} & \ding{51} & & \\ \hline

\end{tabularx}
\end{minipage} 

\end{table*}

\begin{table*}
\tiny
\centering
\newcolumntype{Y}{>{\raggedright\arraybackslash}X} 
\renewcommand\tabularxcolumn[1]{m{#1}} 
\renewcommand{\arraystretch}{1.5} 
\setlength\tabcolsep{3pt} 

\begin{minipage}{0.48\linewidth}
\begin{tabularx}{\linewidth}{|Y|c|c|c|c|c|c|c|c|c|c|c|c|}
    \hline
    
    \center{Paper} & \rotatebox{90}{QoS} & \rotatebox{90}{Cost} & \rotatebox{90}{Energy} & \rotatebox{90}{Bandwidth} & \rotatebox{90}{Security} & \rotatebox{90}{Foundation} & \rotatebox{90}{RAS} & \rotatebox{90}{Mobility} & \rotatebox{90}{Scalability} & \rotatebox{90}{Heterogeneity} & \rotatebox{90}{Management} & \rotatebox{90}{Programmability\hspace{3pt}} \\ \hline
    
    \citet{375} & & & & & & \ding{51} & & & & & \ding{51} & \\ \hline
    
    \citet{376} & \ding{51} & & & & & & \ding{51} & & \ding{51} & \ding{51} & & \\ \hline
    
    \citet{377} & \ding{51} & & & \ding{51} & & & & & \ding{51} & & & \ding{51} \\ \hline
    
    \citet{378} & \ding{51} & & & & & & & & \ding{51} & \ding{51} & \ding{51} & \ding{51} \\ \hline
    
    \citet{379} & \ding{51} & & & & & & & \ding{51} & \ding{51} & & \ding{51} & \\ \hline

\end{tabularx}
\vspace*{0pt}
\end{minipage}
\hspace{10pt}
\begin{minipage}{0.48\linewidth}
\begin{tabularx}{\linewidth}{|Y|c|c|c|c|c|c|c|c|c|c|c|c|}
    \hline
    
    \center{Paper} & \rotatebox{90}{QoS} & \rotatebox{90}{Cost} & \rotatebox{90}{Energy} & \rotatebox{90}{Bandwidth} & \rotatebox{90}{Security} & \rotatebox{90}{Foundation} & \rotatebox{90}{RAS} & \rotatebox{90}{Mobility} & \rotatebox{90}{Scalability} & \rotatebox{90}{Heterogeneity} & \rotatebox{90}{Management} & \rotatebox{90}{Programmability\hspace{3pt}} \\ \hline
    
    \citet{380} & & & & & \ding{51} & & & & \ding{51} & \ding{51} & & \\ \hline
      
	\citet{381} & \ding{51} & & & & & \ding{51} & & & & & & \\ \hline
    
    \citet{382} & & & & \ding{51} & & & & & \ding{51} & & \ding{51} & \\ \hline
    
    \citet{383} & \ding{51} & & & \ding{51} & & & & \ding{51} & \ding{51} & \ding{51} & & \\ \hline
    
    \citet{384} & \ding{51} & & & & & & & \ding{51} & \ding{51} & \ding{51} & & \\ \hline

\end{tabularx}
\end{minipage} 

\end{table*}

\end{document}